\documentclass[10pt,twocolumn,tightenlines,aps,rmp,amsmath,floatfix,notitlepage,nofootinbib,showpacs]{revtex4-1}

\usepackage[T1]{fontenc}        
\usepackage{palatino}           
\usepackage{amsmath}            
\usepackage{amssymb}            
\usepackage{graphicx}           
\usepackage{rotating}
\usepackage{booktabs}
\usepackage{hyperref}           
\usepackage{hypernat}
\usepackage{pslatex}
\usepackage{epsfig}
\usepackage{epsf}
\usepackage[usenames]{xcolor}
\usepackage{mathrsfs}           
\usepackage{mathtools}          

\newcommand{\rem}[1]{}          
\def\app#1#2{%
  \mathrel{%
    \setbox0=\hbox{$#1\sim$}%
    \setbox2=\hbox{%
      \rlap{\hbox{$#1\propto$}}%
      \lower1.1\ht0\box0%
    }%
    \raise0.25\ht2\box2%
  }%
}

\newcommand{\clJ}{\mathcal{J}}
\newcommand{\clD}{\mathcal{D}}
\newcommand{\clK}{\mathcal{K}}

\makeatletter
\def\p@subsection{}
\def\p@subsubsection{}
\makeatother

\setcitestyle{sort&compress,numbers,square,comma}

\begin{document}

\title{TOPICAL REVIEW\\ \ \\ Structure and spin dynamics of multiferroric BiFeO$_3$}

\author{
{\bf Je-Geun Park$^{1,2}$, Manh Duc Le$^{1,2}$, Jaehong Jeong$^{1,2}$, and Sanghyun Lee$^{3}$}
\\ \ \\
$^1$Center for Correlated Electron Systems, Institute for Basic Science (IBS), Seoul 151-747, Korea\\
$^2$Department of Physics and Astronomy, Seoul National University, Seoul 151-747, Korea\\
$^3$Institute of Materials Structure Science and J-PARC Center, KEK, Tsukuba 305-0801, Japan \\
\ \\
E-mail: jgpark10@snu.ac.kr, mducle@snu.ac.kr
}

\date{\today}

\begin{abstract}
{\large
{\bf Abstract}\\ \ \\
Multiferroic materials have attracted much interest due to the unusual coexistence of ferroelectric and (anti-)ferromagnetic ground states in a single compound. They offer an exciting platform for new physics and potentially novel devices. BiFeO$_3$ is one of the most celebrated of multiferroic materials with highly desirable properties. It is the only known room-temperature multiferroic with $T_\mathrm{C}\approx1100$~K and $T_\mathrm{N}\approx650$~K, and exhibits one of the largest spontaneous electric polarisation, $P\approx 80~\mu$C/cm$^2$. At the same time, it has a magnetic cycloid structure with an extremely long period of 630~\AA, which arises from a competition between the usual symmetric exchange interaction and antisymmetric Dzyaloshinskii-Moriya (DM) interaction. There is also an intriguing interplay between the DM interaction and the single ion anisotropy $K$. In this review, we have tried to paint a complete picture of bulk BiFeO$_3$ by summarising the structural and dynamical properties of both spin and lattice parts, and their magneto-electric coupling.
}
\end{abstract}


\maketitle
\tableofcontents


\section{Introduction} \label{sec-upd3-intro}

When a solid is cooled, it often assumes one or another ordered states including ferromagnetism and ferroelectricity amongst others~\cite{ChaikinLubensky}. Usually, these phase transitions are driven by only one of the four fundamental degrees of freedom of a solid: the electronic charge, spin or orbital angular momentum or the structure of the atomic lattice. The ordering is always accompanied by, and sometimes defined by the breaking of some symmetry elements of the solid. For example, ferromagnetism, the parallel ordering of electronic spins, breaks time reversal symmetry whereas ferroelectricity, a net ordering of dipolar charges, is produced by breaking inversion symmetry, such as by shifting charges off their centre of inversion. However, compared with the large number of materials showing only one such type of phase transitions, materials showing two types of transitions, such as both ferromagnetism and ferroelectricity, are few.

These \emph{multiferroics} were first studied in Russia by those interested in explaining the weak ferromagnetic component in Cr$_2$O$_3$, $\alpha$-Fe$_2$O$_3$, MnCO$_3$, and FeBO$_3$~\cite{Dzyaloshinskii1960,Astrov1960,Smolenskii1982,Venevtsev1994}. In doing so, Dzyaloshinskii famously discovered that there is a new type of exchange interaction for materials without inversion symmetry. Moriya adopted the phenomenological form of the interaction proposed by Dzyaloshinskii and derived, from the superexchange theory of Anderson, microscopic expressions for its strength~\cite{Dzyaloshinskii1958,Moriya1960}. This anisotropic exchange interaction, now known as the Dzyaloshinskii-Moriya (DM) interaction, can be written as $\mathcal{H}_{\mathrm{DM}} = \mathbf{D}\cdot\mathbf{S}_i \times\mathbf{S}_j$ with $\mathbf{D}$ being the DM vector and $\mathbf{S}_i$ the spin vector at the $i$ site. The DM interaction is antisymmetric upon exchange of two spins unlike the Heisenberg exchange interaction. While, in retrospect, this was an important point in the development of the theory of magnetism, foreshadowing later discoveries, it failed to attract the attention of the wider community outside Russia due to the relatively few materials having such behaviour, until the issue was taken up again in late 1990's and early 2000's by a number of people, notably Schmid and Spaldin (n\'ee Hill)~\cite{Schmid1994,Hill2000}.

This was the beginning of a renaissance, as noted by Spaldin and Fiebig~\cite{Spaldin2005}, in multiferroic research which is attested to by the ever growing number of new multiferroic materials and sheer volume of papers (For reviews, see the following papers:~\cite{Eerenstein2006,Cheong2007,Ramesh2007,Fiebig2005,Khomskii2006,Kimura2007,Tokura2010,Noda2008}): for example, as of this writing, we counted over 3000 papers published on BiFeO$_3$ alone listed by the Web of Science, most written within the past ten years.

As Moriya showed, the microscopic origin of the DM interaction lies in the relativistic spin-orbit coupling, so its strength should scale with atomic number as $Z^4$. Whilst all multiferroics thus far discovered involves $3d$ transition metal ions, where the spin-orbit interaction is expected to be small, it will be interesting to find what effects a larger spin-orbit coupling could have on multiferroic materials where the magnetism arises from $4d$ or $5d$ electrons, given the strong recent interest in spin-orbit physics~\cite{Khomskii2014}.

The DM interaction is also important to multiferroics as being the microscopic origin of the magnetoelectric coupling which gives rise to the magnetically induced ferroelectricity in perovskite manganites~{\cite{Sergienko2006,Arima2006prl}}. In these materials, the ferroelectric ordering accompanies a magnetic spiral order. Whilst the spiral itself is caused by competition between the superexchange and DM interaction, it is the DM interaction that stabilises the oxygen displacements, causing the net electric polarisation~\cite{Cheong2007}. This mechanism has subsequently been termed the \emph{inverse Dzyaloshinskii-Moriya interaction}, the "inverse" coming from the sense that the DM interaction causes inversion symmetry breaking rather than a non-centrosymmetric bond giving a nonzero DM interaction. 

An equivalent approach, called the \emph{spin current} mechanism~\cite{Katsura2005} also gives a net ferroelectric polarisation in conjunction with a magnetic spiral, but has a different microscopic origin. The spin current $\mathbf{j}\propto \mathbf{S}_i\times \mathbf{S}_j$ arises from the spiral, and this current induces a net polarisation $\mathbf{P}\perp\mathbf{j}$, analogous to the magnetic moment induced by a charge current. In both these mechanisms, a magnetic spiral order breaks inversion symmetry and allows the emergence of an electric polarisation~\cite{Cheong2007}.

The many multiferroic compounds so far discovered can be divided into two groups~\cite{Khomskii2009}: type I (or "proper") multiferroics where there is coexistence of magnetism and ferroelectricity but each arises from different sources; and type II ("improper") multiferroics where a magnetic order causes ferroelectricity~{\cite{note1}}. The first group contains the "older" multiferroics, including BiFeO$_3$, studied by early researchers in Russia, whilst the second group encompasses compounds such as the perovskite manganites (e.g. TbMnO$_3$~\cite{Kimura2003}), or MnWO$_4$~\cite{Heyer2006} which have spiral magnetic structures that break inversion symmetry, or mixed valent materials such as LuFe$_2$O$_4$~\cite{Ikeda2005} or Ca$_3$CoMnO$_6$~\cite{Choi2008}, where the ferroelectric polarisation comes from the magnetically induced polar charge ordering.

Of the many multiferroic systems thus found, several families in particular have drawn intense interests: hexagonal RMnO$_3$, orthorhombic (perovskite) RMnO$_3$ and RMn$_2$O$_5$~\cite{Noda2008}, and BiFeO$_3$. While the orthorhombic RMnO$_3$ and RMn$_2$O$_5$ are attractive because they are type-II multiferroics with strong magnetoelectric couplings, the ordering occurs at quite low temperatures (<77K). BiFeO$_3$ and the hexagonal RMnO$_3$, on the other hand, have much larger transition temperatures. Since we will discuss in detail of physical properties of BiFeO$_3$ later in the article, we should like to make some brief remarks here on the hexagonal RMnO$_3$. The microscopic magnetoelectric coupling was first demonstrated by second harmonic spectroscopy in HoMnO$_3$~\cite{Fiebig2002}, and later found to arise from a large magnetoelastic coupling, which is most likely to be related to the geometrical frustration inherent in the two-dimensional triangular lattice of Mn moments~\cite{Lee2008_RMO}. More recently, it was shown that the spin waves of this triangular lattice exhibit the hitherto unobserved magnon breakdown due to nonlinear magnon-magnon interaction~\cite{Oh2013}.

Whilst the study of multiferroics address many important questions of scientific interest, the key motivation behind many recent efforts has been the possible applications of this class of materials which potentially allow devices to manipulate magnetic moments by electric fields or vice versa~\cite{Ishiwara2012}. For this purpose, however, a material which is multiferroic at room temperature or above is essential. Thus far, only BiFeO$_3$ satisfies this criterion, and over the past decade, there have been many successful realisation of this possibility in the form of thin film devices~\cite{Lebeugle2009,Wang2003}. Another possible candidate for room temperature applications are the hexaferrites, (Ba,Sr)$_2$(Mg,Zn)$_2$Fe$_{12}$O$_{22}$~\cite{Ishiwata2008,Kimura2005,Chun2010}, whose transitions are lower than room temperatures but exhibit noticeable magnetoelectric coupling near room temperature.

\begin{figure}
  \begin{center}
    \includegraphics[width=0.9\columnwidth]{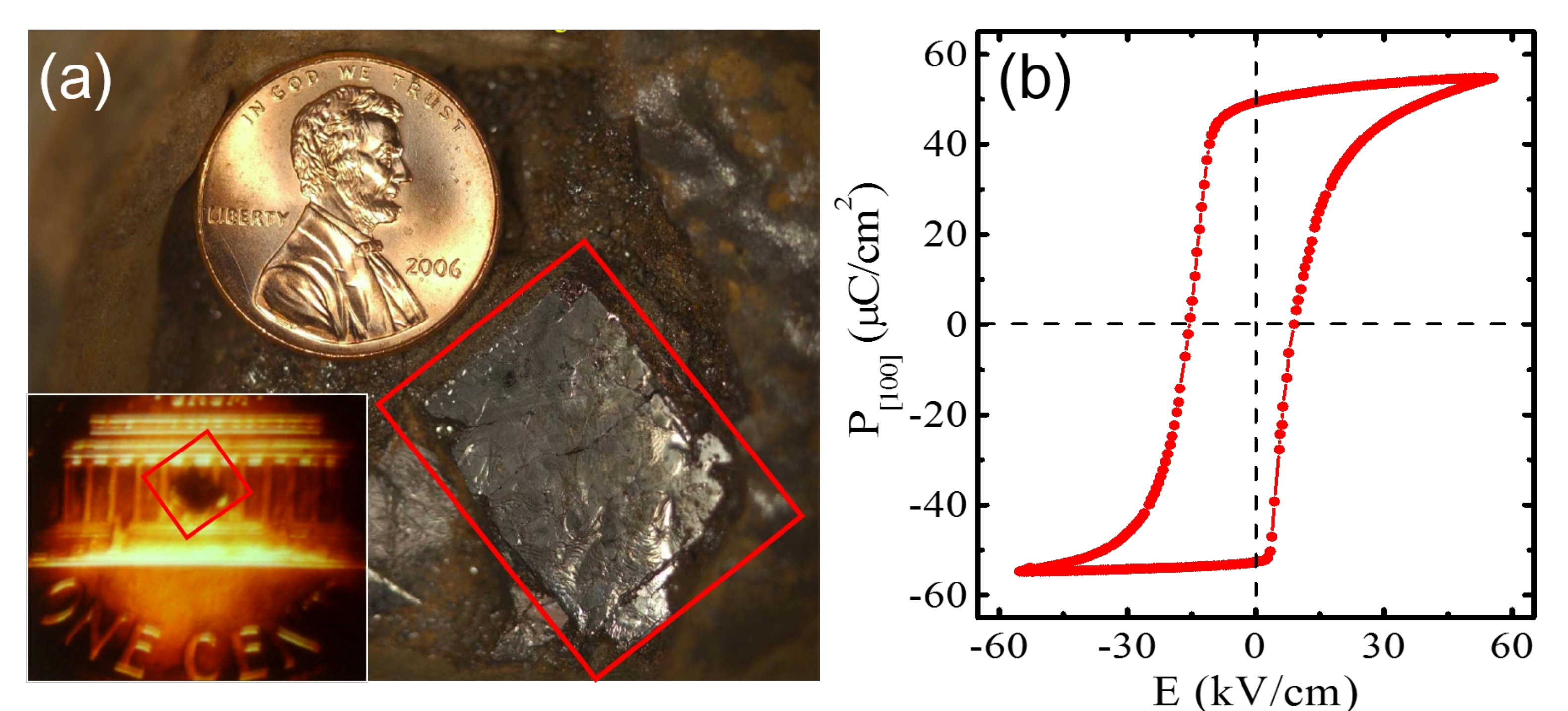}
    \caption{(Color online) (a) Single crystal of BiFeO$_3$ with a volume of $\sim$100~mm$^3$ and a mass of $\sim$1~g grown by flux method~\cite{CheongPC}. Inset: Typical size of crystals available for the last 50 years since 1950s~\cite{Royen1957}. (b) Ferroelectric hysteresis loop~\cite{CheongPC}. 
}
    \label{fg:intro1} \end{center}
\end{figure}

The current strong interest in BiFeO$_3$ can be traced back to two seminal works, the first of which is a thin film study~\cite{Wang2003} by the group of Ramesh, which reported a large electric polarisation $\approx$60~$\mu $C/cm$^2$ in stark contrast to previous studies (P$\lesssim$10~$\mu $C/cm$^2$) and also a large ferromagnetic moment $\approx$1~$\mu_B$/Fe, both attributed to epitaxial strain effects in the thin films. Whilst the large polarisation was subsequently confirmed and also found in bulk single crystals~\cite{Lebeugle2007prb}, the large moment was later revealed to be not intrinsic~\cite{Eerenstein2005,Wang2005}, with the consensus being that there may be much weaker ferromagnetic moment in thin film BiFeO$_3$ of $\approx$0.02~$\mu_B$/Fe~\cite{Bea2007}. The second important contribution is the report by Cheong's group of a large photovoltaic effect in single crystal BiFeO$_3$~\cite{Choi2009}, which also reported the measurement of the currently accepted intrinsic polarisation $P=$86~$\mu $C/cm$^2$, in line with theoretical predictions. Most of the measurements of the main subject of this review, in particular, benefit from the large, high quality single crystals which have only recently been synthesised. Large crystals are especially needed for the inelastic neutron scattering measurements~\cite{Jeong2012} described latter in this review, which simply could not be carried out on the small samples previously available. For example, we show in Figure~\ref{fg:intro1} the crystals grown by Cheong's group compared to those available before since the 1950's.

BiFeO$_3$ adopts the noncentrosymmetric $R3c$ space group at room temperatures with a ferroelectric transition at $T_\mathrm{C}\approx1100$~K and an antiferromagnetic phase transition at $T_\mathrm{N}=650$~K~\cite{Moreau1971,Roginskaya1966,Kaczmarek1974,Kiselev1962}. From early on, the nature of the ferroelectric transition and the space group of the paraelectric phase were examined by several groups with as many as 6 proposals being put forward~\cite{Haumont2006,Selbach2008,Kornev2007,Arnold2009,Haumont2008}. As we will discuss in detail in Section~\ref{sec-struct-3-2}, the general consensus emerging is that the paraelectric phase structure has the orthorhombic $Pbnm$ space group symmetry. The main driving force of the ferroelectric phase transitions seems to be the stereochemical activity of the Bi $6s$ lone-pair, resulting in a displacement of Bi and O sublattices. However, we should also note that high-resolution neutron diffraction studies on high quality single crystals showed that the Fe sublattice is also displaced and gives a sizeable contribution to the total electric polarization~\cite{Lee2013}.

This Fe displacement thus appears to be the natural source of the magnetoelectric coupling in this material, via the exchange-striction mechanism below the onset of the ordering of the Fe moments at $T_N$=650~K. The largest Fe-Fe exchange interaction is the nearest neighbour superexchange which is antiferromagnetic and leads to a simple G-type magnetic ordering~\cite{Kiselev1962}. In addition, there is also an antisymmetric DM interaction due to the broken inversion symmetry in the ferroelectric phase, which produces a cycloid magnetic structure with an exceptionally long period $\approx$630~\AA~\cite{Sosnowska1995}. It is important to note that this cycloid structure causes a cancellation of a linear magnetoelectric effect but that when it is suppressed by epitaxial strain or high magnetic fields, a such a linear effect ensues~\cite{Popov1993}. The presence of a single ion anisotropy (SIA) should render the cycloid anharmonic. However, neutron diffraction measurements found this to be quite small and below the detectability threshold of current instruments~\cite{Sosnowska2011}. A further interesting aspect of the magnetic structure is a spin density wave (SDW) induced by a second effective DM vector alternating along the $c$-axis as seen by polarized SANS measurements~\cite{Ramazanoglu2011}. However, the canting angle estimated from the SANS experiment is at least one order of magnitude larger than that obtained by fitting the spin wave spectrum measured by inelastic neutron scattering (INS), which requires further examination~\cite{Jeong2014}.

These INS measurements, described in Section~\ref{sec-excitations}, took advantage of the large single crystals now available~\cite{Jeong2012,Matsuda2012}. It was found that the high-energy spin dynamics is remarkably consistent with a G-type antiferromagnetic ordering~\cite{Jeong2012}. Only when one begins to measure the low-energy spin dynamics below $\approx$10 meV, can one then see the effects due to the SIA, DM interactions and the cycloid. There is also a rich interplay between the DM vectors and single ion anisotropy to be seen in the low-energy spin dynamics as we will discuss~\cite{Jeong2014}.

This article consists of three main parts: in the first part we will outline the theoretical models used to describe the physical properties of BiFeO$_3$, and the connections between them. The structural aspects of the multiferroic orderings, namely the PE-FE transition, the magnetic transition, and the magnetoelectric coupling, are discussed in the following section. We then move on to describe the spin dynamics, with a summary of the inelastic neutron scattering and optical and THz spectroscopy measurements and the linear spin wave theory used to describe them. We will also comment briefly on the phonon dispersion before discussing the hybrid spin-lattice excitations called \emph{electromagnons} in BiFeO$_3$. In the final section, we will summarize the key results and give an outlook for future work. We note that whilst much current research on BiFeO$_3$ is concentrated on thin films, we shall primarily focus on the properties of bulk BiFeO$_3$ in this review.

\section{Theoretical overview} \label{sec-theory}

The current theoretical understanding of ferroelectricity and magnetism in BiFeO$_3$ is derived from several complementary techniques, including both empirical and first principles approaches. We begin this section by outlining the different types of calculations and their contributions to knowledge of the structure and dynamics of BiFeO$_3$, and the magneto-electric coupling of this multiferroic material. 

Extensive work has been carried out using \emph{ab initio} methods in the framework of density functional theory (DFT), which allows the accurate determination of ground state properties at zero temperature, including the atomic and magnetic structures and origin of the electric polarisation. Metastable phases may also be determined, and thus phase transitions as functions of pressure, epitaxial strain or temperatures predicted. Lattice dynamics may also be calculated using either linear response theory or from the determination of the force constants between the atoms. One important early contribution~\cite{Neaton2005} which \emph{ab initio} calculations made to the study of BiFeO$_3$ is to support the large polarisation measured in thin films~\cite{Wang2003} which was some twenty times larger than the polarisation previously measured in bulk samples~\cite{Teague1970}. The calculations also determined, in the framework of the `modern' (Berry-phase) theory of polarisation, that this large polarisation is intrinsic rather than the result of epitaxial strain. The discrepancy with previously measured values was attributed to either sample quality or different \emph{switching paths}, since the spontaneous polarisation is defined as half the difference in the polarisation of two states of opposite polarity via some pathway. Details of the calculations are reviewed in Ref.~\cite{Picozzi2009}, and are supported by latter works~\cite{Ravindran2006,Dieguez2011}.

DFT calculations may also be used to determine the parameters for \emph{effective Hamiltonian} calculations~\cite{Kornev2007,Kornev2009,Albrecht2010,Rahmedov2012}. In this approach, the total energy is written as a function of the system variables (such as the atomic displacements and magnetic moments) with the important couplings between these variables determined \emph{a priori}. The total energy at finite temperatures and magnetic fields can then be calculated from a large \emph{Monte Carlo} sample of the system variables. Alternatively, phase transitions may be calculated through Landau-Ginzburg theory~\cite{Zhdanov2006,Tokunaga2010,Gareeva2013}, where the free energy is expanded in powers of the system order parameters. 

Finally the spin dynamics can be accurately modelled (1) by linear spin wave theory~\cite{Jeong2012,Fishman2012,Fishman2013,Fishman2013a}, (2) by including the effective Hamiltonian in the Landau-Liftshitz-Gilbert equation of motion~\cite{Wang2012}, or (3) by including the Landau-Ginzburg free energy in the Lagrangian and solving the resulting equations of motion~\cite{DeSousa2008,Zvezdin2009b}. Linear spin wave theory takes as its starting point a microscopic spin Hamiltonian, whose parameters, including the exchange interactions and anisotropies, may be deduced by fitting to experimental spectroscopic measurements. These parameters are equivalent to some of the coefficients used in the Landau-Ginzburg free energy, as we will show, and may also be determined from DFT calculations. The magnitude of the competing interactions governs the physical properties of BiFeO$_3$, but also allows the magneto-electric coupling mechanism and its strength to be elucidated, as discussed in the final part of this section. First, however, we shall discuss in detail the different theoretical methods employed to study BiFeO$_3$.

\subsection{Theoretical methods} \label{sec-theory-methods}

\begin{figure}
  \begin{center}
    \includegraphics[width=0.9\columnwidth,viewport=14 236 590 774]{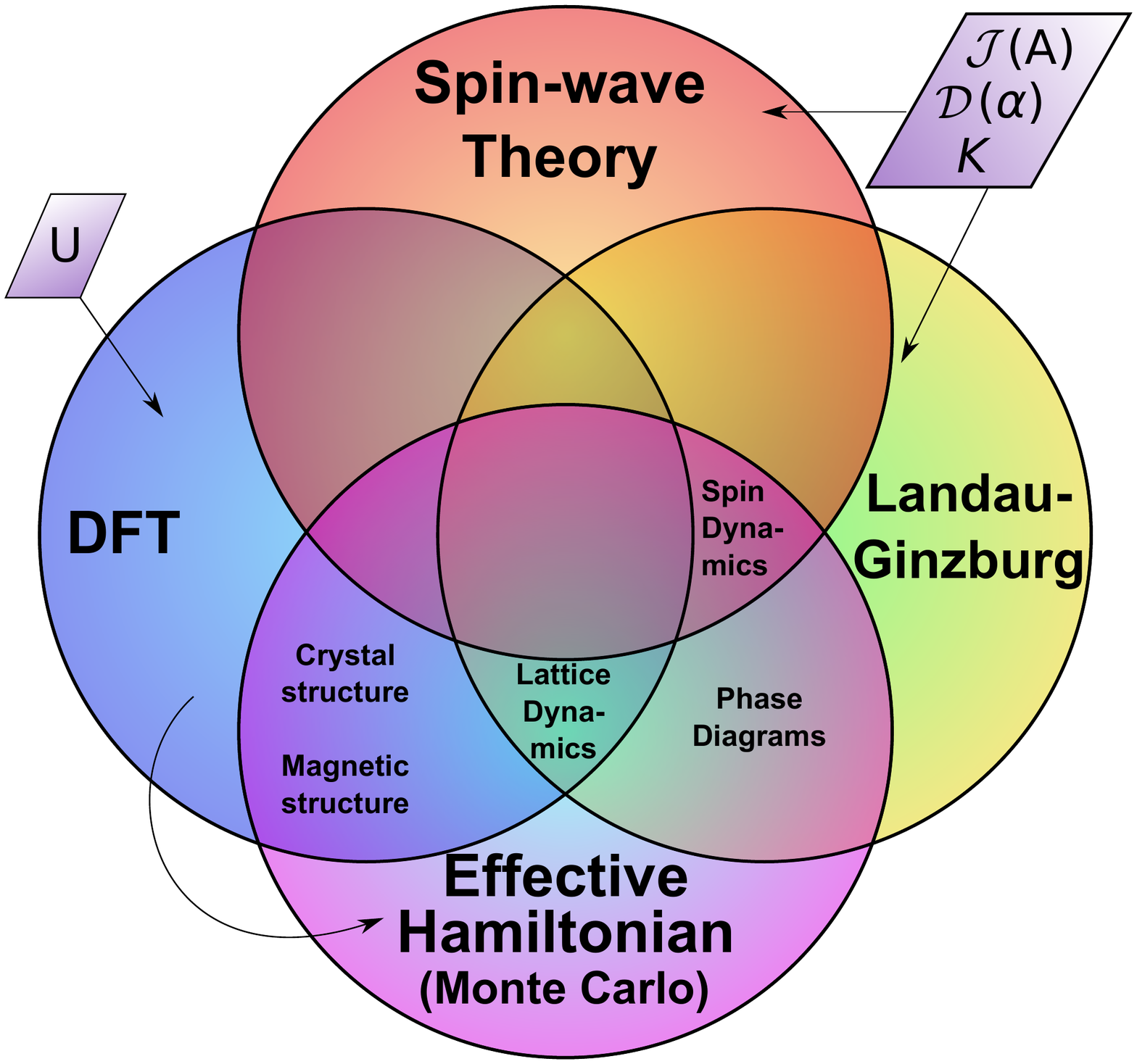}
    \caption{
The different theoretical techniques used to study BiFeO$_3$, indicated by the circles. The physical properties that can be calculated are indicated within the intersections of the techniques, whilst the input required by each techniques are indicated by arrows. For example, often a Coulomb $U$ parameter is required for DFT calculations which can then be used to calculate crystal structures or lattice dynamics and in turn provides input for effective Hamiltonian calculations of phase diagrams.
}
    \label{fg:theorymethods} \end{center}
\end{figure}

The different theoretical techniques, their inputs and what experimental observables may be calculated by them are summarised in Figure~\ref{fg:theorymethods}. Whilst density functional theory (DFT) is \emph{in principle} parameterless, calculations for magnetic insulators invariably have to add an on-site Coulomb energy $U$ in order to account for the strong correlation. A value of $U\approx4$~eV is accepted in the literature for BiFeO$_3$, although values ranging from 2 to 6~eV have been used. A larger $U$ is required to match the calculated bandgap to that measured, but overestimates the ordered moment, whereas a smaller $U$ underestimates the bandgap. Both the local density (LDA+$U$) and generalised gradient (GGA+$U$) approximations to DFT and their spin polarised versions (LSDA+$U$ and $\sigma$GGA+$U$ respectively) have been used to study BiFeO$_3$. For L(S)DA+$U$ calculations, the all-electron projector augmented wave (PAW) method is often used for efficiency and accuracy reasons, whilst for ($\sigma$)GGA+$U$ calculations the PBE (Perdew-Burke-Ernzehoff) functional is most commonly employed. Di\'eguez {\it et al.}~\cite{Dieguez2011} found that the PBE functional adapted to solids (PBEsol) provided the most accurate calculation of the relative energies of the ground state and metastable structures. This sensitivity to small structural differences, which can affect the calculated polarisation, prompted Ravindran {\it et al.}~\cite{Ravindran2006} to use the GGA (PBE) yielding $P=88.7~\mu$C/cm$^2$. It should be pointed out, however, that very similar spontaneous polarisation was also calculated using PAW pseudopotentials in the LDA ($P=84.2~\mu$C/cm$^2$)~\cite{Ederer2005} and LSDA ($P=98.9~\mu$C/cm$^2$)~\cite{Neaton2005}, with the spread in calculated results much smaller than that in the reported experimentally measured polarisation ($P\approx8-153~\mu$C/cm$^2$). On the other hand, Goffinet {\it et al.}\cite{Goffinet2009} found that the Born effective charges, and hence polarisation, is quite sensitive to the optimised crystal structure used in the calculation, which turns out to be different for each functional used. They suggested that a hybrid functional, B1-WC, which mixes an exact exchange term into the Wu-Cohen exchange correlation functional in the GGA, best reproduces the experimental structure and so may be more suitable.

Since DFT is a zero-temperature method and restricted, under computational constraints, to fairly small unit cells, other techniques must be used to calculate the phase diagram. The most successful has been the \emph{effective Hamiltonian} \rem{|\ref{fg:theorymethods})|} method as used by Bellaiche {\it et al.}~\cite{Kornev2007,Kornev2009,Albrecht2010,Rahmedov2012}. In this method, the total energy surface is expressed as an expansion in the system variables (as opposed to an order parameter as in Landau-Ginzburg theory) with DFT calculations of discrete points on this energy surface determining the coefficients in this expansion. For BiFeO$_3$ the variables are the Fe magnetic moment ($\mathbf{m}$), the FeO$_6$ octahedra rotation ($\omega$), and the soft mode displacement ($\mathbf{u}$) and strain ($\eta$) associated with the ferroelectric order. Phase transitions are determined numerically by calculating the total energy $E_{\mathrm{tot}}({\mathbf{m}_i,\omega_i,\mathbf{u}_i,\eta_i})$ for a large lattice of several thousand sites $i$ at particular values of temperatures or applied magnetic or electric fields. $E_{\mathrm{tot}}$ is then minimised using the Metropolis Monte Carlo algorithm in order to determine the stable phase (system configuration) at that field or temperature. With the addition of terms coupling the octahedral (antiferro-distortive) rotation $\omega$~\cite{Albrecht2010} and soft mode displacement $\mathbf{u}$~\cite{Rahmedov2012} with the antisymmetric exchange (Dzyaloshinskii-Moriya) interactions, the effective Hamiltonian method was successfully used to model both the cycloidal magnetic structure~\cite{Rahmedov2012} in bulk BiFeO$_3$ and the weak ferromagnetism due to spin canting in thin films~\cite{Albrecht2010}. Finally, by including the effective magnetic field $\mathbf{B}_{\mathrm{eff}}^i = \partial E_{\mathrm{tot}} / \partial \mathbf{m}_i$ into the Landau-Liftshitz-Gilbert equation of motion, the dynamics of BiFeO$_3$ may also be calculated~\cite{Wang2012}, including the energies of both magnons and phonons.

The other main phenomenological approach used to determine the phase transitions in BiFeO$_3$ is the Landau-Ginzburg (L-G) theory, where the critical temperatures and fields are determined analytically from minimising the free energy, expressed as a series expansion in the order parameters of the system, by means of the Euler-Lagrange equations. This method has been used extensively by Zvezdin {\it et al.}~\cite{Popov1993,Sparavigna1994,Sosnowska1995,Tehranchi1997,Zhdanov2006,Zvezdin2009} to determine the critical electric and magnetic fields for transition from the cycloidal magnetic structure to a canted antiferromagnet~\cite{Popov1993,Sparavigna1994}, the cycloid period~\cite{Sosnowska1995}, and the phase diagram as a function of magnetic field and anisotropy~\cite{Tehranchi1997,Gareeva2013}. The Landau free energy may also be included in the Lagrangian and used to determine the equations of motion of the system and thus to calculate the spin and lattice dynamics~\cite{DeSousa2008,Zvezdin2009b}.

Finally, the magnetic excitations are mostly easily calculated using a microscopic spin Hamiltonian in linear spin wave theory~\cite{Jeong2012,Fishman2012,Xu2012,Fishman2013,Fishman2013a}. The form of the Hamiltonian most commonly used for BiFeO$_3$ is

\begin{widetext}
\begin{equation} \label{eq:lswH}
\mathcal{H} = \mathcal{J} \sum_{\mathbf{r},\mathbf{\alpha}} \mathbf{S}_{\mathbf{r}} \cdot \mathbf{S}_{\mathbf{r}+\mathbf{\alpha}} + \mathcal{J}' \sum_{\mathbf{r},\mathbf{\beta}} \mathbf{S}_{\mathbf{r}} \cdot \mathbf{S}_{\mathbf{r}+\mathbf{\beta}} 
-\mathcal{D}_u \sum_{\mathbf{r}} \hat{\mathbf{u}} \cdot ( \mathbf{S}_{\mathbf{r}} \times \mathbf{S}_{\mathbf{r} + a \hat{\mathbf{v}}} )
-\mathcal{D}_c \sum_{\mathbf{r}} (-1)^{6 \mathbf{r}\cdot\hat{\mathbf{c}}/c}~ \hat{\mathbf{c}} \cdot ( \mathbf{S}_{\mathbf{r}} \times \mathbf{S}_{\mathbf{r} + \frac{c}{2} \hat{\mathbf{c}}} ) 
-\mathcal{K} \sum_{\mathbf{r}} ( \mathbf{S}_{\mathbf{r}} \cdot \hat{\mathbf{c}} )^2 ,
\end{equation}
\end{widetext}

\noindent where $\boldsymbol{\alpha}$ and $\boldsymbol{\beta}$ refer respectively to nearest and next-nearest neighbour iron atoms, and $\bold{u}$, $\bold{v}$ and $\bold{c}$ refer to the Cartesian directions which defines the cycloid with the propagation direction along $\bold{v} || [1\bar{1}0]_{\mathrm{pc}}$ (in pseudocubic (pc) notation, $[110]_{\mathrm{hex}}$ in hexagonal notation), and the moment initially align parallel or antiparallel to $\bold{c} || [111]_{\mathrm{pc}}$ ($[001]_{\mathrm{hex}}$). The first and second terms are the diagonal Heisenberg exchange interactions, the third and fourth are the antisymmetric Dzyaloshinskii-Moriya (DM) exchange terms, and the final terms is the (easy-axis) single ion anisotropy. In order to account for the cycloid structure, the spins are expressed in a rotating coordinate frame such that in this frame they effectively form a ferromagnetic state. Then the usual steps in linear spin wave theory, namely bosonization of the Hamiltonian Eq.~\ref{eq:lswH} by using the Holstein-Primakoff~\cite{HolsteinPrimakoff} transformation and diagonalization in momentum space using the Bogoliubov transformation, are carried out yielding the excitation energies $\hbar\omega(\mathbf{q})$, and for inelastic neutron scattering, the scattering function $S(\mathbf{q},\omega)$. The particular application of spin wave theory to inelastic neutron scattering studies of BiFeO$_3$ will be discussed in more detail in Section~\ref{sec-excitations}, where the theory itself will also be presented in some detail. Finally, the Hamiltonian Eq.~\ref{eq:lswH} can also be used in a classical spin Monte Carlo calculation to obtain physical bulk properties such as the magnetisation and spin flop transition at high fields~\cite{Park2011,Ohoyama2011}.

Let us now compare the spin Hamiltonian Eq.~\ref{eq:lswH} with the usual expression for the Landau-Ginzburg free energy often used to describe BiFeO$_3$,

\begin{widetext}
\begin{equation} \label{eq:LGfe}
\mathit{f}_l = \int d\mathbf{r} \ A(\nabla \mathbf{l})^2 + \lambda\mathbf{l}^2 + G\mathbf{l}^4 - \alpha\mathbf{P}\cdot\left[ \mathbf{l}(\nabla\cdot\mathbf{l})+\mathbf{l}\times(\nabla\times\mathbf{l})\right] - 2\beta M_0 \mathbf{P}\cdot\left(\mathbf{m}\times\mathbf{l}\right) - K_u l_c^2 + aP_c^2 + bP_c^4 + cP_c^6
\end{equation}
\end{widetext}

\noindent where $M_0$ is the magnitude of the magnetization vector $\mathbf{M}_i$ of sublattices, $\mathbf{P}$ is the spontaneous polarization vector, and $\mathbf{m}= \left(\mathbf{M}_1+\mathbf{M}_2\right)/2M_0$ and $\mathbf{l}=\left(\mathbf{M}_1-\mathbf{M}_2\right) /2M_0$ are the magnetization and antiferromagnetic unit vectors, respectively. The first three terms denote the exchange interactions; the fourth term is the Lifshitz invariant which results in the cycloid, the fifth term is the Dzaloshinskii-Moriya-like invariant which gives a weak ferromagnetism~\cite{Kadomtseva2004}, and both terms are related to the microscopic Dzyaloshinskii-Moriya interaction~\cite{Zvezdin2012}; the sixth term is the single ion anisotropy and the last three terms comprise the free energy associated with the ferroelectric polarisation $\mathbf{P}$. In addition to these terms, the coupling of moments to a magnetic field, $\mathbf{M}\cdot\mathbf{H}$, and the polarisation to an electric field, $\mathbf{P}\cdot\mathbf{E}$ may also be included in order to determine the phase diagram.

The equivalence of the parameters $\mathcal{J}$, $\mathcal{D}_u$ and $\mathcal{K}$ in the spin Hamiltonian Eq.~\ref{eq:lswH} with $A$, $\lambda$, $\alpha$ and $K_u$ in the free energy Eq.~\ref{eq:LGfe} may be determined \rem{|~\ref{eq:lswH}|}\rem{|~\ref{eq:LGfe}|} by equating the ground state energy obtained for particular configurations (such as a collinear AFM, harmonic or anharmonic cycloid structure) given by the two approaches as detailed in Appendix~\ref{sec-ap-equivalence}| and are given by,

\begin{align} \label{eq:equiv_lambda} \lambda &=  \frac{6S^2}{V} \left(-3\mathcal{J}+6\mathcal{J'}\right) \\ 
\label{eq:equiv_AJ} A &= \frac{6S^2}{V} a_{hex}^2 \left(\frac{\mathcal{J}-4\mathcal{J'}}{4}\right) \\ 
\label{eq:equiv_aDu} \alpha &= \frac{6S^2}{V} \frac{1}{P_s} a_{hex}  \mathcal{D}_u \\  
\label{eq:equiv_bDc} \beta &= \frac{6S^2}{V} \frac{1}{M_0 P_s} \mathcal{D}_c \\  
\label{eq:equiv_K} K_u &= \frac{6S^2}{V} \mathcal{K} ,
\end{align}

\noindent where $V$ is the volume of the unit cell with six Fe ions and $P_s$ is the saturated ferroelectric polarization.

\section{Crystal and magnetic structure of BiF\lowercase{e}O$_3$} \label{sec-structure} 

\subsection{Introduction to structure} \label{sec-struc-intro} 
 
BiFeO$_3$ forms in an $R3c$ rhombohedral structure at room temperature~\cite{Moreau1971, Kubel1990}: this rhombohedral structure is also usually described in a hexagonal lattice setting. Because of the three-fold rotational symmetry parallel to the $c$-axis in the hexagonal setting, a spontaneous electric polarization is only possible along the $c$-axis. The rhombohedral structure may be derived from the ideal cubic ($Pm\bar{3}m$) perovskite (ABX$_3$) structure, which consists of corner-sharing BX$_6$ octahedra, by a distortion of the octahedron. This usually occurs when the A and B cations have different ionic radii, and can be described by the Goldschmidt tolerance factor $t=\frac{r_A + r_X}{\sqrt{2}(r_B+r_X)}$, which determines when such tilting is likely~\cite{Goldschmidt1926}. The cubic $Pm\bar{3}m$ structure is adopted by many perovskite compounds at high temperatures and is also theoretically expected in BiFeO$_3$ at high temperatures~\cite{Kornev2007}.

In the 1970s, Mike Glazer categorized 23 possible tilts within the perovskite structure under the assumption that the BX$_6$ octahedron is rigid and only in-phase or out-of-phase tilting is allowed with respects to neighbouring octahedra~\cite{Glazer1972, Glazer1975}. This assumption works remarkably well for most oxide materials, so that it is only recently that it has been challenged in a few cases, such as in SrRuO$_3$~\cite{Lee2013SRO}, where a plastic deformation of the RuO$_6$ octahedron is experimentally reported and found to be closely related to the anomalous temperature dependence of the lattice constants. The original 23 Glazer tilting systems may be reduced to 15 by a group theoretical analysis~\cite{Howard1998, Howard2002}. In addition, atomic displacements and the possibility of broken inversion symmetry to describe the ferroelectric transitions known in perovskite materials may also be included~\cite{Stokes2002}. Taken together, the octahedral tilting and atomic displacement can produce 61 different crystal space groups with 15 paraelectric and 46 ferroelectric space groups starting from the ideal perovskite structure. Over the past four decades, this approach, based on the so-called Glazer notations, has proven to be very successful in modelling the complicated structural changes relating a lower symmetry, low temperature phase to a more symmetric high temperature phase~\cite{Roger2003,Islam2013,Johnsson2007}.

\begin{figure} \begin{center}
  \includegraphics[width=0.9\columnwidth]{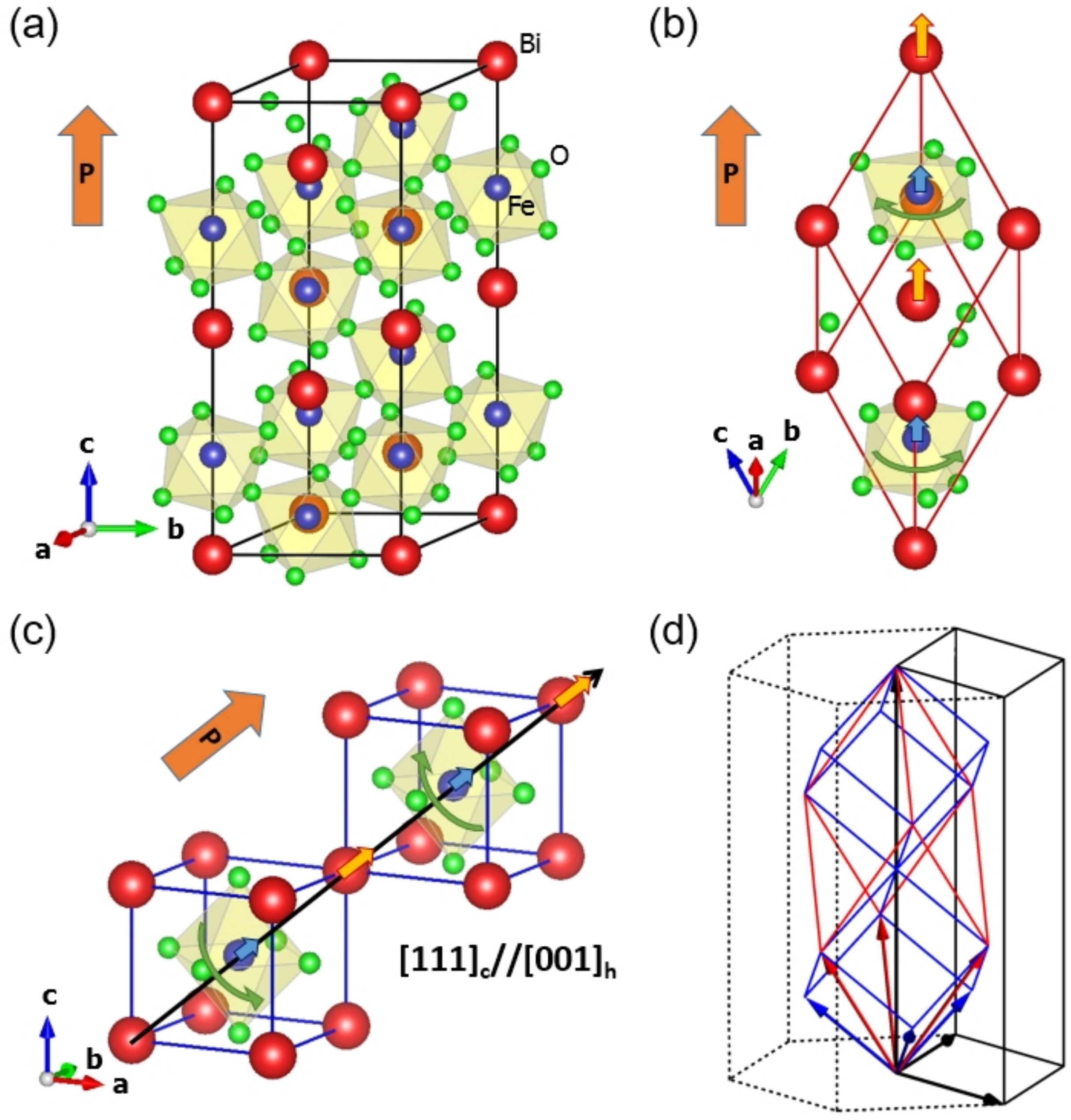}
  \caption{(a) Hexagonal unit cell including 6 formula units. (b) Rhombohedral unit cell including 2 formula units. (c) Pseudocubic unit cell including 1 formula unit. [111] direction in the rhombohedral and the pseudocubic notation is the same to [001] direction in the hexagonal notation. (d) Unit vectors and unit cell in the hexagonal (black), rhombohedral (red) and pseudocubic (blue) notations.
} \label{sf1}
\end{center} \end{figure}

For instance, in this notation the paraelectric $Pbnm$ space group (SG. 62) can be described by $a_0^- a_0^- c_0^+$ while the ferroelectric $R3c$ space group (SG. 161) is denoted by $a_+^- a_+^- a_+^-$. The first, second and third characters correspond to tilting angles along the principal $x$, $y$ and $z$-axes in the ideal cubic perovskite structure. The subscripts and superscripts denote atomic displacements and octahedral tilting along each axis, respectively. Subscripts 0 and + mean zero displacement and non-zero displacement while superscripts 0, $+$ and $-$ means no tilting, in-phase tilting and out-of-phase tilting. Thus, the paraelectric $Pbnm$ space group has its crystal symmetry reduced from the ideal perovskite by out-of-phase tilts along [110]$_{\mathrm{cubic}}$ and in-phase tilts along [001]$_{\mathrm{cubic}}$ without breaking inversion symmetry. On the other hand, the ferroelectric $R3c$ space group has out-of-phase tilts with atomic displacement along the [111]$_{\mathrm{cubic}}$ with broken inversion symmetry as shown in Figure~\ref{sf1}.

Which tilting or displacement mode is more favourable may be determined theoretically by first principles calculations. One elegant way is to compute the phonon dispersion relation in the perovskite structure and to look for unstable soft modes with imaginary energies~\cite{Zinenko2009}. The motions of atoms corresponding to this phonon mode then points to the structure distortion (effectively the freezing in of the phonon) favoured by the crystal. The total energy of structures with atoms displaced in these directions may then be computed and an energy minimum, corresponding to the stable distortion, found. Alternatively, a series of ground state energy calculations may be carried for each symmetry allowed distortion, refining the atomic positions by minimising the forces on each atom, to obtain the relevant total energy. In this way a series of (meta-)stable phases lying close in energy may be found~\cite{Dieguez2011}, with the thermodynamically favoured state having the lowest total energy.

In the case of BiFeO$_3$ at room temperature, the FeO$_6$ octahedron is found experimentally to be tilted out-of-phase along [111]$_{\mathrm{pc}}$ (in pseudocubic notation) and the Bi and Fe cations are shifted along the same [111]$_{\mathrm{pc}}$ direction breaking the inversion symmetry as shown in Figure~\ref{sf1}\rem{|or ~\ref{sf3} ?|}. These cation shifts correspond to a ferroelectric (FE) distortion, whilst the octahedral tilting is called an \emph{antiferrodistortive} (AFD) rotation. The FE distortion is due to the stereochemical activity of the Bi$^{3+}$ 6$s^2$ lone pair, which hybridise with 6$p$ orbitals resulting in an asymmetric wavefunction. This causes the bonding between Bi and O to be asymmetric and stabilises an off-centering displacement of the Bi ions with respects to the oxygen sublattice. Electrostatic repulsion then causes a cooperative displacement of Fe$^{3+}$ ions resulting in a finite electric polarisation~\cite{Ravindran2006}. The AFD octahedron rotation may also arise from the same cause, as the shortening of certain Bi-O bonds with respects to others forces the rigid FeO$_6$ octahedra to rotate to accommodate this~\cite{Dieguez2011}.

Theoretically, both AFD and FE distortions should be treated on an equal footing~\cite{Dieguez2013}. Early work, however, focussed principally on the FE displacements so that it is only recently that the importance of the AFD rotations to the magnetism and magnetoelectric coupling was realised. This is due to the competition between the two distortions, as each distortion lowers the total energy of BiFeO$_3$ with respects to the cubic structure by $\approx$700~meV, but their combination is only $\approx$1~eV lower in energy than the perovskite structure. The AFD affects the Fe-O-Fe bond angle, and thus affects the superexchange and DM interactions, whilst the FE distortion enhances the single ion anisotropy~\cite{Weingart2012}. In the ideal perovskite structure, the Fe-O-Fe bond angle is 180$^{\circ}$ giving a strongly antiferromagnetic superexchange interaction between nearest neighbour Fe spins. Whilst this is reduced in the $R3c$ structure, one can still obtain a G-type AFM magnetic structure with a high $T_N$.

As hydrostatic pressure is applied at low temperatures, the rhombohedral $R3c$ structure is expected theoretically to transform into the non-polar orthorhombic $Pnma$ structure~\cite{Ravindran2006,Zinenko2008,Gonzalez-Vazquez2009,Feng2010,Dieguez2011}. The critical pressure was estimated variously between 3 and 20~GPa, whereas experimentally it was found to be $\approx$10~GPa~\cite{Haumont2009,Guennou2011,Zhang2013}. At lower pressures, however, different groups identified a series of different phase transitions from the low pressure $R3c$ to several orthorhombic or monoclinic structures. The different number of structures and critical pressures in this intermediate pressure region from $\approx$3-10~GPa was attributed to the use of different pressure transmitting media which may each result in some small additional non-hydrostatic stress being applied. Since the different competing metastable phases have only relatively small energy barriers between them~\cite{Dieguez2011}, slight differences in stresses can result in different phases being stabilised under pressure.

Under large epitaxial compression, both \emph{ab initio}~\cite{Hatt2010} and effective Hamiltonian~\cite{Albrecht2010} calculations found a monoclinic $Cc$ phase with a large $c/a$ ratio. The resemblance between this and the tetragonal ($P4mm$) distortion of the perovskite lattice led to the $Cc$ structure being named `T-like'. Unlike the true tetragonal structure, however, which has only a ferroelectric displacement along $[001]_{\mathrm{pc}}$, the T-like phase includes in addition further FE displacements along $[110]_{\mathrm{pc}}$ and an in-phase rotation of the FeO$_6$ octahedron around $[110]_{\mathrm{pc}}$. At low epitaxial strains, a phase with a small $c/a$ ratio is experimentally found, which although still having $Cc$ symmetry, is similar to the unstrained rhombohedral structure, and thus is called `R-like'. The FE polarisation is calculated to remain constant but rotates from the $[111]_{\mathrm{pc}}$ to the $[001]_{\mathrm{pc}}$ direction with increasing strain~\cite{Daumont2012,Sando2014}. The magnetic cycloid, however is suppressed under epitaxial strain allowing a weak ferromagnetic moment due to spin canting (as a result of the Dzyaloshinskii-Moriya interaction) to exist~\cite{Albrecht2010}. It was found theoretically~\cite{Dieguez2011} that the T-phase should have $C$-type AFM ordering (with ferromagnetic $ab_{\mathrm{pc}}$ planes), rather than the $G$-type found in the bulk structures (with ferromagnetic $(111)_{\mathrm{pc}}$ planes). However, experiments~\cite{MacDougall2012} detect mostly $G$-type order, with the possible coexistence of $C$-type at low temperatures.

Finally, a word on the three notations often used in the literatures for BiFeO$_3$: because of its ease of use, many bulk measurements are described in a pseudocubic (pc) notation while the rhombohedral structure is, in some cases, denoted using a hexagonal notation. Here we adopt the both pseudocubic and hexagonal notations throughout, and note that one can convert the two most important axes into the pseudocubic or rhombohedral structures by the following relations~\cite{Megaw1975}: $[0~0~1]_{\mathrm{hex}} = 2[1~1~1]_{\mathrm{pc}} = [1~1~1]_{\mathrm{rhom}}$ and $[1~1~0]_{\mathrm{hex}} = [1~\bar{1}~0]_{\mathrm{pc}} = [\bar{1}~1~0]_{\mathrm{rhom}}$. 

\subsection{Ferroelectric transition} \label{sec-struct-3-1} 

\begin{figure} \begin{center}
  \includegraphics[width=0.9\columnwidth]{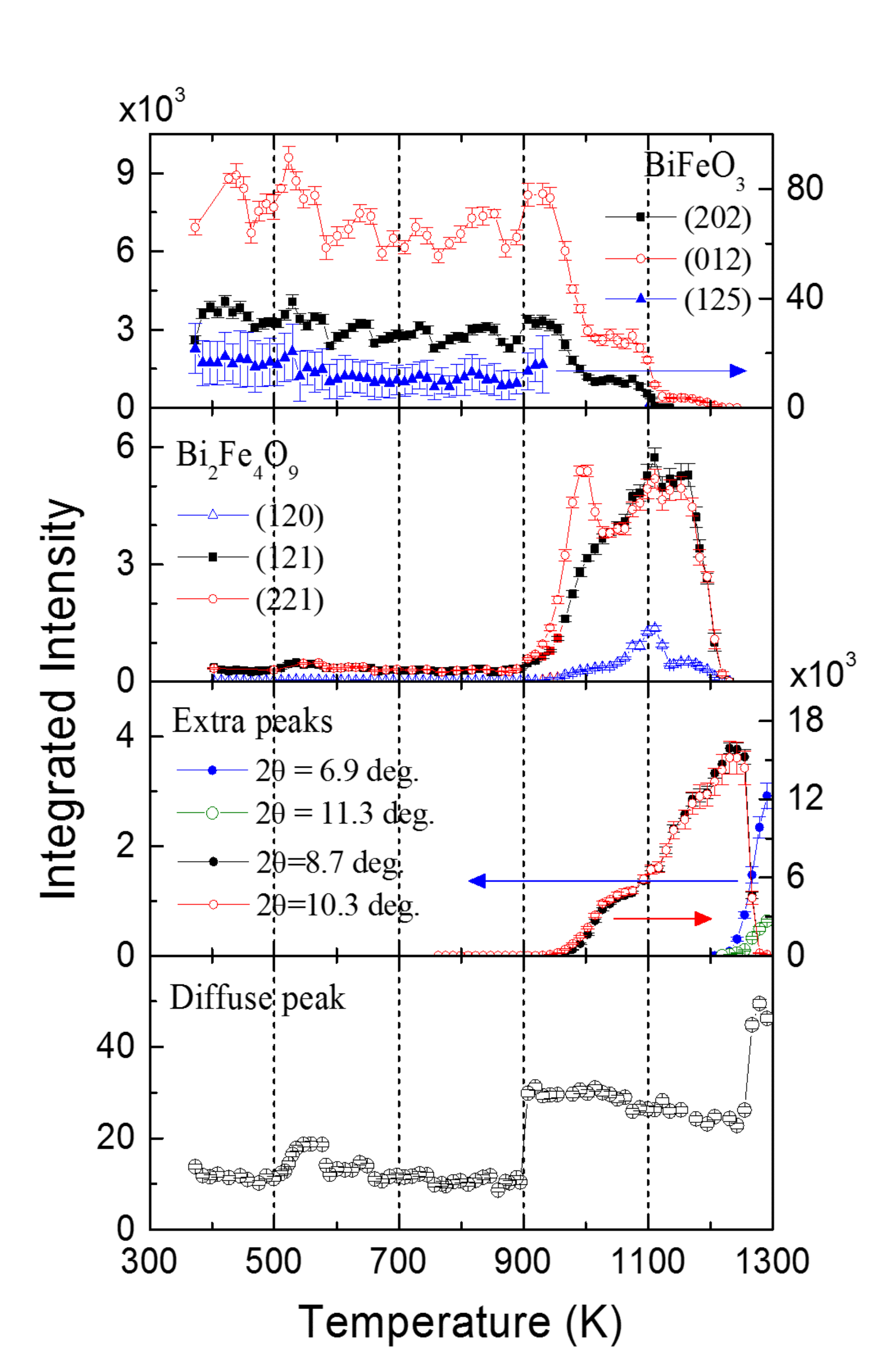}
  \caption{Decomposition of BiFeO$_3$ into Bi$_2$Fe$_4$O$_9$ under heating measured by synchrotron 
  XRD~\cite{ParkUnpub}.
} \label{sf4}
\end{center} \end{figure}

BiFeO$_3$ has been known to have the non-centrosymmetric crystal structure $R3c$ since the 1970s~\cite{Moreau1971}. Despite this long history and its relative importance, the nature of the paraelectric-ferroelectric transition, particularly the symmetry of the paraelectric structure, has remained unclear. Over the years, many attempts to address this issue have been made with several contending crystal structural models: $Pm\bar{3}m$ by Raman spectroscopy~\cite{Haumont2006}; $I4/mcm$ by theoretical calculations~\cite{Kornev2007}; $C2/c$~\cite{Kornev2007}, $P2_1/m$~\cite{Haumont2008} or $R\bar{3}c$~\cite{Selbach2008} by XRD, and $Pbnm$ by powder neutron diffraction~\cite{Arnold2009}. Much of this confusion is related to material issues: First, care must be taken to prepare high-quality single crystals using the flux method. Second, as we will discuss below, BiFeO$_3$ easily decomposes into Bi$_2$Fe$_4$O$_9$ upon heating and avoiding this secondary phase at high temperatures is a major experimental challenge. For example, we show in Figure~\ref{sf4} how a BiFeO$_3$ powder sample prepared by grinding a single crystal of BiFeO$_3$ decomposes into Bi$_2$Fe$_4$O$_9$ gradually upon heating.

The full Bi$_2$O$_3$-Fe$_2$O$_3$ phase diagram was investigated~\cite{Palai2008, Haumont2008}, according to which BiFeO$_3$ undergoes an incongruent melting from a liquid phase. More importantly, Bi$_2$Fe$_4$O$_9$ and other, more complex, Bi-Fe-O phases are found to be easily formed as one of several secondary phases in an inhomogeneous mixture when grown by the flux method. In fact, all flux method growth has been conducted by synthesizing BiFeO$_3$ crystals just below the incongruent melting point. This growth technique has a very advantageous consequence on the samples. By virtue of the low-temperature growth condition, in particular below the ferroelectric transition temperature, the crystals grown by the flux method are usually found to be in a single ferroelectric-ferroelastic domain. This single ferroelastic domain, in particular, turns out to be very useful in co-aligning more than 10 small crystals for inelastic neutron scattering experiments~\cite{Jeong2012}. 
According to the phase diagram, BiFeO$_3$ undergoes a $\gamma$-$\beta$ transition near 1200~K followed by another $\beta$-$\alpha$ transition around 1100~K.

It was suggested that this high temperature $\gamma$ phase is the cubic perovskite structure~\cite{Palai2008} on the basis of Raman scattering~\cite{Haumont2006} and IR spectroscopy~\cite{Massa2010} studies. The structure is stable in only a very narrow temperature range between 1198(5)~K and 1206(5)~K. Above this, BiFeO$_3$ decomposes first into Bi$_2$Fe$_4$O$_9$ and eventually into Fe$_2$O$_3$ before melting around 1240~K~\cite{Haumont2008}. The assignment of cubic symmetry comes from the identification of only three phonon modes in the Raman and IR spectra, which is consistent with the group theoretical representations of the cubic point group, and the observation of polarised light microscope that the crystal surface becomes optically isotropic with the disappearance of the ferroelastic domains above 1198~K~\cite{Haumont2006}. However, at such high temperatures, thermal broadening of the phonon modes is likely to have smeared out other possible modes, whilst chemical disorder from possible decomposition products may erode the ferroelastic domains, rendering a definitive symmetry classification difficult. Later X-ray diffraction measurements by the same group, on the other hand, could not find evidence of the $\gamma$ phase, but rather from extrapolating the trend of the pseudocubic lattice parameters, posit that a cubic phase could occur around 1290~K if the material had not decomposed by this temperature~\cite{Haumont2008}. Thus, further experimental verification is needed, but the narrow temperature range of this possible phase which is also very close to the decomposition temperature of BiFeO$_3$ makes this a challenging endeavour.

\begin{figure} \begin{center}
  \includegraphics[width=0.9\columnwidth]{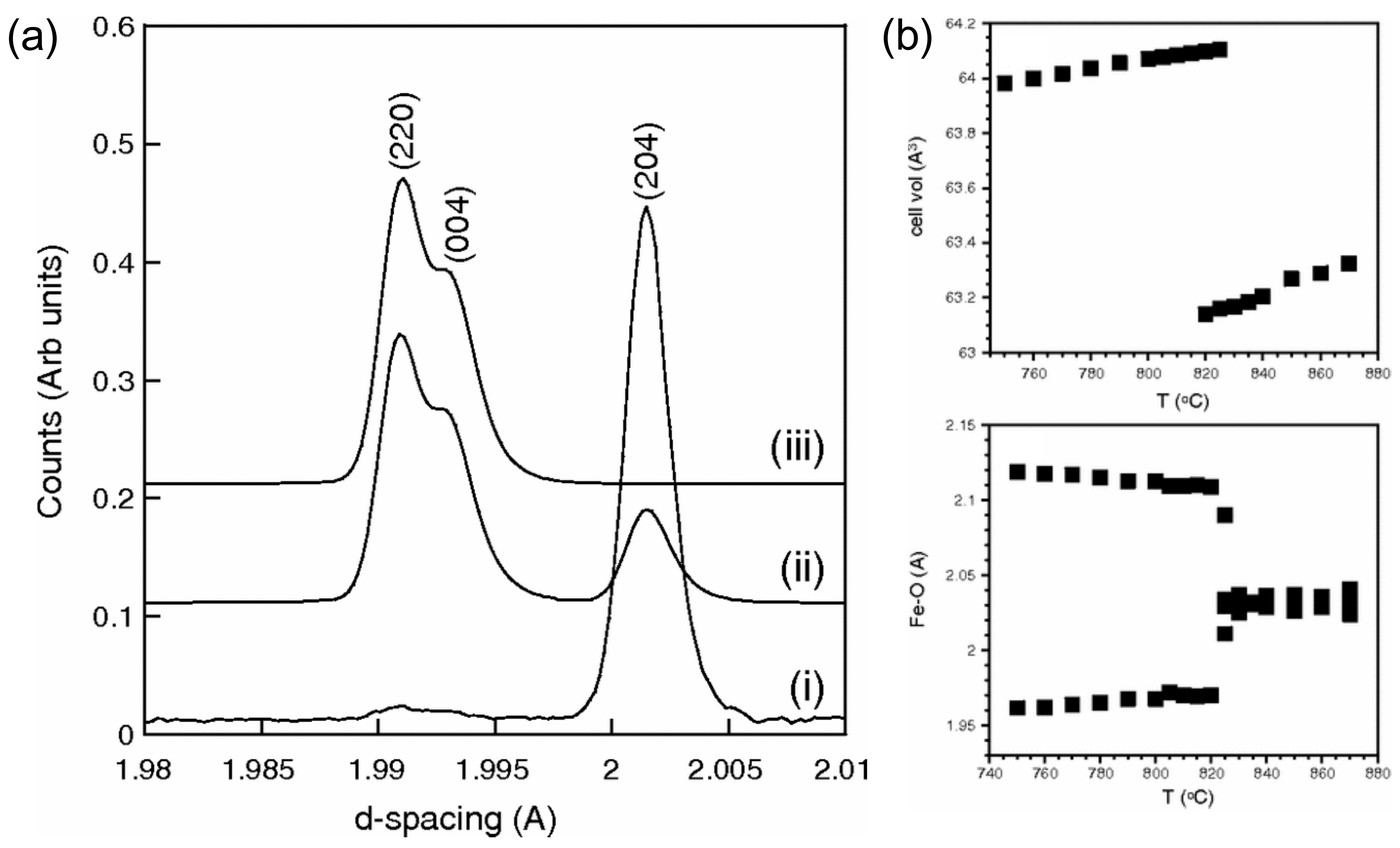}
  \caption{(a) Variation of the pseudocubic (200) peak passing through the ferroelectric transition. (i) Rhombohedral (204) peak (in hexagonal notation) at 1093~K with $R3c$ symmetry, (ii) intermediate state at 1098~K and (iii) orthorhombic (220) and (004) peaks at 1103~K with $Pbnm$ symmetry.  (b) Abrupt change of the unit cell volume and Fe-O bond length at the FE transition temperature. Reprinted with permission from~\cite{Arnold2009}. Copyright 2009 by American Physical Society.
} \label{sf2}
\end{center} \end{figure}

As regards the ferroelectric-paraelectric transition, there is much experimental evidence suggesting that it is of first order. For example, there is a clear hysteresis in the calorimetry data~\cite{Haumont2008} while domain structures studied using polished single crystal of less than 9~$\mu$m thickness~\cite{Palai2008} show the ferroelectric $\alpha$-phase and paraelectric $\beta$-phase coexisting at 820$^{\circ}$C (1090 K) together with a very sharp domain boundary. Furthermore, powder neutron diffraction experiments also show the $\alpha$-$\beta$ phase coexistence, and a volume discontinuity at the transition temperature, which are clear signs of a first order phase transition~\cite{Arnold2009}, as shown in Figure~\ref{sf2}. 

The nature of this first order phase transition is still a puzzle, in part due to the difficulty of conducting experiments above 1100~K. However, temperature dependent studies of phonon frequencies in the ferroelectric phase by various techniques~\cite{Haumont2006,Fukumura2007,Kamba2007,Delaire2012,Zbiri2012,Borissenko2013}, discussed in mode detail in section~\ref{sec-phonon}, indicate that the two lowest energy phonon modes, corresponding the polar and AFD distortions do play a role. Competition between these two types of displacements may then lead to a single first order transition rather than two sucessive second order transition. Intriguingly, an alternative explanation of the first order in terms of chemical disorder due to the decomposition of BiFeO$_3$ and vacancies arising from the loss of oxygen at high temperatures was offered by~\cite{Massa2010} and supported by the observation of an irreversible suppression of the XANES fine structure and shifts of the Bi $L_3$ and Fe $K$ absorption edges to lower binding energy on heating above 773~K.   Nonetheless in the following discussion we shall be guided by the group-subgroup relations of a displacive transition in order to narrow down the possible space group of the paraelectric phase, on the evidence from the phonon studies that these displacements are important to the ferroelectric transition.

As H. Stokes {\it et al.}~\cite{Stokes2002} pointed out, in a displacive phase transition the combinations of the primary distortion modes of the ideal perovskite structure (in-phase and out-of-phase octahedral tilts and cation displacements) can lead to one of 61 possible space groups altogether. Following their notations, these distortion modes can also be denoted by the wavevector of the phonon mode corresponding to the distortion. These modes occur at the high symmetry zone centre ($\Gamma$) and zone boundary points ($M$ and $R$) in the Brillouin zone for the ideal perovskite structure~\cite{Aroyo2014}. Respectively, the in-phase tilting is denoted by $M_3^+$, out-of-phase tilting by $R_4^+$, and cation displacements by $\Gamma_4^-$. The zone boundary modes, $M_3^+$ and $R_4^+$, change the number of formula unit per unit cell, thus introducing superlattice peaks in a diffraction pattern, while the zone centre $\Gamma_4^-$ distortion, responsible for the ferroelectric transition, splits nuclear Bragg peaks and changes their relative intensities. Both effects are readily observed in an X-ray or neutron diffraction measurement, the evidence from which we shall now use to determine the nature of the paraelectric $\beta$ phase of BiFeO$_3$.

\begin{figure} \begin{center}
  \includegraphics[width=0.9\columnwidth]{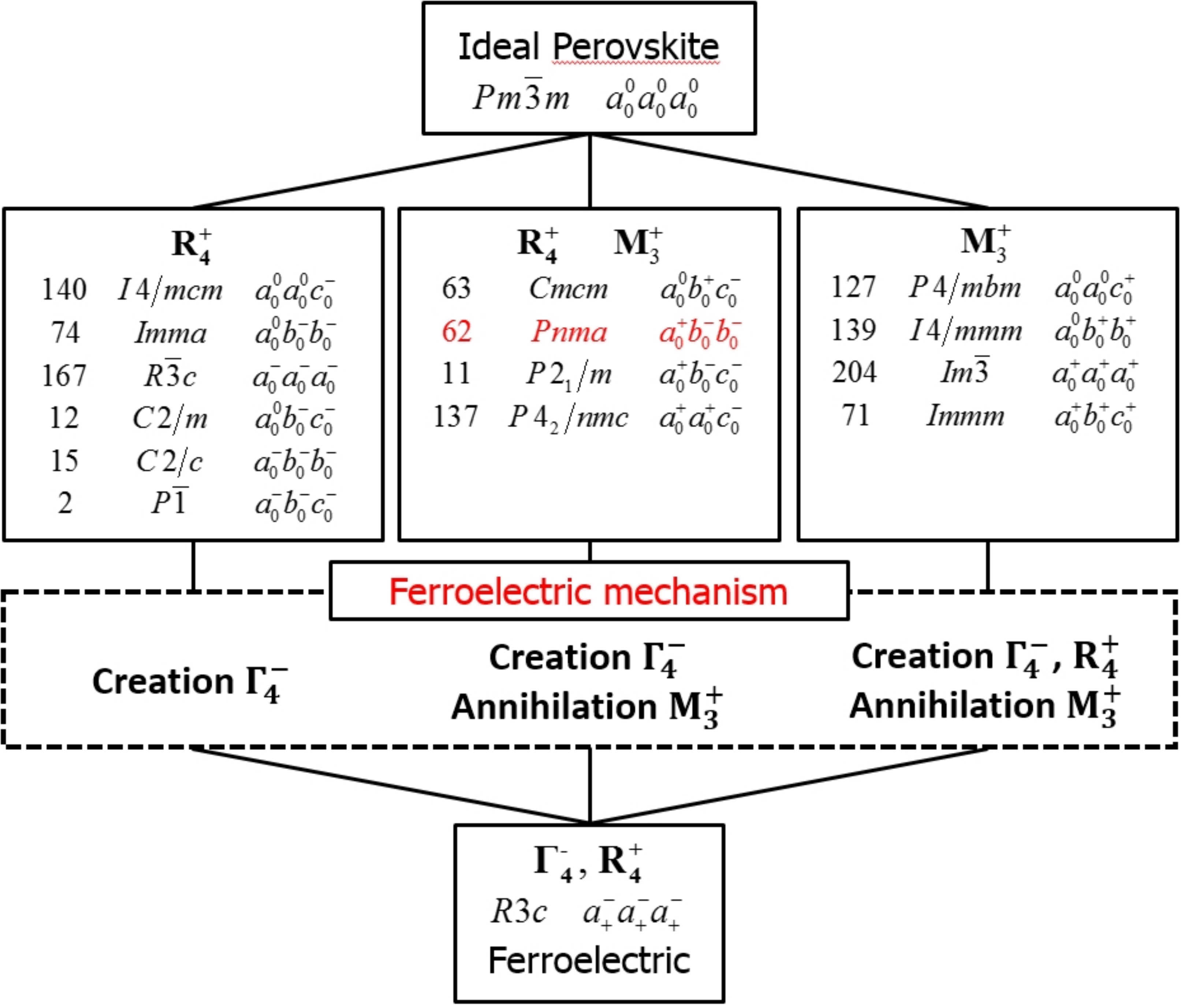}
  \caption{Group-subgroup relation through the FE transition. $\rm{R}_4^+$, $\rm{M}_3^+$ and $\Gamma_4^-$ are symmetry operators for the out-of-phase tilting, in-phase tilting and ferroelectric distortion, respectively.} \label{sf3}
\end{center} \end{figure}

\begin{figure} \begin{center}
  \includegraphics[width=0.9\columnwidth]{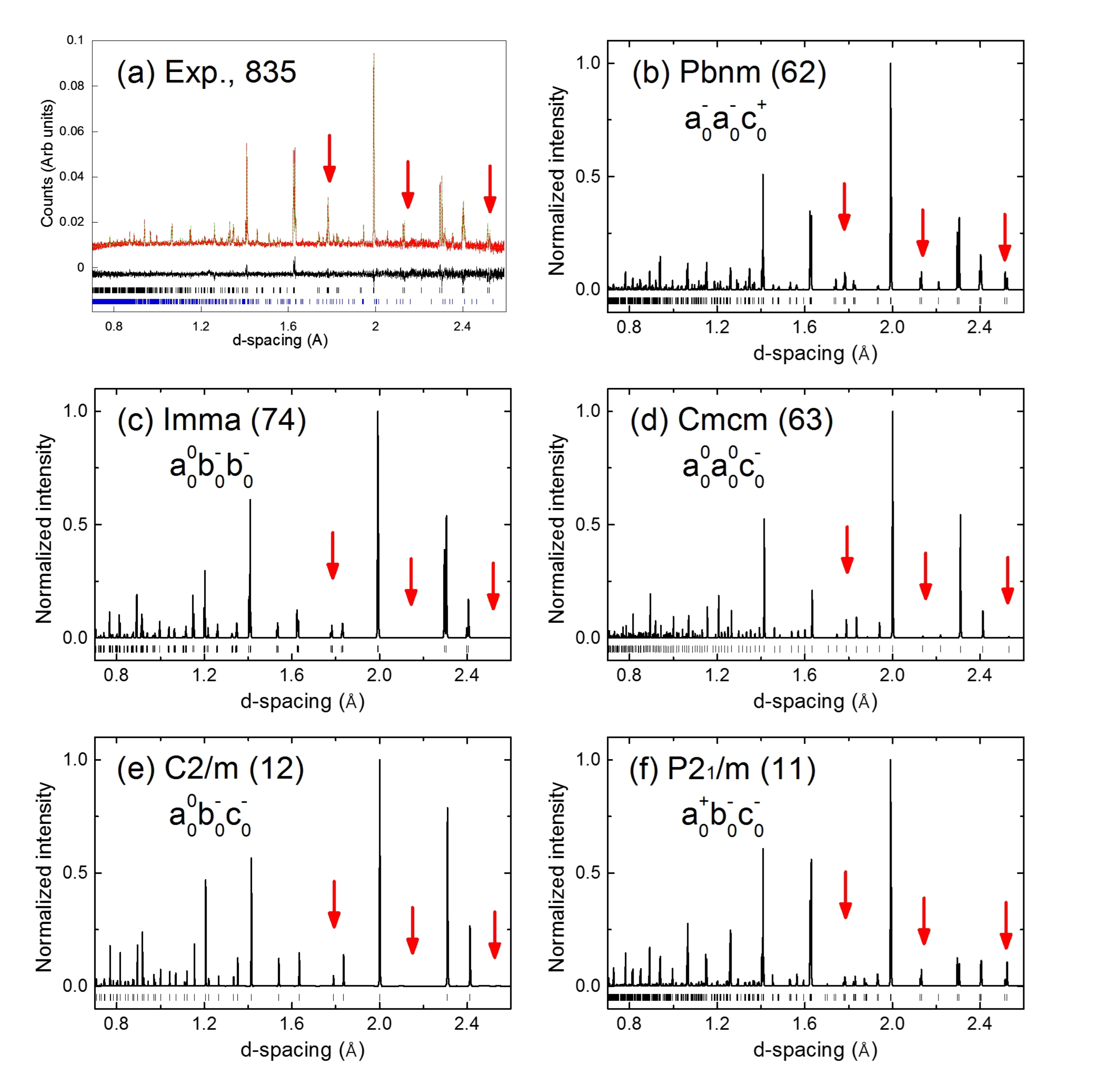}
  \caption{(a) Neutron powder diffraction data collected at 1108 K. Reprinted with permission from \cite{Arnold2009}. Copyright 2009 American Physical Society. (b-f) Simulated profiles using $Pbnm$, $Imma$, $Cmcm$, $C2/m$ and $P2_1/m$, respectively. For the simulation, we employed the Fullprof program~\cite{Rodriguez1993} and Bilbao Crystallography Server~\cite{Orobengoa2009,PerezMato2010}.\rem{|[We can put this in the text.]|} Distinguishable peaks are highlighted by the arrows.} \label{sf5}
\end{center} \end{figure}

The 15 paraelectric candidate space groups that can be related to a $R3c$ ferroelectric structure are shown in Figure~\ref{sf3}. These 15 space groups can be further categorized into four classes: 1) a direct structure transition from the ideal cubic perovskite $Pm\bar{3}m$ straight to the $R3c$ structure; 2) the transition proceeds through an intermediate paraelectric structure between the ideal perovskite structure and the $R3c$ structure which involves only out-of-phase tilting $R_4^+$; 3) the intermediate phase is characterised by both out-of-phase tilts $R_4^+$ and in-phase tilts $M_3^+$ together; 4) the last ones involving only the in-phase tilting mode $M_3^+$. If one does not start from the cubic perovskite space group, a natural alternative is a paraelectric phase possessing just the $R_4^+$ mode since the $R3c$ space group has both an out-of-phase tilting $R_4^+$ and a cation displacement mode $\Gamma_4^-$. On the other hand, J. M. Perez-Mato {\it et al.} pointed out that the paraelectric phase can have both $R_4^+$ and $M_3^+$ in the case of a strongly first order phase transition~\cite{PerezMato2010}. With this view in mind, below we made a thorough comparison of the existing data against all possible space groups, as summarised in Figure~\ref{sf5}.

One important study is the work of Arnold {\it et al.}~\cite{Arnold2009}, which was a neutron powder diffraction measurement using powder sample sealed in quartz at the HRPD beamline of ISIS. Their key observations are that a single rhombohedral peak (204) splits into two peaks that can be indexed as orthorhombic (220) and (004) (See Figure~\ref{sf2}). Since the cubic $Pm\bar{3}m$ structure has the same $d$-spacing for both (220) and (004), it can be easily ruled out. For a similar reason, the $R\bar{3}c$ rhombohedral structure can also be ignored.  We note that both structures were reported as being the $\beta$ phase on the basis of Raman scattering and X-ray diffraction measurements~\cite{Haumont2006, Selbach2008}. Thus, we need to check all the other possible (orthorhombic, monoclinic or triclinic) structures among the 15 paraelectric space groups according to the three classes defined above (involving $R_4^+$, $R_4^+$ and $M_3^+$, or just $M_3^+$ distortions respectively). Since Arnold {\it et al.} could index their data in an orthorhombic cell with no further splitting, in Figure~\ref{sf5} we compare our simulation results for specific orthorhombic and monoclinic space groups with their experimental results. As one can see in the figure, the simulation is in good agreement with the $Pbnm$ paraelectric space group claimed by Arnold {\it et al.}~\cite{Arnold2009}. 

The transition from the paraelectric $Pbnm(a_0^-a_0^-c_0^+$) to the ferroelectric $R3c(a_+^-a_+^-a_+^-$) structure requires both the simultaneous removal of an in-phase tilting $M_3^+$ mode and the activation of a $\Gamma_4^-$ displacement mode. At a first sight, these processes appear to be much less favourable than an $R_4^+$ tilt alone. It is thus interesting to ask why this route is preferred over a hypothetical direct route involving both $R_4^+$ and $\Gamma_4^-$ modes, but not the in-phase tilting $M_3^+$ mode. According to the structural studies at room temperature~\cite{Moreau1971, Arnold2009}, a Bi shift is 2.5 times larger than an Fe shift, which means that the Bi-O bonds and Bi environment undergo a much more drastic change than their Fe counterparts and so are important to the ferroelectric phase transition. Of further interest, new analysis based on a maximum entropy method reveals that Bi-O bond has a more covalent character than Fe-O~\cite{Fujii2013}. Having said that, it is fair to say that there is no clear-cut intuitive understanding of the presence of the Pbnm paraelectric phase, which to our view warrants further studies.

\subsection{Magnetic transition} \label{sec-struct-3-2}
\subsubsection{Magnetic structure} \label{sec-struct-3-2-1}

\begin{figure} \begin{center}
  \includegraphics[width=0.9\columnwidth]{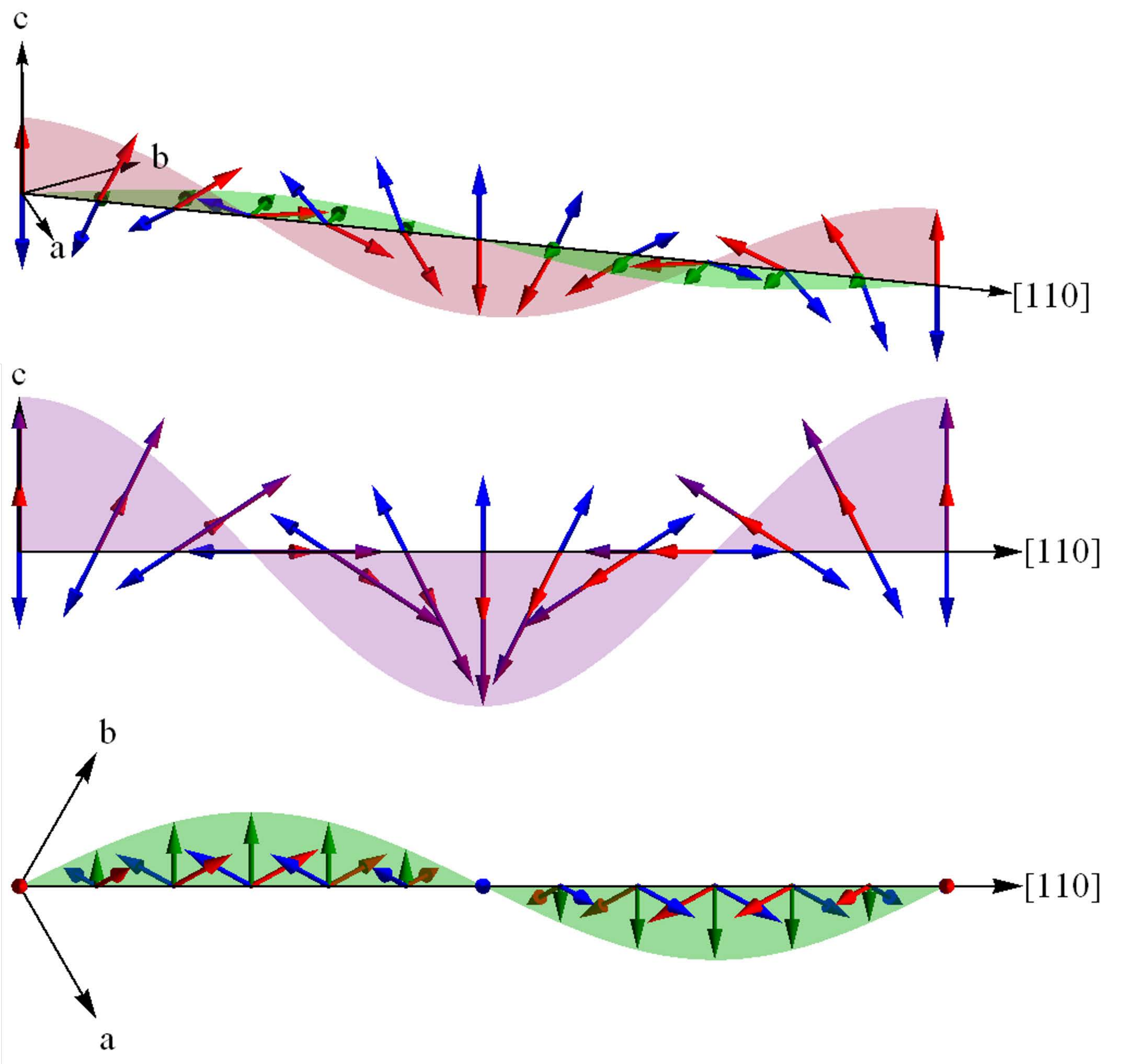}
  \caption{Magnetic structure of BiFeO$_3$. Red and blue arrows correspond to the opposite magnetic moments $\mathbf{M}_i$ on two sublattices. Purple and green arrows represent an antiferromagnetic vector $\mathbf{L}=\mathbf{M}_1-\mathbf{M}_2$ and a local ferromagnetic moment $\mathbf{M}=\mathbf{M}_1+\mathbf{M}_2$ induced by canting out of the cycloid plane, respectively. To clarify them, a modulation of $\mathbf{L}$ and $\mathbf{M}$ along $[1~1~0]_\mathrm{hex}$ direction is shown in the cycloid plane and the perpendicular plane, respectively.} \label{sf6}
\end{center} \end{figure}

As we discussed above, the 180$^{\circ}$ Fe-O-Fe bond favors an antiferromagnetic super-exchange interaction leading to an almost G-type AFM structure~\cite{Kiselev1962}. Then a DM interaction, an order of magnitude smaller than the AF interaction~\cite{Jeong2012}, stabilizes spin canting and induces an incommensurate magnetic structure with a spin cycloid propagating along [110]$_{\mathrm{hex}}$ with an extremely long periodicity of 620--780~\AA~\cite{Sosnowska1982, Sosnowska2011, Lebeugle2008, Lee2008a, Herrero-Albillos2010, Ramazanoglu2011a}. This spin cycloid rotates along the chiral vector $[\bar{1}10]_{\mathrm{hex}}$. A second DM interaction, a further order of magnitude smaller than the primary DM interaction~\cite{Jeong2014}, causes the cycloid to be canted about 1$^{\circ}$ perpendicular to the spontaneous electric polarization vector [001]$_{\mathrm{hex}}$ and the magnetic propagation vector [110]$_{\mathrm{hex}}$, inducing a spin density wave~\cite{Ramazanoglu2011}. In Figure~\ref{sf6}, we plot the magnetic structure given by a polarization direction of [001]$_{\mathrm{hex}}$, a spin cycloid propagation vector [110]$_{\mathrm{hex}}$, and a spin chiral vector $[\bar{1}10]_{\mathrm{hex}}$.

Neutron diffraction is a unique tool to investigate magnetic structures. In fact, the correct incommensurate magnetic structure of BiFeO$_3$ was first obtained from high-resolution time-of-flight neutron diffraction experiments~\cite{Sosnowska1982}. By observing the splitting of the magnetic (101)$_{\mathrm{hex}}$ peaks, they obtained the correct magnetic model of the spin cycloid structure with a period of 620~\AA~and propagating along [110]$_{\mathrm{hex}}$ superimposed on the G-type AFM (See Fig. 9). This long periodicity is one of the distinguishing feature of BiFeO$_3$. Furthermore, a spin cycloid structure can induce (additional) electric polarization through an inverse DM interaction while a spin density wave cannot. Finally, we note that the period of the cycloid has some dependence on sample quality, as observed by Sosnowska {\it et al.}~\cite{Sosnowska2011}. Samples of poor crystallinity show much broader peak profiles, indicative of inhomogeneous local field distribution. Nonetheless, a very long period was observed in all samples.

\begin{figure} \begin{center}
  \includegraphics[width=0.9\columnwidth]{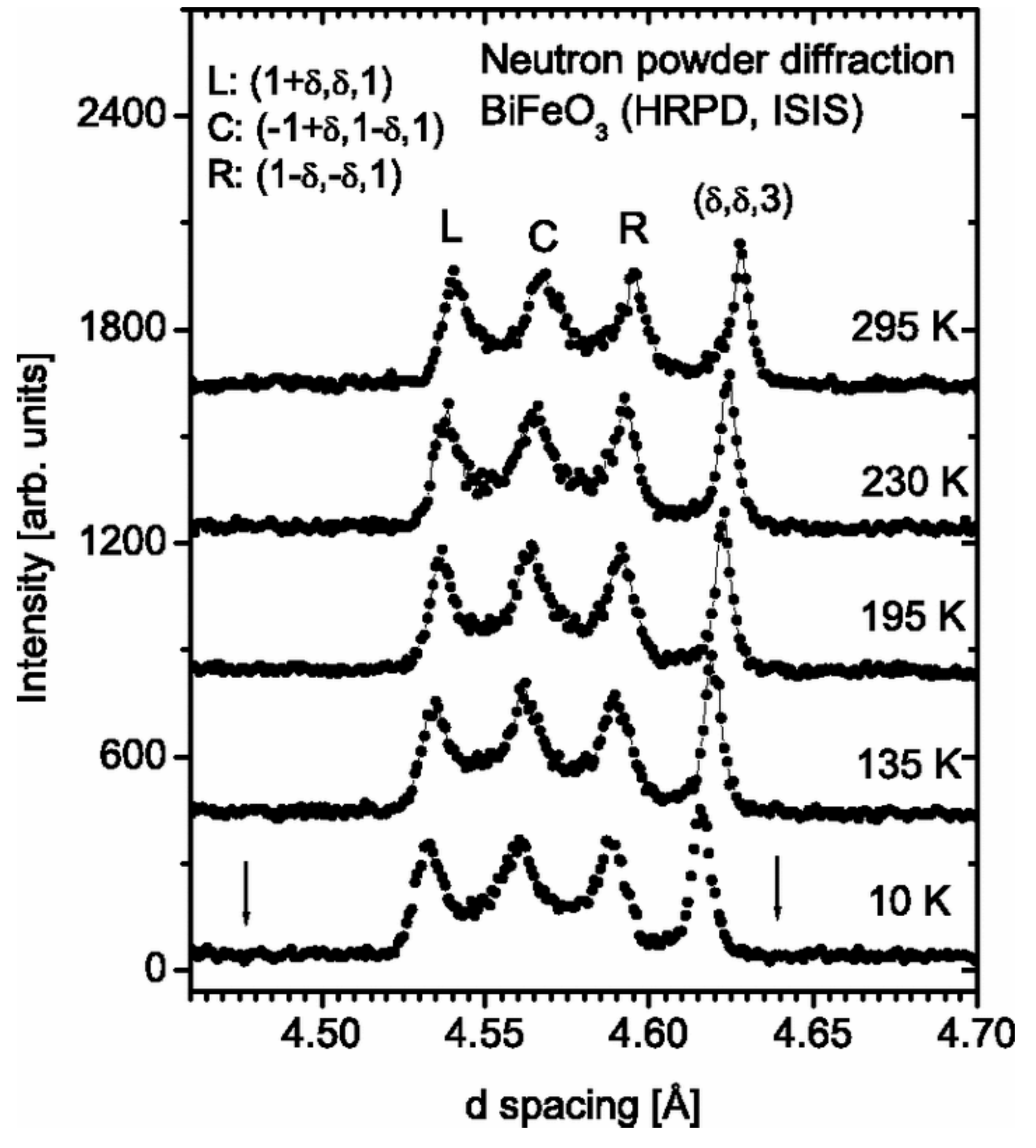}
  \caption{Neutron powder diffraction data collected at various temperatures. The symbols L, C and R represent the left, central and right satellite peaks according to their positions. Vertical arrows denote the positions of the third-order satellites for the anharmonic cycloid. Reprinted with permission from~\cite{Sosnowska2011}. Copyright 2011 American Physical Society.
} \label{sf9}
\end{center} \end{figure}

The spin cycloid magnetic structure was again recently confirmed in a detailed single crystal neutron diffraction measurement~\cite{Lebeugle2008}, taking advantage of the low-temperature flux-method growth conditions described above to obtain samples with a single ferroelectric domain. One important result from this study is the observation of magnetic satellite peaks around $(\frac{1}{2}~\bar{\frac{1}{2}}~\frac{1}{2})_{\mathrm{pc}}$, in the pseudocubic notation. In the $R3c$ crystal structure this corresponds to a magnetic propagation vector along $[1~\bar{1}~0 ]_{\mathrm{pc}}=[1~1~0]_\mathrm{hex}$, with three equivalent magnetic domains, namely $\tau_1=\left[\delta~\bar{\delta}~0\right]_{\mathrm{pc}}$, $\tau_2=\left[\delta~0~\bar{\delta}\right]_{\mathrm{pc}}$ and $\tau_3=\left[0~\bar{\delta}~\delta\right]_{\mathrm{pc}}$ with $\delta$=0.00437. Lebeugle {\it et al.}~\cite{Lebeugle2008} concluded that these data are only consistent with a spin cycloid model but not ellipsoid nor SDW magnetic models. In addition, these magnetic domains show a strong sensitivity to external pressure: a relatively small applied uniaxial pressure of up 15~MPa along the [001]$_{\mathrm{pc}}$ axis on Nd doped BiFeO$_3$  can change the population of magnetic domain drastically. At around 7.2~MPa, a change occurs from the $\tau_1$ to $\tau_2$ magnetic domains, denoting a 60$^{\circ}$ rotation of the easy-plane~\cite{Ramazanoglu2011b}. However, it is important to note that the spin cycloid structure still persists up to much greater pressures, around 3~GPa~\cite{Guennou2011}.
Ramazanoglu {\it et al.} further investigated the temperature dependence of the incommensurate magnetic peak~\cite{Ramazanoglu2011a} to find that the spin cycloid periodicity increases from 629(5)~\AA~at 5~K to 780(30)~\AA~at 615~K upon heating with a critical exponent $\beta$=0.34(3).

The change in the cycloid periodicity can be explained by the variation of the DM interaction, as discussed in Section~\ref{sec-lowE}, which is, in turn, dependent on the Fe-O-Fe bond angle, since $\mathbf{D}\propto\mathbf{r}_{\mathrm{Fe}}\times\mathbf{r}_{\mathrm{O}}$~\cite{Keffer1962}: $\mathbf{r}_{\mathrm{Fe}}, \mathbf{r}_{\mathrm{O}}$ are the positions of Fe and O atoms with respect to the center of the Fe-Fe direction. The pressure dependence, on the other hand, is mostly due to changes in the single-ion anisotropy (SIA) which is very sensitive to small changes in the crystal structure as shown by the temperature dependence of the spin wave spectrum shown in Section~\ref{sec-lowE}. BiFeO$_3$ was found to have a very small easy-axis anisotropy along [111]$_{\mathrm{pc}}$~\cite{Jeong2014}, a result of competition between the AFD rotation, which favours an easy-plane anisotropy, and the FE distortion, which stabilises the easy-axis term~\cite{Weingart2012}. When the easy axis anisotropy is enhanced, for example by epitaxial strain, the cycloid can be suppressed, as observed at relatively small strains of around 1\%~\cite{Sando2014}. 

\begin{figure} \begin{center}
  \includegraphics[width=0.9\columnwidth]{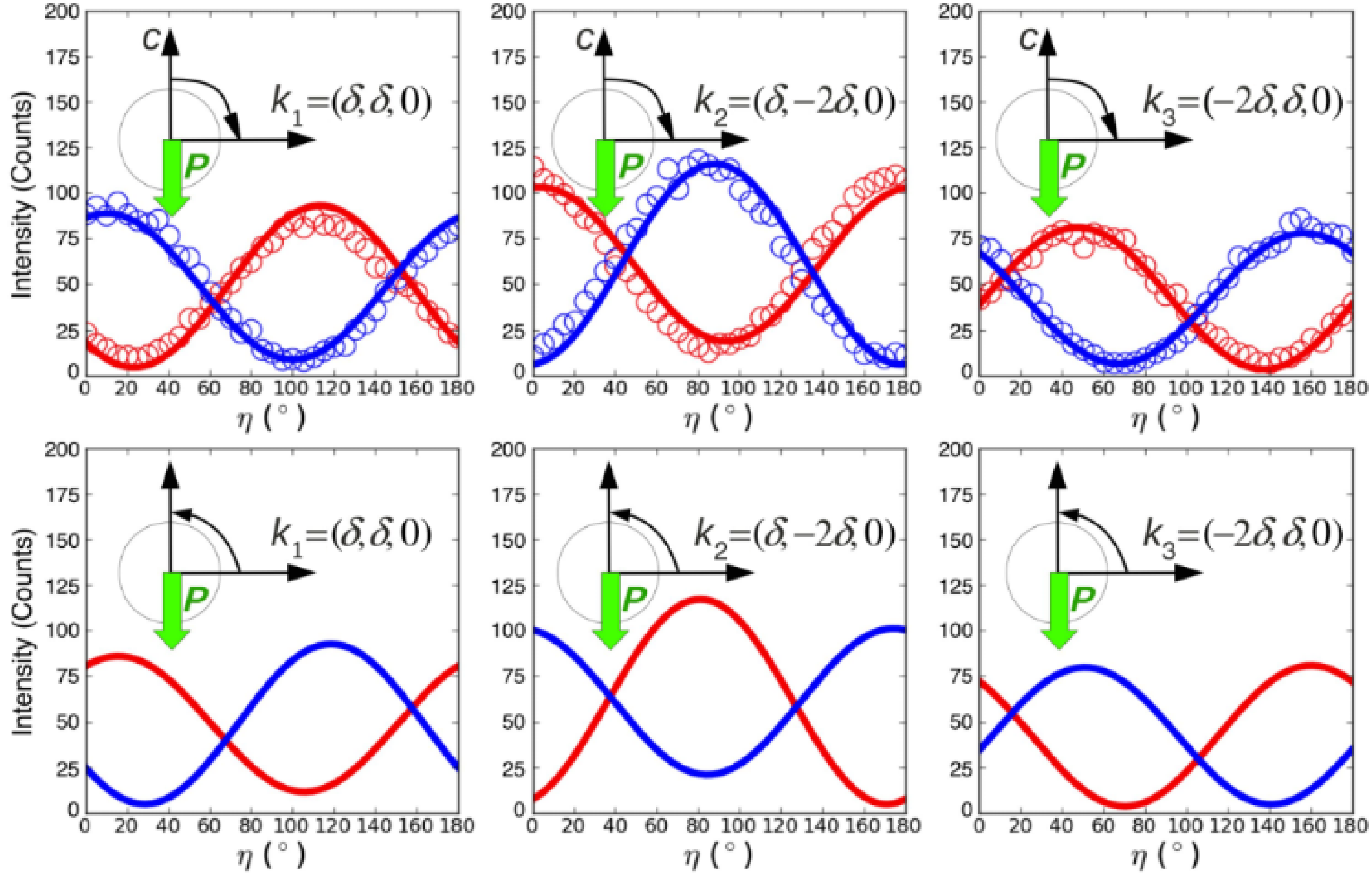}
  \caption{Top: Measured X-ray intensity as a function of the analyzer angle $\eta$ (circle) with a refinement (solid line) assuming CW rotation of the spin cycloid. Red and blue colors correspond to a right-handed and left-handed polarization of incident X-ray. Bottom: Calculated X-ray intensity assuming CCW. Reprinted with permission from \cite{Johnson2013}. Copyright 2013 American Physical Society.
} \label{sf8}
\end{center} \end{figure}

For a given polarization and magnetic propagation vector, the spins in the cycloid can choose either a clockwise or anticlockwise rotation, defining a chiral vector. 
This chiral vector cannot be uniquely determined by unpolarised neutron diffraction data, and while it is possible with polarised neutrons as demonstrated in TbMnO$_3$~\cite{Yamasaki2007}, no such measurement of the chiral vector by polarised neutron diffraction have been reported to our knowledge.
However, by using a circularly polarized X-ray beam in a diffraction measurement on a single magnetic domain sample, Johnson {\it et al.}~\cite{Johnson2013} determined that the spin rotation should be clockwise when viewed with polarisation down and the magnetic propagating vector to the right (see Figure~\ref{sf8}).

More recently, Ramazanoglu {\it et al.} conducted SANS experiments using polarized neutrons to find that the cycloid magnetic structure is tilted away from the hexagonal $c$-axis at an angle of 1$^{\circ}$, corresponding to a SDW with 0.09(1)~$\mu_B$ amplitude~\cite{Ramazanoglu2011} (See Fig. 11).  As we will discuss later in Section~\ref{sec-lowE}, this canting implies a second DM term which is much larger than one found from the analysis of the low energy inelastic neutron scattering spectrum.

\subsubsection{Spin anharmonicity} \label{sec-struct-3-2-2}

The L-G free energy Eq.~\ref{eq:LGfe} discussed in Section~\ref{sec-theory} can be written in the following form, where a magnetic anisotropic energy $K_u$ can distort an ideal spin cycloid structure, as explained in appendix~\ref{sec-ap-anharmonicity}:

\begin{align} \label{eq:structfe}
f &= f_{\mathrm{exch}}+f_{\mathrm{L}}+f_{\mathrm{an}} \nonumber \\
 &= A\sum_{i=x,y,z}{\left( \nabla l_i \right)^2} - \alpha\mathbf{P}\cdot\left[\mathbf{l}\left(\nabla\cdot\mathbf{l}\right) + \mathbf{l}\times\left(\nabla\times\mathbf{l}\right)\right] - K_u  l_z^2
\end{align}

\noindent This distorted spin cycloid structure can be characterized by using Jacobi elliptic functions with a parameter $m$, which quantifies the degree of anharmonicity from an ideal spin cycloid ($m=0$) to a square spin modulated structure ($m=1$). When the cycloid structure deviates from an ideal cycloid with $m=0$, third and higher odd order satellite peaks should appear in addition to the first order satellite peak seen in the neutron powder diffraction. This possibility of spin anharmonicity in BiFeO$_3$ has drawn some interest with conflicting reports over the past few years. As we note in Section~\ref{sec-electromagnon} the anharmonicity allows higher order zone-centre magnons to become dipole active, to become \emph{electromagnons}, in other words. Initially, Zalesski\u{i} {\it et al.}~\cite{Zalesskii2000} reported a possible distorted spin cycloid structure from $^{57}$Fe NMR experiments: NMR peak shapes give information about spatially modulated spin structures. The authors observed highly anisotropic profiles in BiFeO$_3$ samples and interpreted it as evidence of distorted spin cycloid structure with an elliptic parameter of $m=0.83$ at 77 K~\cite{Zalesskii2000}.

\begin{figure} \begin{center}
  \includegraphics[width=0.9\columnwidth]{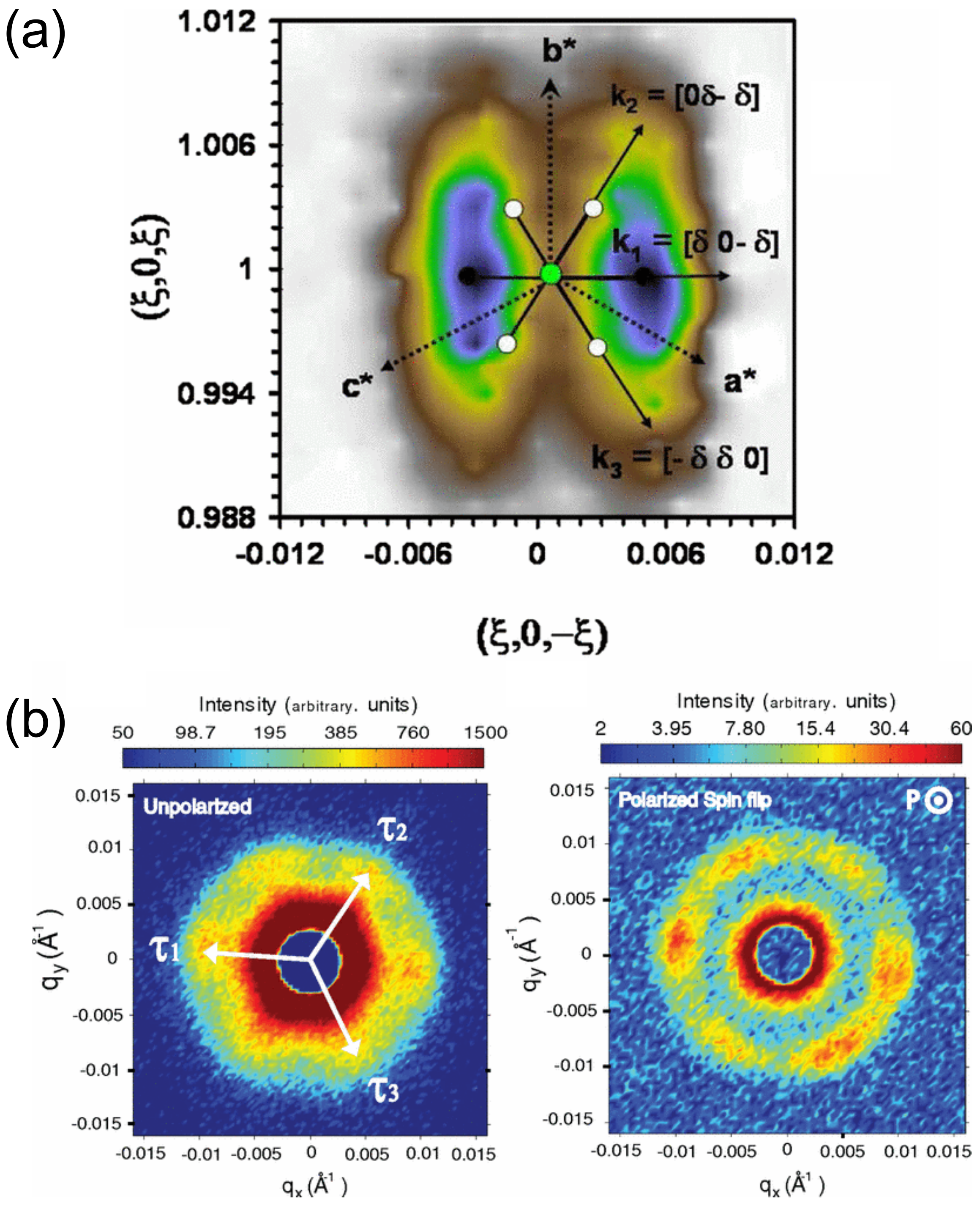}
  \caption{(a) Two diffraction satellites collected by single crystal neutron diffraction around the $(1/2,-1/2,1/2)_\mathrm{pc}$ Bragg reflection. Reprinted with permission from~\cite{Lebeugle2008}. Copyright 2008 by American Physical Society. (b) Diffraction pattern in the $(1~1~1)_\mathrm{pc}$ scattering plane collected by SANS using unpolarized neutrons (left) and neutrons polarized with $\mathbf{P} ||$beam (right). Reprinted with permissions from \cite{Ramazanoglu2011}. Copyright 2011 American Physical Society.
} \label{sf7}
\end{center} \end{figure}

On the other hand, Sosnowska {\it et al.}~\cite{Sosnowska2011} carried out new high-resolution powder neutron diffraction experiment specifically to address the issue of spin anharmonicity, finding no additional satellite peaks beyond first order that was distinguishable from background. This gave an upper bound of $m\lesssim 0.25$~\cite{Sosnowska2011}. However, a later single crystal neutron diffraction study reported $m=0.5$, larger than the powder diffraction results~\cite{Ramazanoglu2011a}. 

Interestingly, a recent NMR study found that the NMR spectrum shape and $T_2$ relaxation time are both strongly dependent on the amount of $^{57}$Fe isotope and the high frequency magnetic amplitude $h$~\cite{Pokatilov2010}. Including dynamical effects in their analysis for samples with 95\% $^{57}$Fe, Pokatilov {\it et al.} concluded that their results can be described by an undistorted spin cycloid structure. Therefore, there are four conflicting experimental data on the spin anharmonicity. As we will discuss in Section~\ref{sec-excitations}, we have carried out high-resolution inelastic neutron scattering experiments at low energies around the Brillouin zone centre, which enable a determination of parameters (such as the DM interaction and single ion anisotropy) of a microscopic spin Hamiltonian. When these parameters are used to determine $m$ as detailed in the Appendix~\ref{sec-ap-anharmonicity}, we conclude that the spin anharmonicity cannot be bigger than $m=0.65-0.71$, in agreement with the powder diffraction data and the latest NMR results~\cite{Catalan2009}. Finally, there is also a temperature dependence of the anharmonicity, with the cycloid become more harmonic at high temperatures. This is supported by the observed disappearance of a higher order electromagnon at high temperatures in Raman scattering measurements~\cite{Talbayev2011}.

\subsubsection{Additional low-temperature phase transitions} \label{sec-struct-3-2-3}

In addition to the well established N\'eel and ferroelectric transitions at high temperatures, several anomalies were observed in the temperature dependence of phonon frequencies measured by Raman scattering~\cite{Singh2008,Cazayous2008,Rovillain2009}, the dielectric constant and Young's modulus~\cite{Redfern2008} were ascribed to a low temperature spin reorientation or glass transitions in analogy with the rare-earth orthoferrites. Initial reports assigned a spin reorientation transition between $\approx$140 and $\approx$210~K, the temperatures when the spins undergo and complete their rotation~\cite{Singh2008,Cazayous2008}. Later field cooled magnetisation measurements~\cite{Singh2008prb} supported instead a view that spin glass behaviour sets in below $\approx$250~K. Notably, however, neutron diffraction measurements show no abrupt changes to the bulk cycloidal structure at these temperatures~\cite{Palewicz2010, Herrero-Albillos2010, Ramazanoglu2011a}: only a gradual and slight change of the cycloid period was observed~\cite{Ramazanoglu2011a}, and single crystal heat capacity and dielectric measurements~\cite{Lu2010} found no evidence of a low temperature phase transition so that if one does occurs in BiFeO$_3$, it does not affect the bulk magnetic structure.

A more comprehensive study by Jarrier {\it et al.}~\cite{Jarrier2012}, instead, points to a structural phase transition which only affects a $\approx$10~nm surface layer. The same group had previously reported that such a 'skin' behaves distinctly differently from the bulk, showing a sharp, $\approx$0.2~\%, thermal expansion of the out-of-plane lattice parameter at around 550~K~\cite{Marti2011} which correlates with the stiffening of phonon modes above this temperature. The authors speculate that the surface relaxation energies might be enough to overcome the small energy barriers between the many metastable states found by DFT calculations~\cite{Dieguez2011} and one of these may be stabilised in the surface below this 550~K transition.

The same grazing-incidence X-ray diffraction measurements also showed a large, $\approx$1~\%, thermal expansion in the out-of-plane lattice parameters between 140 and 180~K~\cite{Jarrier2012}, which corresponds to the phonon frequency anomaly observed by Raman scattering. The authors also observed peaks in the pyroelectric current at $\approx$140 and $\approx$210~K in an initial zero-field cooled measurement. On warming again, only the anomaly at $\approx$140~K was seen, whilst this is shifted to lower temperatures, and the $\approx$200~K anomaly reappears, albeit weakly, on subsequent electric field-cooled runs. This poling dependent shift of the peak current position to lower temperature was attributed to a current generated by emission of trapped charges within the bandgap. After the surface undergoes a structure phase transition, the Fermi level is abruptly altered, allowing interface defect states to cross above $E_F$ and hence release their charge. This is supported by electron paramagnetic resonance measurements, which shows an increase in the asymmetry, which is directly related to the conductivity, at $\approx$140~K, and DFT calculations by the same authors which showed that Bi vacancies can produce the required defect states trapped within the bulk bandgap~\cite{Jarrier2012}. Thus the 140~K surface phase transition may be due to the strains imposed by Bi vacancies in the `skin'. Since the magnetic domains are extremely sensitive to even small uniaxial strains~\cite{Ramazanoglu2011b} this may even account for the glassy state between 140 and 250~K seen in the magnetisation measurements~\cite{Singh2008prb}.

\subsection{Magneto-electric coupling} \label{sec-struct-3-3}

The magneto-electric coupling between the electric polarization and ordered magnetic moment is one of most interesting issues in BiFeO$_3$. In this section, we discuss the magneto-electric coupling by examining how the magnetic structure varies with magnetic and electric fields and temperature. 

\subsubsection{Applied magnetic field} \label{sec-struct-3-3-1}

\begin{figure} \begin{center}
  \includegraphics[width=0.9\columnwidth]{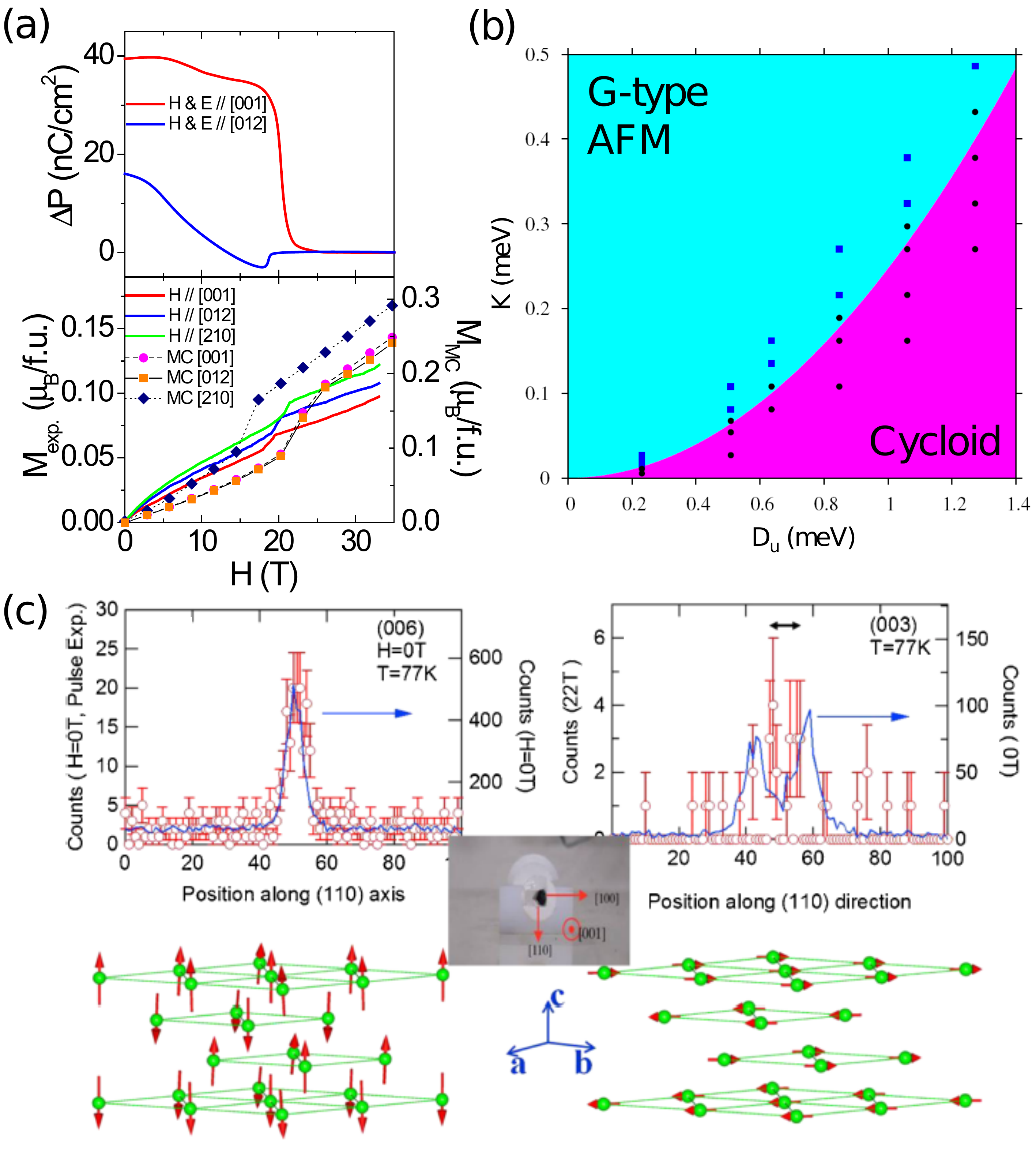}
  \caption{(a) Magnetoelectric current (top) and magnetization data with the MC simulation (bottom). (b) Magnetic phase diagram obtained from the Monte-Carlo simulation. (c) Top: Neutron diffraction data for the $(0,0,6)$ nuclear Bragg peak (left) at zero field  and for the $(0,0,3)$ magnetic Bragg peak (right) at zero field (line) and 22T (symbol) (right). The inset shows the picture of the sample. Bottom: Magnetic structures below (left) and above (right) the metamagnetic transition~\cite{Ohoyama2011}.} \label{sf10}
\end{center} \end{figure}

Similarly to a large easy-axis single ion anisotropy, a large applied magnetic field also suppresses the cycloid.
Experimentally, this was found to occur at around 20~T with $\mathbf{H}\parallel$[111]$_{\mathrm{pc}}$, yielding a homogeneous (canted) antiferromagnetic structure in which the moments flop from being approximately parallel or anti-parallel to [111]$_{\mathrm{pc}}$ to lying in the hexagonal basal plane~\cite{Popov1993,Tokunaga2010,Park2011,Ohoyama2011}. The transition occurs at slightly higher fields when $\mathbf{H}$ is applied in the hexagonal basal plane. This behaviour was satisfactorily described by Monte-Carlo modelling~\cite{Ohoyama2011} and Landau-Ginzburg theory~\cite{Popov1993}. As the SIA increases, the critical field is expected to decrease, which behaviour was found by Ref.~\cite{Gareeva2013} from Landau-Ginzburg theory. The authors actually found that before the homogeneous easy plane phase is reached, a conical cycloid is formed in which the spins now cant out of the cycloidal plane but maintain a spiral arrangement. Whereas the easy-axis anisotropy term reduces the critical field, and leads to an easy-plane structure at high fields, an \emph{easy-plane anisotropy} or equivalently a large Dzyaloshinskii-Moriya interaction since the two are treated within the same reduced anisotropy variable in the phase diagram of Ref.~\cite{Gareeva2013}, causes a phase transition to an \emph{easy-axis} phase at higher critical fields. Furthermore, when this term is large, then a homogeneous tilted phase is stabilized at high fields, in which the spins are collinearly aligned but at an angle to the symmetry axes. 

The meta-magnetic transition around 20~T from the incommensurate (IC) cycloid to commensurate (C) canted antiferromagnet was first observed in magnetisation measurements~\cite{Popov1993}. Later ESR measurements at 4.2~K up to 25~T found a clear change in the electron spin resonance frequency, where previous magnetization measurements reported an anomaly, and explained it as the IC to C transition from a Landau-Ginzburg free energy analysis~\cite{Ruette2004}. The authors also observed that the transition is hysteretic, and thus of first order, which was also seen in pulsed-field magnetometry~\cite{Wardecki2008}. Single crystal neutron diffraction measurements in pulsed fields up to 28~T also found that the incommensurate satellite peaks merged into a single commensurate magnetic peak within experimental resolution~\cite{Ohoyama2011}. Wardecki {\it et al.}~\cite{Wardecki2008} also found that the critical magnetic field decreases from 18.1~T at 80~K to 16.1~T at 235~K. More recently, Fishman analysed the directional dependence of the critical field~\cite{Fishman2013} to find that the critical field $H_c$ is dependent on the competition between magnetic domains.

When BiFeO$_3$ undergoes a field-induced transition from an incommensurate to a commensurate magnetic structure, it also exhibits a clear change in its electric polarization values. For example, Tokunaga {\it et al.} and Park {\it et al.} measured an induced polarization and magnetization simultaneously under magnetic field~\cite{Park2011,Tokunaga2010}, with a relative change of about ~40 nC/cm$^2$ in the electric polarization above the commensurate magnetic transition (See Fig. 12). Although it appears to be somewhat consistent with the spin current model~\cite{Katsura2005}, an increase of the polarization is expected of the correct spin chirality determined by circularly polarised x-ray measurements~\cite{Johnson2013}. We note, however, that because of the experimental configuration it was not determined whether the change increases or decreases the total polarization.

Finally it should be noted that in the high field homogeneous phase, a linear magneto-electric effect is permitted by theory. This was experimentally observed in measurements of the spontaneous polarisation as a function of field in both polycrystalline~\cite{Popov2001} and single crystal samples~\cite{Tokunaga2010,Park2011}. Below the meta-magnetic transition, in the presence of the cycloid, the magneto-electric coupling is expected to be quadratic. This was confirmed by Tabares-Mu\~noz {\it et al.} who measured the induced electric polarization in magnetic fields up to 1 Tesla using single crystal samples~\cite{Tabares-Munoz1985}. They observed that the induced polarization is proportional to the square of applied magnetic field, $H^2$, and determined the nonlinear magnetoelectric susceptibility tensor.

\subsubsection{Applied electric field} \label{sec-struct-3-3-2}
 
An applied electric field can switch the electric polarisation between one of the four equivalent $[111]_{\mathrm{pc}}$ directions~\cite{Kubel1990, Lebeugle2008}. In addition, single crystal neutron diffraction measurements showed that while applying an electric field along $[010]_{\mathrm{pc}}$ on an as-grown sample with a single ferroelectric domain changes the polarisation by 71$^{\circ}$, the magnetic propagation vector remains parallel to the $[10\bar{1}]_{\mathrm{cubic}}$ direction~\cite{Lebeugle2008}. Since the spin cycloid rotation plane is defined by the polarization direction and the magnetic propagation vector, this means that the electric field changes the spin rotation plane. At the same time, Lee {\it et al.} reported similar changes in the magnetic domain population induced by electric field~\cite{Lee2008a, Lee2008} (See Fig. 13).  They also discovered, using polarized neutron diffraction, that a single chiral vector belongs to a single magnetic domain.

\begin{figure} \begin{center}
  \includegraphics[width=0.9\columnwidth]{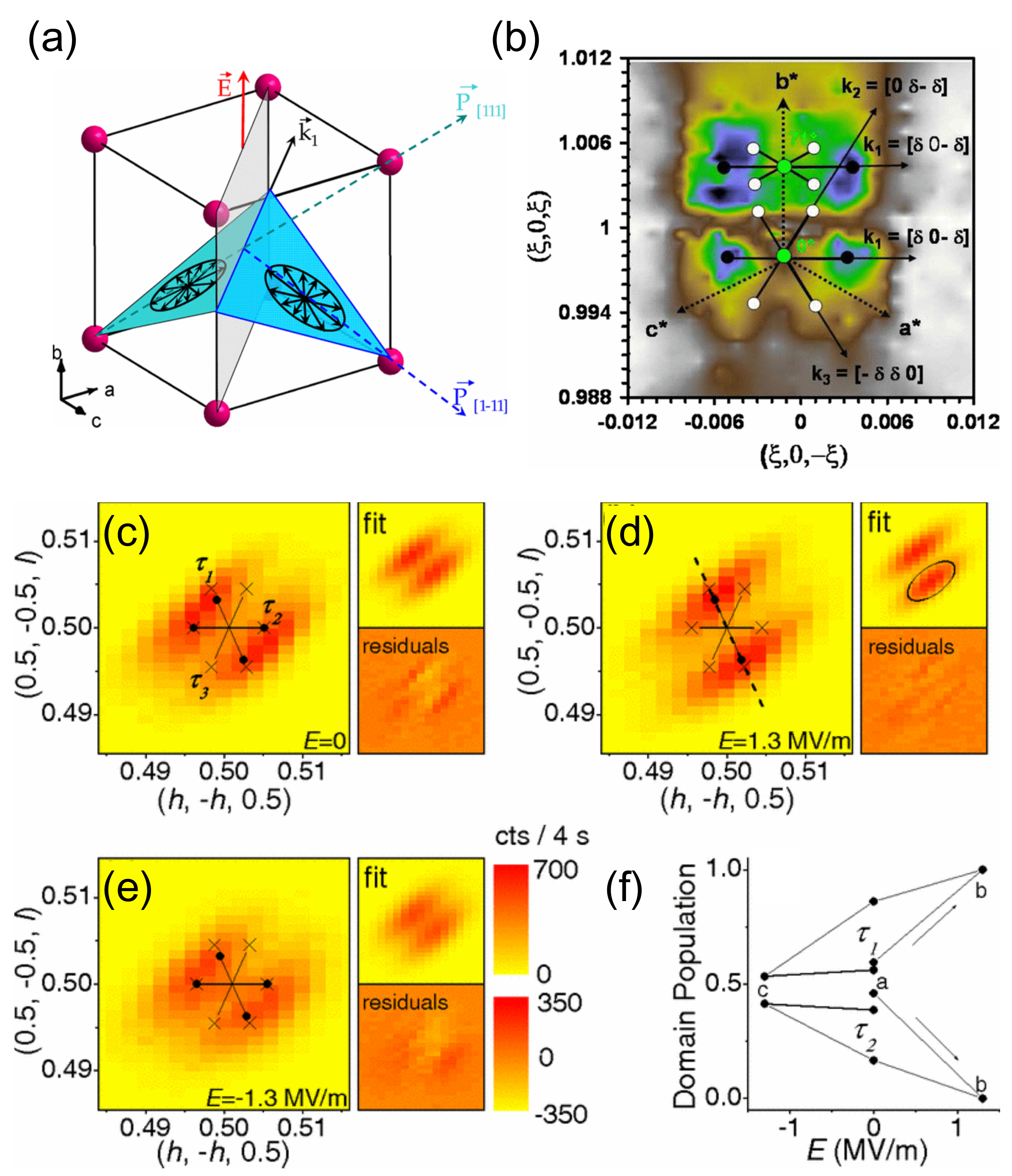}
  \caption{(a) The spin cycloid plane for the two ferroelectric domain separated by a domain wall (light gray). (b) Diffraction satellites in the multidomain state around the $(1/2,-1/2,1/2)_{pc}$ Bragg peak. The black and white spots indicate theoretical positions. Reprinted with permission from \cite{Lebeugle2008}. Copyright 2008 American Physical Society. Magnetic peaks near the $(1/2,−1/2,1/2)_\mathrm{pc}$ Bragg peak for (c) $E=$ 0 in the as-prepared sample, for (d) $E=$ 1.3 MV/m, and for (e) $E=-1.3$ MV/m. (f) Populations of the magnetic domains 1 and 2 as functions of $E$. Reprinted with permission from~\cite{Lee2008}. Copyright 2008 by American Physical Society.} \label{sf11}
\end{center} \end{figure} 

Whilst a linear magnetoelectric coupling would result in the electric field enhancing the easy-axis anisotropy in the Landau free energy~\cite{Sparavigna1994}, a quadratic magnetoelectric term contributes to the Lifshitz invariant term and thus enhances the cycloid~\cite{Zhdanov2006}. In the former case, a transition from the cycloid structure to a homogeneous antiferromagnet would occur at zero applied magnetic field for some critical electric field estimated as of the order of 10$^8$~V/cm~\cite{Sparavigna1994}. However, the linear magnetoelectric coupling is globally cancelled out in the cycloidal structure, and it is more likely that the second case occurs, wherein the electric field acts to maintain the cycloid. Thus no $\mathbf{H}=0$ transition should be expected as a function of electric field, and finite electric fields should increase the critical magnetic field at which the cycloid is destroyed~\cite{Zhdanov2006}. However, de Sousa {\it et al.}~\cite{DeSousa2013} has recently proposed a different linear magneto-electric coupling mechanism derived from a tight-binding model in which the single ion anisotropy of the Fe spins has a large contribution from the hybridisation between Fe $3d$ and Bi $6p$ electrons and is thus enhanced by the large spin-orbit coupling of the Bi ions. An electric field increases this hybridisation and thus enhances the easy-axis anisotropy, leading to a phase transition at finite $\mathbf{E}$ in particular applied field directions. This model was used to describe the electric field behaviour of magnons observed by Raman spectroscopy where a $\approx$5~cm$^{-1}$ linear shift of the magnons was observed in fields up to 100~kV/cm~\cite{Rovillain2010}.

A much larger electric field, on the other hand, is predicted to cause a structure transition. Effective Hamiltonian calculations~\cite{Lisenkov2009} found that a transition to the tetragonal $P4mm$ structure occurs with $\mathbf{E}\parallel [001]_{\mathrm{pc}} >$ 22.8~MV/cm in BiFeO$_3$ thin films, whereupon the AFD octahedral rotation is suppressed. This high theoretically calculated field is expected to be decreased in real world measurements due to inhomogeneities in samples which should enhance the effect of the field on the polarisation~\cite{Lisenkov2009}. At fields below this transition, a first order isostructural transition between two $Cc$ phases is predicted which is identified with the experimentally determined flop of the cycloidal plane when the applied field ($E>E_c\approx$12~kV/cm) changes the polarisation direction in BiFeO$_3$ single crystals mentioned above~\cite{Lee2008,Lebeugle2008}. In addition, the cycloidal structure itself can give rise to an additional electronic contribution to the polarisation under the spin-current model proposed by Katsura, Nagaosa and Balatsky~\cite{Katsura2005} (and thus also termed the KNB model), where $P\propto \mathbf{r}_{ij}\times (\mathbf{S}_i\times\mathbf{S}_j )$. A similar contribution was attributed to the flexomagnetoelectric effect through an analogy between the Lifshitz invariant describing the cycloid in BiFeO$_3$ and that describing the director of liquid crystals by Zvezdin {\it et al.}~\cite{Zvezdin2009,Zvezdin2012}. The Lifshitz invariant was found to arise from the DM interaction, so that one may think of the KNB model and the flexomagnetoelectric effect as the microscopic and macroscopic views of the same physical phenomenon, which results from the microscopic coupling of the cycloid to the polarisation. This means that the energy of the cycloid is minimised when the cycloidal plane is perpendicular to $\mathbf{P}$. Thus when the applied electric field switches the polarisation direction, the cycloid plane also flops as a result. This effect was subsequently used to image the change in magnetic domains by an applied magnetic field in polarised neutron diffraction measurements~\cite{Ratcliff2013}. In addition to affecting the polarisation and magnetic cycloid plane, the electric field is also predicted to change the AFD rotation and the combination of these effects leads to a rich (calculated) phase diagram with up to four different phases as a function of field direction~\cite{Lisenkov2009}.

\subsubsection{Evolution with temperature} \label{sec-struct-3-3-3}

\begin{figure} \begin{center}
  \includegraphics[width=0.9\columnwidth]{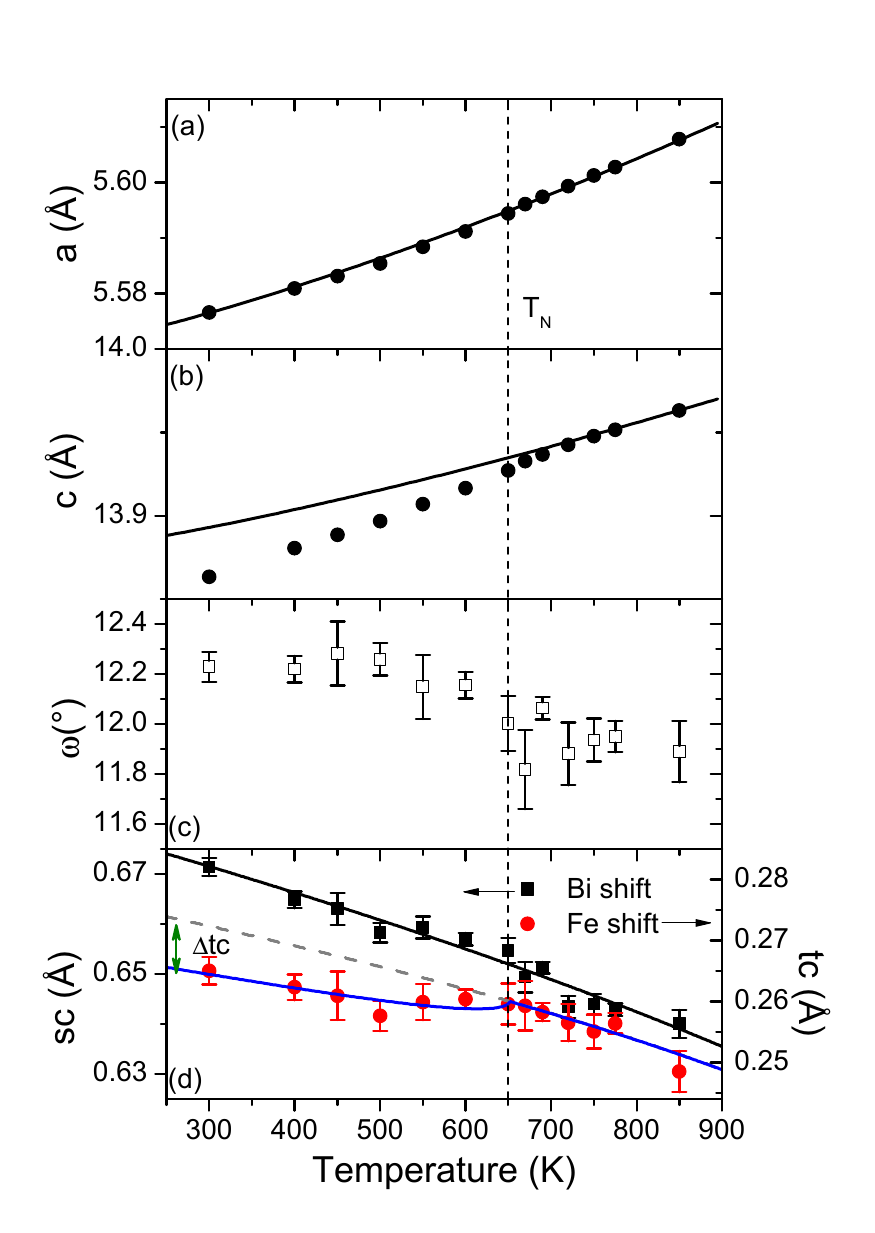}
  \caption{(a) and (b) Temperature dependence of the lattice constants. The solid lines represent theoretical calculations using a Debye-Gr\"{u}neisen formula. (c) The anti-phase rotation angle $\omega$ of the oxygen octahedron (d) The Bi (sc) and Fe (tc) shifts along the $c$-axis. The solid (dashed) lines are the theoretical curves for the first-order ferroelectric order parameter based on the Ginzburg-Landau free energy analysis with (without) the magnetoelectric coupling term~\cite{Lee2013}.} \label{sf12}
\end{center} \end{figure}

Apart from the external magnetic and electric field, there is another way to examine the magnetoelectric coupling, namely via the change in polarisation as BiFeO$_3$ becomes magnetically ordered. The N\'eel transition is expected to be accompanied by ionic displacements, which can contribute to the electric polarisation. This kind of magnetoelectric coupling through a spin-lattice coupling has been shown to be rather strong in hexagonal {\it R}MnO$_3$~\cite{Lee2008, Fabreges2009, Park2010, Oh2013}. Therefore, it is also natural to look for such a possibility in BiFeO$_3$, which requires an accurate measurement of atomic positions. Some of earliest measurements were reported by several groups including Palewicz {\it et al.}, who conducted powder neutron diffraction measurements to obtain structural information such as the lattice constant, rhombohedral angle and cation shifts~\cite{Palewicz2007}. According to their results, there is a clear change in the slope of the $c$ lattice constant and rhombohedral angle near $T_N$. Park {\it et al.} also carried out synchrotron X-ray diffraction from 373 to 895~K (in 6~K steps) in addition to single crystal neutron diffraction studies and confirmed a similar anomaly in the temperature dependence of $c$ and that the rhombohedral angle has a minimum slightly above $T_N$~\cite{Park2011}. Smirnova {\it et al.} employed an ultrasonic pulsed-echo technique and found that the longitudinal elastic constant $C_L$ shows an anomaly slightly above $T_N$~\cite{Smirnova2011}. These results, taken together, constitute strong evidence of a spin-lattice coupling. However, it is fair to say that despite these studies, the details of how the electric polarization is coupled to the antiferromagnetic order remains to be determined. To gain insights into the microscopic details of the magnetoelectric coupling, Lee {\it et al.} carried out extensive studies of the temperature dependence of the crystal structure using the high-resolution single crystal neutron diffractometer, D9 at ILL, with a neutron wavelength of 0.835~\AA~that is much shorter than usual in neutron diffraction experiments~\cite{Lee2013}. Equipped with a wide area detector, this enabled us to examine the atomic positions with extremely high accuracy, which was then used to calculate the cation shift directly and ultimately the electric polarization induced by the magnetic ordering.
We also ought to stress that our findings of the noticeable magnetoelastic coupling below T$_N$ but with a very weak temperature dependence in the atomic positions is at variance with the claim of a strong temperature dependence in the atomic position made for 0.9BiFeO$_3$--0.1BaTiO$_3$~\cite{Singh2008prl}.

As we discussed above, BiFeO$_3$ has the $R3c$ structure denoted $a_+^-a_+^-a_+^-$ in the Glazer tilting system. The anti-phase octahedral tilting does not break inversion symmetry, so cannot induce a polarization itself, while the Bi and Fe cation shifts are directly related to electric polarization. Therefore these cation shifts can be taken as the order parameter of a Landau phase transition. It is important to note that one needs to distinguish between the nonmagnetic Bi shift (sc) and magnetic Fe shift (tc), and thus their respective contributions to a total electric polarization, in order to determine the magnetically induced polarisation change, which we achieved in the high-resolution neutron diffraction studies. s and t denote the relative coordinates of Bi or Fe ions along $c$ in hexagonal notation and sc and tc denote the shift with respects to their ideal positions in the cubic perovskite structure.

Since the ferroelectric transition of BiFeO$_3$ is of first order, one can write down the free energy by including terms up to 6$^{\mathrm{th}}$ order in the order parameter without a magnetoelectric coupling term:

\begin{equation} \label{eq:structME}
F_{\mathrm{Bi,Fe}} = \frac{\alpha_2}{2} P^2_{\mathrm{Bi,Fe}} + \frac{\alpha_4}{4} P^4_{\mathrm{Bi,Fe}} + \frac{\alpha_6}{6} P^6_{\mathrm{Bi,Fe}} \ ,
\end{equation}

\noindent where $P_{\mathrm{Bi,Fe}}$ represents the electric polarization associated with Bi and Fe ions individually.

The free energy can then be minimized for a given order parameter to obtain its theoretical temperature dependence, which can be compared with experimental values estimated by the measured Bi and Fe shifts. The accuracy of the measurements can be tested by calculating the electric polarization from the experimental results. If we assume that the ionic charges correspond to the nominal valences of Bi and Fe (+3), then our estimate of the total electric polarization is $P=73~\mu$ C/cm$^{2}$, which is comparable to the electric polarization measured on single crystals~\cite{Choi2009}. Of further interest, as shown in Figure~\ref{sf12}, the Bi shift follows the theoretical line of a first order ferroelectric transition remarkably well whereas the Fe shift deviates clearly from the theoretical line below the Neel temperature. It was found that the Fe shift is reduced by about ~0.007~\AA~at 300 K from the theoretical line without the magnetoelectric coupling. Most interestingly, the reduced Fe shift amounts to a reduction in the polarization by as much as $\Delta P\sim-0.4 \mu$ C/cm$^{2}$ or $\Delta P/P\sim-0.55$~\%. We should stress that this polarization reduction is one order larger than what is expected from the inverse DM interaction. Moreover, the contribution due to the inverse DM interaction should be positive for the spin chirality determined from the circularly polarised X-ray scattering~\cite{Johnson2013}, which is opposite to that found in our single crystal neutron diffraction studies. Therefore, one can conclude that the spin-lattice coupling through magnetic exchange-striction gives rise to a magnetoelectric coupling in BiFeO$_3$. Recent theoretical studies also confirm such a magnetoelectric coupling via a spin-lattice coupling and their results seem to be in good agreement with our conclusions~\cite{Lee_Fishman}.

Finally, this spin-lattice coupling may also explain the reduction in the photostrictive effect observed in BiFeO$_3$ when a magnetic field is applied~\cite{Kundys2010}. The visible light induced lattice change is thought to be due to a combination of the photovoltaic and piezoelectric effects, where the electric field from an electron-hole pair excited by incident photons creates a strain in the crystal. An applied magnetic field, however, may cause a counter-strain opposing the photostriction via the spin-lattice coupling. Further theoretical work is required to establish this, however.

\section{Magnons, phonons and electromagnons} \label{sec-excitations}

Having extensively covered the crystal and magnetic structure of BiFeO$_3$ we now turn to the dynamical behaviour of the atoms and electrons. The quantised excitations of the lattice or magnetic subsystem, namely phonons and magnons, are completely described by their dispersion relation $\omega(\mathbf{q})$,~and a measurement of this can yield the interaction strengths (force constants and exchange constants) of the atoms and magnetic moments involved. These measurements may also be exquisitely sensitive to the symmetry of the atoms involved and thus may also contribute to knowledge of the atomic structure~\cite{Palai2008,Zbiri2012,Jarrier2012}. Finally, the magneto-electric coupling can also be studied by measuring the quantised excitation unique to multiferroics, namely \emph{electromagnons}, which are hybrid excitations of the spins and lattice that have characteristics of both.

The advantage of inelastic neutron scattering in measuring these disparate excitations is in its ability to measure the full momentum dependence of the phonons or magnons, whereas the wavelength of light at the frequencies of these excitations is much larger than the distance between atoms, so that the momentum transfered by the photon to the sample is small enough to restrict optical spectroscopies to near the Brillouin zone centre. In addition, the large neutron-magnetic scattering cross-section compared with that of light is advantageous for measurements of the spin dynamics~\cite{JeongJKMS2012}. Thus the exchange interactions were determined by neutron scattering as we describe below. However, the energy resolution of even state-of-the-art inelastic neutron spectrometers compares relatively poorly to THz spectroscopy at low energies, where the anisotropic Dzyaloshinskii-Moriya (DM) interaction and single-ion anisotropy (SIA) dominates. Thus optical spectroscopic techniques, especially terahertz spectroscopy has proved valuable in determining these interactions in BiFeO$_3$. One aspect, unique to BiFeO$_3$, which facilitates this is the long period of the magnetic cycloid structure giving a small modulation wavevector $q$, and the mixing of modes at integer harmonics $nq$ due to the DM interactions and SIA allowing such modes to probed by optical techniques~\cite{Talbayev2011,Fishman2012,Nagel2013,Fishman2013a}. Furthermore, the hybrid electromagnons have only been seen with optical techniques in BiFeO$_3$ because the polarised neutron measurements which would separate the spin and ionic components of these excitations are difficult to conduct requiring quite large single crystals.
We shall now discuss the magnetic excitations, followed by the lattice dynamics and ending this section with the electromagnons.

\subsection{Magnon} \label{sec-magnon}

Spin waves are the collective excitations of a magnetic fluctuation in the spin structure, which can be quantized as a quasi-particle called a \emph{magnon} with definite energy and momentum and possessing spin $S=1$. The excitation is fully described by the magnon dispersion relation $\omega(\mathbf{q})$, a measurement of which is sufficient to determine the underlying interactions governing the spin dynamics. These interactions include the nearest and next-nearest exchange, and Dzyaloshinskii-Moriya interactions and the single-ion anisotropy, and were first included in a Hamiltonian to describe the high field spin structure in a Monte Carlo calculation~\cite{Ohoyama2011}. They were subsequently determined in detail by spectroscopic measurements~\cite{Jeong2012,Matsuda2012,Fishman2012,Fishman2013a}, as described in the next two subsections which cover, respectively, the high and low energy spin dynamics.

\subsubsection{High energy spin dynamics: Super-exchange interaction} \label{sec-highE}

\begin{figure} \begin{center}
  \includegraphics[width=0.9\columnwidth]{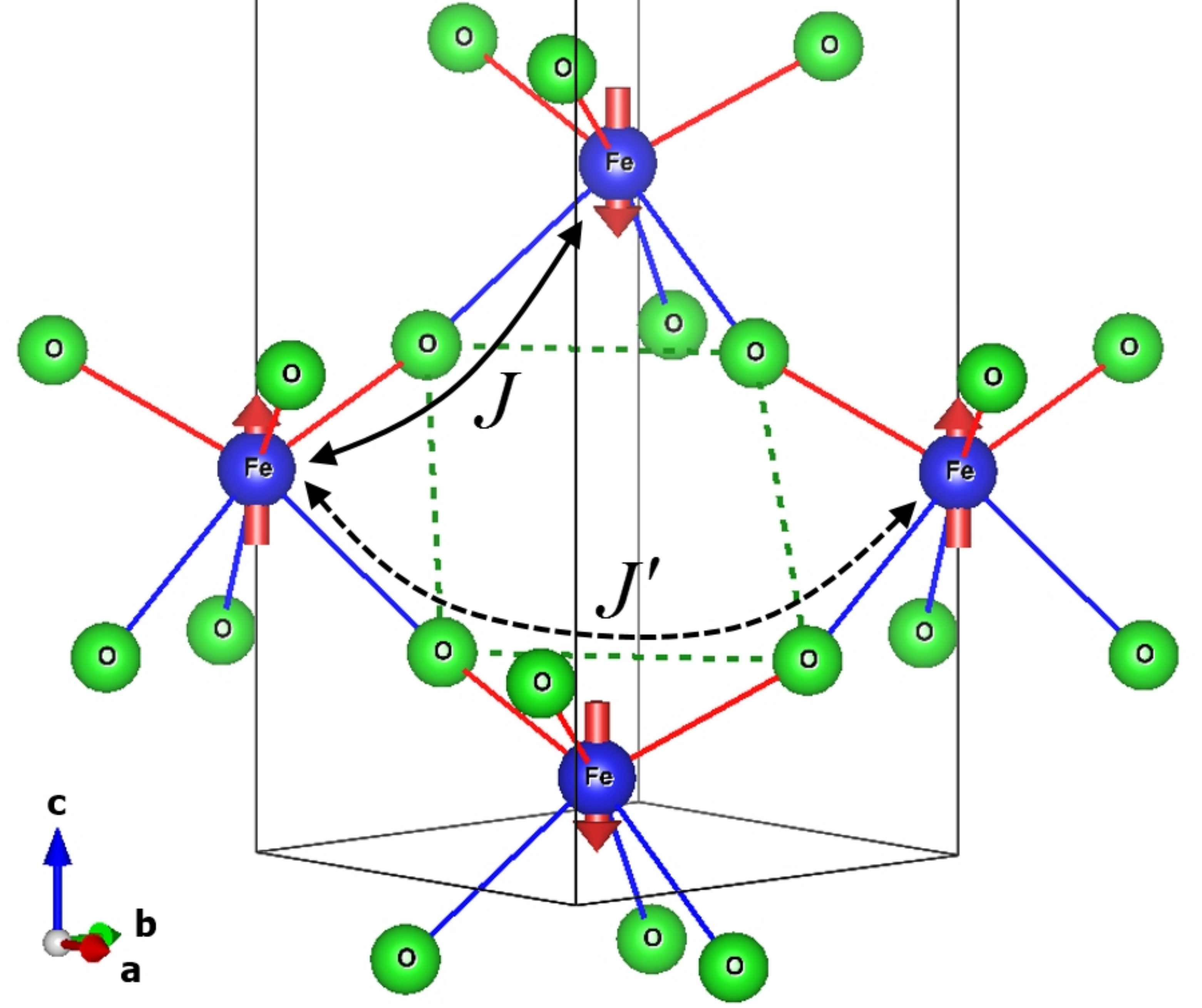} 
  \caption{The Fe-O-Fe and Fe-O-O-Fe paths for the superexchange interaction between the NN and NNN in the distorted FeO$_6$ octahedra. Blue and red dashed lines represent a short and long Fe-O bond, respectively.} \label{jf1}
\end{center} \end{figure}

The principal interactions, which determines the local G-type antiferromagnetic structure and high N\'eel temperature ($T_\mathrm{N}\approx650$~K) in BiFeO$_3$, are the nearest- and next-nearest neighbour exchange interactions (denoted NN and NNN in the subsequent discussion respectively). Since the NN Fe-O-Fe bond has an angle close 180$^{\circ}$, this interaction is strongly antiferromagnetic resulting in a structure in which nearest neighbouring spins align anti-parallel. The NNN interaction is also antiferromagnetic, and both interactions are shown in Figure~\ref{jf1}. Due to the requirement of sizeable single crystals, the magnon dispersion measurement which could directly determine these interactions was not completed until recently, when Jeong {\it et al.}~\cite{Jeong2012}, for the first time, reported the magnon spectrum across the full Brillouin zone using inelastic neutron scattering with a time-of-flight spectrometer, and Matsuda {\it et al.}~\cite{Matsuda2012} reported a similar study using a triple-axis spectrometer subsequently. 

The spin Hamiltonian including only the Heisenberg exchange interactions is given by: 
\begin{equation} \label{eq:Hexch}
\mathcal{H}_{\mathrm{exch}} = \sum_{\mathbf{r},\Delta} \mathcal{J}_{\Delta} \mathbf{S}_{\mathbf{r}}\cdot\mathbf{S}_{\mathbf{r}+\Delta} = \mathcal{J} \sum_{\mathbf{r},\mathbf{\alpha}} \mathbf{S}_{\mathbf{r}}\cdot\mathbf{S}_{\mathbf{r}+\mathbf{\alpha}} + \mathcal{J'} \sum_{\mathbf{r},\mathbf{\beta}} \mathbf{S}_{\mathbf{r}}\cdot\mathbf{S}_{\mathbf{r}+\mathbf{\beta}} \ ,
\end{equation}
where $\mathcal{J}_\Delta$ is the interaction parameter for neighbouring spins with the displacement vector $\Delta$ (See Fig. 15).  Here, $\mathcal{J}=\mathcal{J}_\mathbf{\alpha}$ and $\mathcal{J'}=\mathcal{J}_\mathbf{\beta}$ are for the NN and NNN with the displacement vectors $\mathbf{\alpha}$ and $\mathbf{\beta}$, respectively. 
The magnon dispersion relation $\omega(\mathbf{q})$ and structure factor $S(\mathbf{q},\omega)$ can be calculated using the Holstein-Primakoff boson operators. Assuming a G-type antiferromagnetic (AFM) structure, the spin operators at two AFM sublattices can be written as 
\begin{equation} \label{eq:HPoper}
\begin{array}{l}
S_1^z = S - b_1^{\dagger}b_1 \\
S_1^+ = \sqrt{2S} b_1 \\
S_1^- = \sqrt{2S} b_1^{\dagger}	\end{array} , ~~
\begin{array}{l}
S_2^z = -S + b_2^{\dagger}b_2 \\
S_2^+ = \sqrt{2S} b_2^{\dagger} \\
S_2^- = \sqrt{2S} b_2			\end{array}
\end{equation}
with the boson operators $b_i$ and $b_i^\dagger$ obeying $[ b_i, b_j^\dagger ]=\delta_{ij}$. In other words, the spin operator vector at the position $\mathbf{r}$ can be expressed as
\begin{equation} \label{eq:Soper}
\mathbf{S}_\mathbf{r}= \mathrm{R}_x \left( \mathbf{Q}_G \cdot \mathbf{r} \right) \mathbf{T}_\mathbf{r} ~ , ~~
\begin{array}{l}
T_{\mathbf{r}}^x =   \sqrt{S/2}\left(b_\mathbf{r} + b_\mathbf{r}^{\dagger}\right) \\
T_{\mathbf{r}}^y = -i\sqrt{S/2}\left(b_\mathbf{r} - b_\mathbf{r}^{\dagger}\right) \\
T_{\mathbf{r}}^z = S - b_\mathbf{r}^{\dagger}b_\mathbf{r} 	\end{array}
\end{equation}
where $R_x$ is the rotation matrix about the $x$-axis, $\mathbf{T}_\mathbf{r}$ is the spin operator vector defined in a local frame, in which spins are directed to the positive $z$-direction, and $\mathbf{Q}_G$ is the magnetic propagation vector for the collinear G-type AFM state in BiFeO$_3$: 
\begin{equation} \label{eq:QG}
\mathbf{Q}_G = \left[0~0~3\right]_{\mathrm{hex}} 
           = \left[0.5~0.5~0.5\right]_{\mathrm{pc}}.
\end{equation}
The dot product in the Hamiltonian is simply rewritten as $\mathbf{S}_{\mathbf{r}}\cdot\mathbf{S}_{\mathbf{r'}} = \mathbf{T}_{\mathbf{r}}^T \mathrm{R}_x \left(\mathbf{Q}_G\cdot\Delta\right) \mathbf{T}_{\mathbf{r'}}$ and the Hamiltonian can be rewritten as

\begin{multline} \label{eq:Hex-r}
\mathcal{H}_\mathrm{exch} = \sum_{\mathbf{r},\Delta} \mathcal{J}_\Delta\mathrm{cos} \left(\mathbf{Q}_G\cdot\Delta\right) \left[ S^2-2S b_\mathbf{r}^{\dagger}b_{\mathbf{r}} \right] \\
+\frac{S}{2}\sum_{\mathbf{r},\Delta} \mathcal{J}_\Delta \left[ 1+\mathrm{cos} \left(\mathbf{Q}_G\cdot\Delta\right) \right] K_{\mathbf{r},\mathbf{r}+\Delta} \\
+\frac{S}{2}\sum_{\mathbf{r},\Delta} \mathcal{J}_\Delta \left[ 1-\mathrm{cos} \left(\mathbf{Q}_G\cdot\Delta\right) \right] P_{\mathbf{r},\mathbf{r}+\Delta} 
\end{multline}

\noindent where $K_{\mathbf{r},\mathbf{r'}}=b_\mathbf{r}b_{\mathbf{r'}}^{\dagger}+b_\mathbf{r}^{\dagger}b_{\mathbf{r'}}$ and $P_{\mathbf{r},\mathbf{r'}}=b_\mathbf{r}b_{\mathbf{r'}}+b_\mathbf{r}^{\dagger}b_{\mathbf{r'}}^{\dagger}$. Taking the Fourier transform

\begin{gather} \label{eq:FourierT}
b_{\mathbf{r}}=\frac{1}{\sqrt{N}} \sum_{\mathbf{k}} e^{i\mathbf{k}\cdot\mathbf{r}}b_{\mathbf{k}} ~, ~~ b_{\mathbf{r}}^{\dagger}=\frac{1}{\sqrt{N}} \sum_{\mathbf{k}} e^{-i\mathbf{k}\cdot\mathbf{r}}b_{\mathbf{k}}^{\dagger} \ ,
\end{gather}

\noindent the Hamiltonian can be rewritten as
\begin{equation} \label{eq:Hex_k}
\mathcal{H}_\mathrm{exch} = NS^2 I_0 + S \sum_\mathbf{k} \left(\begin{array}{cc} b_{\mathbf{k}}^{\dagger} & b_{\mathbf{-k}} \end{array}\right) \left(\begin{array}{cc} A_{\mathbf{k}} & B_{\mathbf{k}} \\ B_{\mathbf{k}} & A_{\mathbf{k}} \end{array}\right) \left(\begin{array}{cc} b_{\mathbf{k}} \\ b_{\mathbf{-k}}^{\dagger} \end{array}\right) \ ,
\end{equation}
where 
\begin{gather}
A_\mathbf{k} = \frac{1}{2}\left(I_\mathbf{k}+J_\mathbf{k}\right)-I_0 ~,~~ B_\mathbf{k} = \frac{1}{2}\left(J_\mathbf{k}-I_\mathbf{k}\right) , \\
I_\mathbf{k} = \sum_{\Delta} \mathcal{J}_\Delta \mathrm{cos}\left(\mathbf{Q}_G\cdot\Delta\right) \mathrm{cos}\left(\mathbf{k}\cdot\Delta\right) , \label{eq:Ik} \\
J_\mathbf{k} = \sum_{\Delta} \mathcal{J}_\Delta \mathrm{cos}\left(\mathbf{k}\cdot\Delta\right) . \label{eq:Jk}
\end{gather}
Performing the Bogoliubov transformation
\begin{equation} \label{eq:BogT}
\left(\begin{array}{c} b_{\mathbf{k}} \\ b_{\mathbf{-k}}^{\dagger} \end{array}\right) = \left(\begin{array}{cc} \mathrm{cosh}~\lambda_{\mathbf{k}} & \mathrm{cosh}~\lambda_{\mathbf{k}} \\ \mathrm{cosh}~\lambda_{\mathbf{k}} & \mathrm{cosh}~\lambda_{\mathbf{k}} \end{array}\right) \left(\begin{array}{cc} \gamma_{\mathbf{k}} \\ \gamma_{\mathbf{-k}}^{\dagger} \end{array}\right)
\end{equation}
with $\mathrm{tanh}~2\lambda_\mathbf{k} = -B_\mathbf{k}/A_\mathbf{k}$, the Hamiltonian is diagonalized as
\begin{gather} \label{eq:Hex_final}
\mathcal{H}_\mathrm{exch} = E_{cl} + S\sum_\mathbf{k}\omega_\mathbf{k} + \sum_\mathbf{k} 2S\omega_\mathbf{k} \gamma_{\mathbf{k}}^{\dagger} \gamma_{\mathbf{k}}~,\\
\omega_\mathbf{k} = \sqrt{A_{\mathbf{k}}^2 - B_{\mathbf{k}}^2}=\sqrt{\left(J_{\mathbf{k}}-I_0\right)\left(I_{\mathbf{k}}-I_0\right)}~,
\end{gather}
where $E_{cl}=NS^2I_0$ is the classical ground-state energy, the second term gives an $O(1/S)$ quantum correction to the energy, and $2S\omega_\mathbf{k}$ is the magnon energy.

\begin{figure} \begin{center}
  \includegraphics[width=0.9\columnwidth]{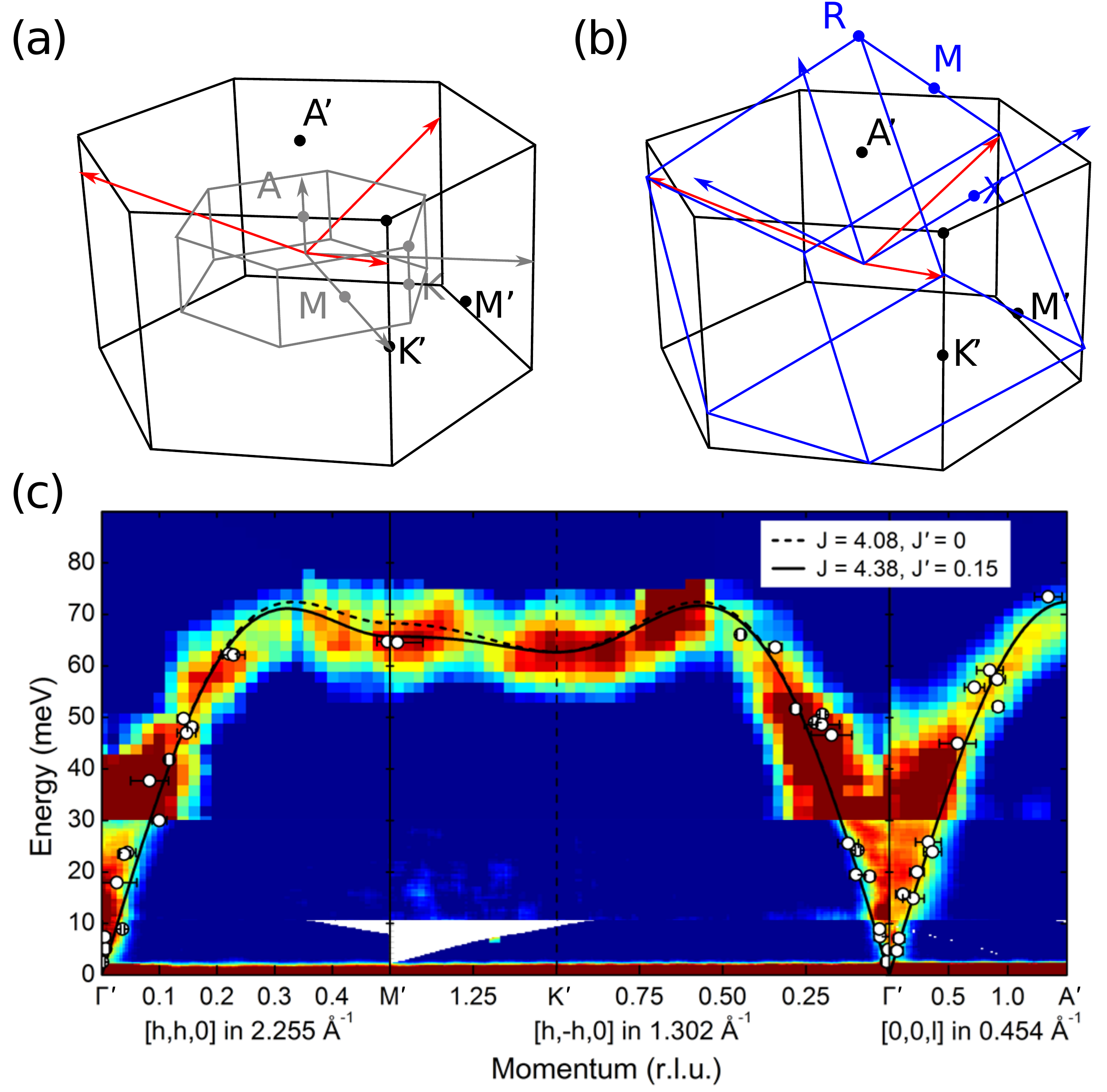} 
  \caption{(a) The usual (gray) and conventional (black) hexagonal Brillouin zone. (b) Comparison with pseudocubic Brillouin zone (blue). Red arrows denote the rhombohedral reciprocal vectors. (c) Full experimental magnon dispersion with several $\mathcal{J}$-$\mathcal{J'}$ models~\cite{Jeong2012}.} \label{jf2}
\end{center} \end{figure}

Simple algebraic expressions can be obtained for the high symmetry points, shown in Figure~\ref{jf2}(a), and given by

\begin{gather}
\omega_{M'} = 4S\sqrt{2\left({\mathcal{J}_0}^2 - 2\mathcal{J}_0\mathcal{J'}\right)} \ , \\
\omega_{K'} = 3S\sqrt{3\left({\mathcal{J}_0}^2 - \mathcal{J'}^2\right)} \ , \\
\omega_{A'} = 6S(\mathcal{J}-2\mathcal{J'}) \ ,
\end{gather}
where the adapted high symmetry points $M'=(\frac{1}{2},\frac{1}{2},0)_{\mathrm{hex}}=(\frac{1}{2},\bar{\frac{1}{2}},0)_{\mathrm{pc}}$, $K'=(\frac{1}{2},0,0)_{\mathrm{hex}}=(\frac{1}{3},\bar{\frac{2}{3}},\frac{1}{3})_{\mathrm{pc}}$ and $A'=(0,0,\frac{3}{2})_{\mathrm{hex}}=(\frac{1}{4},\frac{1}{4},\frac{1}{4})_{\mathrm{pc}}$ suitable to the G-type Brillouin zone (shown in Figure~\ref{jf2}(a), are used rather than the usual hexagonal points $M=(\frac{1}{2}$,0,0), $K=(\frac{1}{2},\frac{1}{2},0)$, $A=(0,0,\frac{1}{2})$.~\footnote{Note that Ref.~\cite{Jeong2012} uses the adapted G-type high symmetry points but without the prime notation.} The energy of the magnon at $A'$ is proportional to the total magnetic energy and thus is determined by the width of the magnon spectrum. It should also determine the N\'eel temperature. From the data, ignoring the NNN interaction, a value of $\mathcal{J}_0=4.08$~meV (with $S=\sqrt{\frac{5}{2}(\frac{5}{2}+1)}$) was determined which results in a Monte-Carlo estimate of $T_N^{\mathrm{MC}}=620$~K that is consistent with the measured N\'eel temperature~\cite{Ohoyama2011}.

However, there are clear discrepancies between the measured and calculated dispersion in certain regions including near the $M'$ point which may be accounted for by including the NNN interaction, as shown in Figure~\ref{jf2}(c). In particular, the energy at the $M'$ point is the most crucial to determine $\mathcal{J'}$ with a fixed $\mathcal{J}_0$.

The best fitting result is given by $\mathcal{J}=4.38$ and $\mathcal{J'}=0.15$~meV~\cite{Jeong2012, Matsuda2012} with an antiferromagnetic NNN interaction. When the NN and NNN interaction are both antiferromagnetic, magnetic frustration can arise if the NNN is sufficiently strong. However, in this case the G-type AFM state is stable since the condition $\mathcal{J'}<\mathcal{J}/4$ is satisfied. Recent DFT calculation~\cite{Weingart2012,HSKimPC} obtained a similar value for the NN interaction and confirmed that the ground state is the G-type AFM state with a NN to NNN interaction ratio $\approx\frac{1}{10}$ close to that determined by INS.

\subsubsection{Low energy spin dynamics: DM interaction and single-ion anisotropy} \label{sec-lowE}

Although the simple Heisenberg Hamiltonian discussed above describes the measured magnon dispersion quite well, the complex magnetic cycloid structure cannot be stabilised without the Dzyaloshinskii-Moriya (DM) interaction. Moreover, the single-ion anisotropy (SIA) is important in that a large uniaxial anisotropy is expected to destroy the cycloid, whilst smaller values causes the cycloid to become anharmonic. Both these interactions come in part from perturbations induced by a spin orbit coupling, and have a much smaller magnitude than the exchange interactions discussed above. Their effects are thus only apparent in the low energy magnon spectrum. Finally, the DM interaction, in particular, is also the microscopic origin of the Lifshitz invariant term in the free energy density~\cite{Zvezdin2012} which is responsible for the non-linear magnetoelectric coupling between the cycloid and electric polarisation called by some authors the {\it flexomagnetoelectric} interaction~\cite{Zvezdin2009}.

\begin{figure} \begin{center}
  \includegraphics[width=0.9\columnwidth]{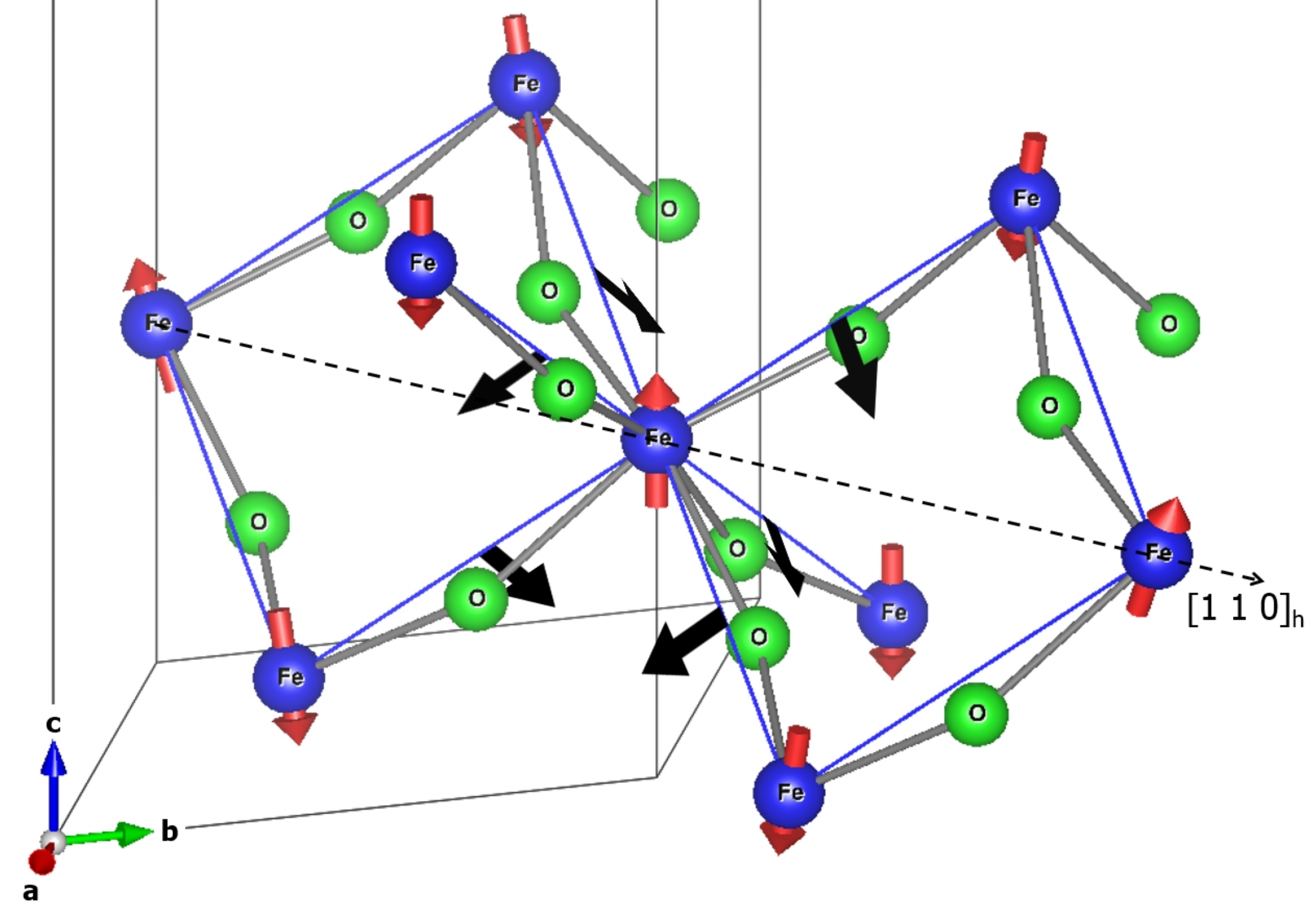} 
  \caption{The DM vectors on distorted octahedra with the spin cycloid. The arrows in the midle of the Fe-O-Fe bond indicate the local direction of the DM vectors for that particular bond, and the resuling total DM vector can be effectively separated into two directions: $\left[\bar{1}~1~0\right]$ and $\left[0~0~1\right]$.} \label{jf3}
\end{center} \end{figure}

The DM interaction arises from the spin-orbit coupling which permits spin-flip hopping between neighbouring ions that leads, in a second order perturbation theory, to the antisymmetric exchange term $\mathbf{D}_{ij}\cdot\left(\mathbf{S}_i\times\mathbf{S}_j\right)$~\cite{Moriya1960}. The $\mathbf{D}$ coefficient is only non-zero when the mid-point between spins $\mathbf{S}_i$ and $\mathbf{S}_j$ is not at an inversion centre. This is satisfied in BiFeO$_3$ below the ferroelectric transition due to the ferroelectric distortion which causes a displacement of the oxygen ligand in the Fe-O-Fe bond. The DM interaction tends to cause neighbouring spins to align perpendicular to each other, and thus competes with the exchange interaction which favours their (anti-)parallel alignment, the result often being either a canting of the spins, or in the case of BiFeO$_3$ a long range cycloid.

In BiFeO$_3$, the DM interaction between the nearest neighbours can be described by two effective DM terms, though the summation of the DM vectors is along the $c$-axis due to the 3-fold symmetry of the $R3c$ crystal structure. As shown in Figure~\ref{jf3}, the DM interaction can be effectively rewritten as

\begin{align} \label{eq:Hdm}
\mathcal{H}_{\mathrm{DM}} &= -\sum_{\mathbf{r},\Delta} \mathbf{D}_\Delta \cdot \left(\mathbf{S}_\mathbf{r}\times\mathbf{S}_{\mathbf{r}+\Delta}\right) \nonumber \\
&= -\sum_\mathbf{r} \mathbf{D}_u \cdot \left(\mathbf{S}_\mathbf{r}\times\mathbf{S}_{\mathbf{r}+a\hat{\mathbf{v}}}\right) - \sum_\mathbf{r} \mathbf{D}_c \cdot \left(\mathbf{S}_\mathbf{r}\times\mathbf{S}_{\mathbf{r}+\frac{c}{2}\hat{\mathbf{c}}}\right)
\end{align}

\noindent with $\mathbf{D}_u=\mathcal{D}_u\hat{\mathbf{u}}$ working between the next nearest neighbors along $v$-axis and $\mathbf{D}_c= \mathcal{D}_c(-1)^{\frac{6\mathbf{r}\cdot\mathbf{c}}{c}} \hat{\mathbf{c}}$ working alternately on different layers along the $c$-axis, where $\hat{\mathbf{u}}$, $\hat{\mathbf{v}}$ and $\hat{\mathbf{c}}$ are the unit vectors along $\left[1~\bar{1}~0\right]_\mathrm{hex}$, $\left[1~1~0\right]_\mathrm{hex}$ and $\left[0~0~1\right]_\mathrm{hex}$, respectively. 
The first term mainly produces the spin cycloid structure~\cite{Sosnowska1995} propagating along the $v$-axis on the $vc$-plane, and the second term produces a spin density wave perpendicular to the cycloid plane \cite{Ramazanoglu2011}. The latter also induces weak ferromagnetism by canting the spins in the $uv$-plane when the spin cycloid is suppressed.

The single ion anisotropy Hamiltonian is given by

\begin{equation} \label{eq:Hsia}
\mathcal{H}_{\mathrm{SIA}} = -\mathcal{K}\sum_{\mathbf{r}} \left(\mathbf{S}_\mathbf{r}\cdot\hat{\mathbf{c}}\right)^2 \ ,
\end{equation}

\noindent where the sign of $\mathcal{K}$ determines whether it is of an easy-axis ($\mathcal{K}>0$) or easy-plane ($\mathcal{K}<0$) form. Whilst the DM interaction is due mainly to the spin orbit coupling, the SIA comes mainly from an anisotropic deformation of the structure due to the spontaneous polarization in BiFeO$_3$. Recent DFT calculations~\cite{Weingart2012} suggests that the competing antiferrodistortive (AFD) and ferroelectric (FE) distortions produces, respectively, easy-plane (AFD) and easy-axis (FE) anisotropies. The final form of the SIA is then determined by the relative amplitudes of these two distortions, which Weingart {\it et al.} concluded to be of the easy-plane type with a magnitude of $-1.3~\mu$eV~\cite{Weingart2012}, in contrast to much of the literature which assumes an easy-axis type anisotropy: a more recent DFT calculation found a magnetic easy axis with $\clK = 3.5~\mu$eV~\cite{Lee_Fishman}. Moreover, the spin cycloid is no longer harmonic when the SIA is included as discussed in Section~\ref{sec-structure}. Whilst NMR and M\"{o}ssbauer measurements~\cite{Zalesskii2000,Zalesskii2002,Palewicz2006} indicated that a large anharmonicity is present, this could not be measured by neutron diffraction~\cite{Przenioslo2006,Przenioslo2006b,Sosnowska2011,Ramazanoglu2011a}. A precise measurement of the strength of the DM interaction and SIA would thus resolve this inconsistency. Finally, the measurements of Rovillain {\it et al.}~\cite{Rovillain2010} and theory of de Sousa {\it et al.}~\cite{DeSousa2013} suggests that the SIA may be altered by an applied electric field in a linear magnetoelectric effect, as discussed in Section~\ref{sec-struct-3-3-2}.

There is thus a need for detailed measurements of the low energy spin dynamics, which has been fulfilled by studies using THz spectroscopy~\cite{Fishman2013a} and inelastic neutron scattering~\cite{Jeong2014}. First, though, we discuss the linear spin wave theory needed to determine the effect of the DM interaction and SIA on the magnon dispersion.

While the magnon dispersion for the full Hamiltonian Eq.~\ref{eq:lswH} may be solved numerically in a linear spin wave theory by considering a large enough unit cell to encompass the full cycloid~\cite{Fishman2012} using a spin rotation and Green function technique~\cite{Haraldsen2010}, this requires solving equations of motions for over 444 spins at each momentum transfer. On the other hand, if a harmonic cycloid is assumed (e.g. the SIA and DM term along $c$ are ignored) the much simpler rotating frame operators Eq.~\ref{eq:Soper} may be used to find an analytic solution.
The corresponding spin operator vector can be written as $\mathbf{S}_{\mathbf{r}} = R_u\left(\mathbf{Q}_\mathrm{cyc}\cdot\mathbf{r}\right) \mathbf{T}_{\mathbf{r}}$, where $R_u(\theta)$ is the rotation matrix about the $u$-axis perpendicular to the cycloid plane and $\mathbf{Q}_\mathrm{cyc} = \mathbf{Q}_{G} + \mathbf{Q}_{m}$ is the modulated magnetic propagation vector for the cycloid structure with

\begin{equation} \label{eq:Qm}
\mathbf{Q}_{m} = \left[\delta~\delta~0\right]_{\mathrm{hex}} = \left[\delta~\bar{\delta}~0\right]_{\mathrm{pc}}
\end{equation}

\noindent where $\delta=0.0045$ in BiFeO$_3$. The first term of Eq.~\ref{eq:Hdm} can be rewritten as

\begin{multline} \label{eq:Hdm-r}
\mathcal{H}_\mathrm{DM}^u = -\sum_{\mathbf{r}} \mathcal{D}_u\mathrm{sin} \left(\mathbf{Q}_\mathrm{cyc}\cdot\lambda \right) \left[ S^2-2S b_\mathbf{r}^{\dagger}b_{\mathbf{r}} \right] \\
+\frac{S}{2}\sum_{\mathbf{r}} \mathcal{D}_u\mathrm{sin} \left(\mathbf{Q}_\mathrm{cyc}\cdot\lambda\right)
 \left( K_{\mathbf{r},\mathbf{r}+\lambda} - P_{\mathbf{r},\mathbf{r}+\lambda} \right)
\end{multline}

\noindent where $\lambda=a\hat{\mathbf{v}}$ is the displacement vector between adjacent spins along the cycloid direction. The $\omega_\mathbf{k}$ and $S(\omega,\mathbf{k})$ can be calculated following the same way in the previous section by replacing Eq.~\ref{eq:Ik} with 

\begin{multline} \label{eq:Ik2}
I_\mathbf{k} = \sum_{\Delta} \mathcal{J}_\Delta \mathrm{cos}\left(\mathbf{Q}_\mathrm{cyc}\cdot\Delta\right) \mathrm{cos}\left(\mathbf{k}\cdot\Delta\right) \\ - \mathcal{D}_u \mathrm{sin}\left(\mathbf{Q}_\mathrm{cyc}\cdot\lambda\right) \mathrm{cos}\left(\mathbf{k}\cdot\lambda\right) .
\end{multline}

A non-zero $\mathbf{D}_c$ introduces a canting of the cycloid with a classical ground state given by

\begin{equation} \label{eq:GS3}
\mathbf{S}_{\mathbf{r}} = S\left( \begin{array}{c} (-1)^{m_r}\mathrm{sin}~\Phi~\mathrm{sin}\left(\mathbf{Q}_\mathrm{cyc}\cdot\mathbf{r}\right) \\
\mathrm{cos}~\Phi~\mathrm{sin}\left(\mathbf{Q}_\mathrm{cyc}\cdot\mathbf{r}\right) \\
\mathrm{cos}\left(\mathbf{Q}_\mathrm{cyc}\cdot\mathbf{r}\right) \end{array} \right) \ ,
\end{equation}

\noindent where $\Phi$ is the canting angle out of the cycloid plane, which is responsible of the SDW with $m_r=6\mathbf{r}\cdot\hat{\mathbf{c}}/c$. The corresponding spin operator vector is 

\begin{equation} \label{eq:Srot2}
\mathbf{S}_{\mathbf{r}} = R_c \left((-1)^{m_r}\Phi\right) R_u\left(\mathbf{Q}\cdot\mathbf{r}\right) \mathbf{T}_{\mathbf{r}} \ ,
\end{equation}

\noindent where $R_c(\phi)$ and $R_u(\theta)$ are the rotation matrices about the $c$- and $u$-axis, respectively.

If instead of considering a non-zero $\mathbf{D}_c$, we consider the SIA in conjunction with the cycloid, then the spin operator vector becomes $\mathbf{S}_{\mathbf{r}} = R_u\left(\mathbf{Q}\cdot\mathbf{r}+\phi_\mathbf{r}\right) \mathbf{T}_{\mathbf{r}}$ where $\phi_{\mathbf{r}}$ is the modulation angle responsible for the anharmonicity of the cycloid caused by the SIA term.

\begin{figure*} \begin{center}
  \includegraphics[width=\textwidth]{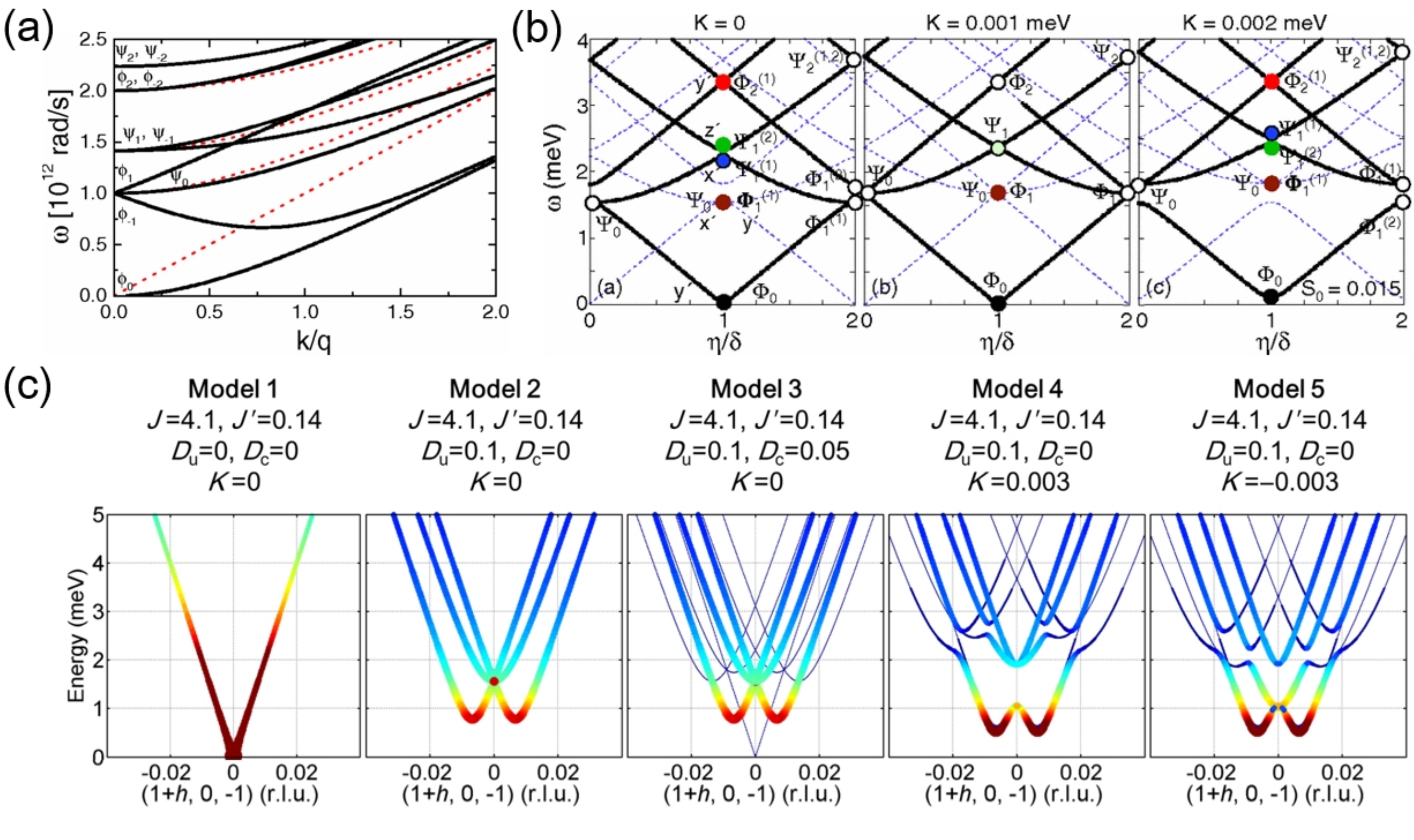} 
  \caption{
(a) Theoretical spin wave spectra for a cycloid multiferroic without the anisotropy term. The solid and dashed lines are for propagation along the direction perpendicular to the cycloid plane and along the polarization direction, respectively. The modes at $nq$ are labeled by $\phi_n$ (cyclon) and $\psi_n$ (extracyclon) for the in-plane and out-of-plane excitations, respectively. Reprinted with permission from~\cite{DeSousa2008}. Copyright 2008 American Physical Society. 
(b) The spin wave modes calculated with the full spin Hamiltonian Eq.~\ref{eq:lswH} along $[\eta, -\eta,0]_\mathrm{pc}$ from $(0.5, 0.5, 0.5)_\mathrm{pc}$. The dashed lines show all possible excitations. Reprinted with permission from~\cite{Fishman2013}. Copyright 2013 American Physical Society. 
(c) Calculated magnon dispersion with several models~\cite{Jeong2014}.} \label{jf4}
\end{center} \end{figure*}

These approximations were termed models 1-5 by Ref.~\cite{Jeong2014} with the corresponding magnon dispersion illustrated in Figure~\ref{jf4}(c). Model~1 (exchange interactions only) has a typical V-shape and the three possible modes which are detected by different components of $S(\mathbf{q},\omega)$ are degenerate. The induced spin cycloid in model~2 (exchange and $\mathbf{D}_u$ nonzero) breaks this symmetry, so that the degenerate modes split into a phason mode and two spin-flip modes corresponding to the in-plane and out-of-plane modes, respectively. This splitting of low energy magnon modes was theoretically expected~\cite{DeSousa2008}. Including the alternate DM interaction along the $c$-axis in model~3 mixes the modes at the wave vectors $\mathbf{q}$ and $\mathbf{q}\pm\mathbf{Q}_m$ such that the dispersion lines are folded in a complex manner. However, the mixing amplitude becomes very small in the low-energy limit, so that there is little noticeable difference~\cite{Fishman2012,Fishman2013a,Jeong2014}. Finally, if this effect is ignored ($\mathbf{D}_c=0$) and the SIA is included instead ($\mathcal{K}\neq 0$), as in model~4 ($\mathcal{K}>0$, easy-axis) and 5 ($\mathcal{K}<0$, easy-plane), the modes at $\mathbf{q}$ and $\mathbf{q}\pm 2\mathbf{Q}_m$ are mixed, which changes the magnon dispersion significantly~\cite{Fishman2012,Fishman2013a,Jeong2014}. The folded spectrum shows a large energy gap at the zone center. Moreover, there is a large difference between the forms of the SIA: easy-axis or easy-plane with respects to the $c$-axis~\cite{Jeong2014}.

As one can see, the significant changes caused by the DM and SIA terms are apparent only at the zone center at low energies. Because the mode splitting corresponds to the extremely small incommensurate modulation vector, it is very close to the original mode at the zone center. Although the full dispersion and exchange interactions could be determined by high-energy measurements, the effects of the DM interactions or the SIA were invisible within the limits of instrument resolution~\cite{Jeong2012,Matsuda2012}. Therefore, low-energy measurements at the zone center are critical to determine such small interactions.

\begin{figure} \begin{center}
  \includegraphics[width=0.9\columnwidth]{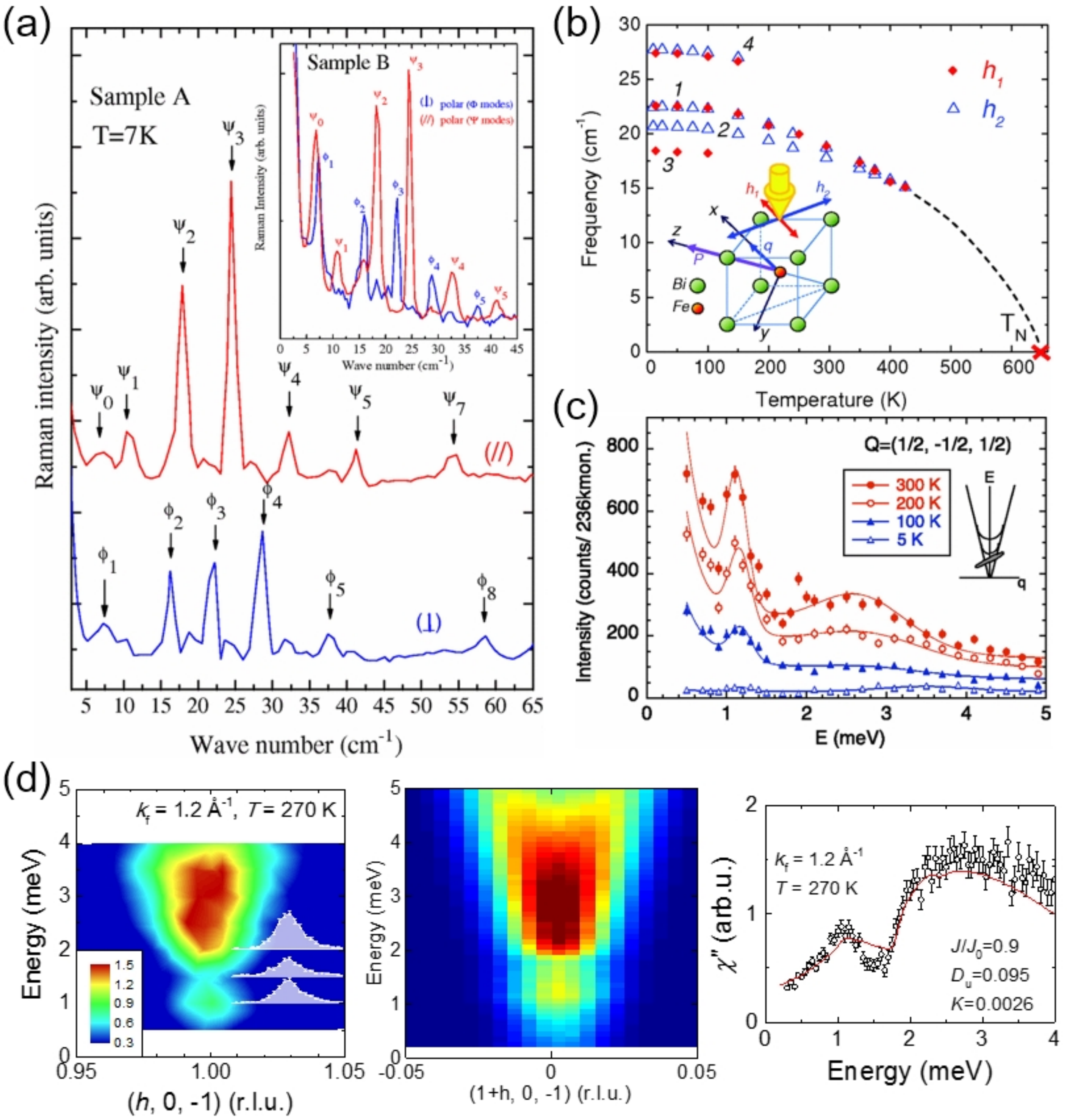} 
  \caption{
(a) Raman spectra of spin excitations. The $\Psi$ and $\Phi$ cycloidal modes were selected using parallel (red) and crossed (blue) polarizations. Reprinted with permission from~\cite{Cazayous2008}. Copyright 2008 American Physical Society.
(b) Temperature dependence of the magnetic resonance modes measured by the THz transimission. Inset: geometry of  the measurement. Reprinted with permission from~\cite{Talbayev2011}. Copyright 2011 American Physical Society.
(c) Low-energy magnon spectra at the magnetic zone center $(1/2, -1/2,1/2)_{pc}$ using cold neutron TAS. Reprinted with permission from~\cite{Matsuda2012}. Copyright 2012 American Physical Society.
(d) Measured (left) and simulated (middle) low-energy magnon dispersions along $[1~0~0]_\mathrm{hex}$ direction near the magnetic zone center $(1,0,-1)_\mathrm{hex}$. Right: Measured constant $q$-scan at $(1,0,-1)_\mathrm{hex}$  with the best fit~\cite{Jeong2014}. } \label{jf5}
\end{center} \end{figure}

Several magnon modes at low frequency were measured using polarized Raman spectroscopy as shown in Figure~\ref{jf5}~(a)~\cite{Cazayous2008,Rovillain2009}. In principle, optical spectroscopy experiments only probes the modes close to the zone center because the velocity of light is large compared to the magnon velocity. However, magnon excitations in BiFeO$_3$ at finite $\mathbf{q}$ can be measured due to the magnon zone folding which results from the DM interaction and SIA and thus the resultant small cycloid modulation wavevector. By choosing parallel and crossed polarizations, Cazayous {\it et al.} successfully obtained two distinct species of spin excitations, called the cyclon ($\Phi$, with spins oscillating within the cycloid plane) mode and the extra-cyclon ($\Psi$, spins oscillating out of the cycloid plane) mode~\cite{Cazayous2008}. As shown in the inset of Figure~\ref{jf5} (a), the $\Phi$ modes are equally spaced starting at zero energy, while the $\Psi$ modes is not regularly spaced as expected by~\cite{DeSousa2008}. The cyclon and extra-cyclon modes correspond to the gapless phason and gapped spin-flip modes mentioned above. The small gap of the out-of-plane modes can be explained by the energy cost for pinning of the cycloid plane. The mode splitting was confirmed and its temperature dependence was studied by THz absorption measurements~\cite{Talbayev2011} with the additional mode splitting attributed to an additional DM term $F_\mathrm{DM} = D P_z (M_yL_x-M_xL_y)$ corresponding the second effective DM term in Eq.~\ref{eq:Hdm}. Combining the high and low energy measurement, the full set of parameters for the Hamiltonian~Eq.~\ref{eq:lswH} were determined using both INS and THz spectroscopy~\cite{Matsuda2012,Fishman2012,Nagel2013,Fishman2013a,Jeong2014} as: $\mathcal{J}=4.38$, $\mathcal{J'}=0.15$, $\mathcal{D}_u=0.11$, $\mathcal{D}_c=0.05$, $\mathcal{K}=0.003$~meV. 

Thus the SIA is measured to be small but of the easy-axis form ($\mathcal{K}>0$) in contrast to the DFT calculation of~\cite{Weingart2012}, albeit that the magnitude of the anisotropy constant is consistent. In addition, Weingart {\it et al.} also assumed a DM vector parallel to the $c$-axis which is also inconsistent with the experimental findings. We note that the magnitudes of these terms are quite small, so that even extremely precise DFT calculations may not accurately determine them. Furthermore the SIA is, in the calculations, sensitive to the competition of the AFD and FE distortions which results in an almost cancellation of the SIA. Thus it may be that the calculations slightly overestimates the AFD and underestimates the FE distortions compared to the physical system.

\begin{figure} \begin{center}
  \includegraphics[width=0.9\columnwidth]{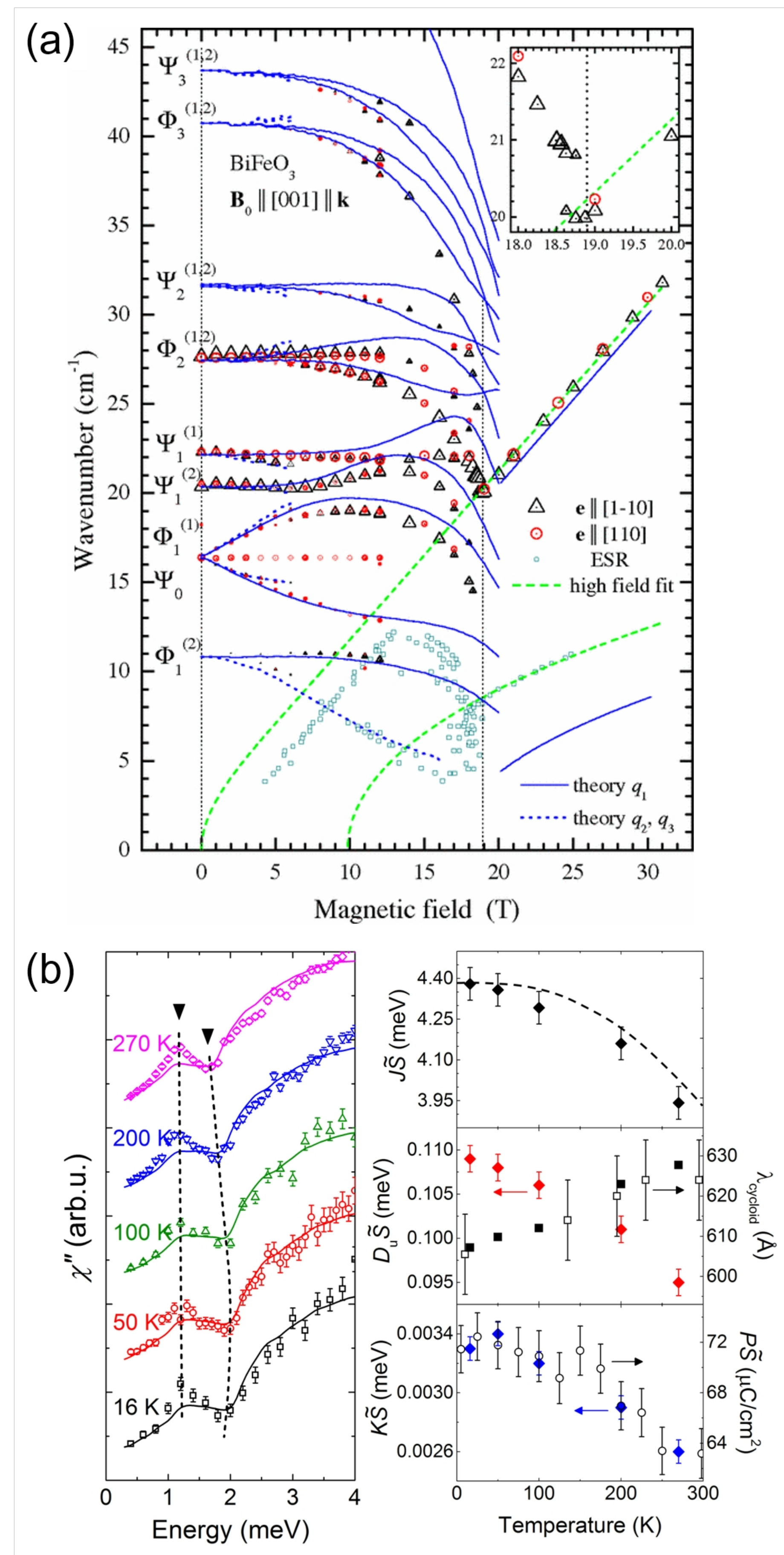} 
  \caption{(a) Magnetic field dependence of spin waves measured by the THz absorption spectroscopy at low temperature. The vertical dashed line denotes the metamagnetic transition at $H_c=$~18.8 T. Reprinted with permission from \cite{Nagel2013}. Copyright 2013 American Physical Society.
(b) Temperature dependence of low-energy magnon excitations with best fit curves (left). Temperature dependence of interaction parameters and physical properties (right).~\cite{Jeong2014}} \label{jf6}
\end{center} \end{figure}

The magnetic field~\cite{Nagel2013,Fishman2013} and temperature~\cite{Jeong2014} dependence of the magnon modes have also been reported. The modes, identified from IR measurements, soften as a function of applied field until the critical field where the cycloid is destroyed. Thereafter, a linear field dependence of the magnon energies was observed. Both observations were well described by the full Hamiltonian~Eq.~\ref{eq:lswH} and in large part fixes the parameters noted above. 
In particular, whilst the second DM ($\mathbf{D}_c$) term has little effect on the inelastic neutron spectrum, it is very important in determining the field dependence of the magnon modes above the critical field, and also in understanding the meta-magnetic phase above the spin-flop transition~\cite{Ohoyama2011,Ramazanoglu2011,Fishman2013}.

As a function of temperature, inelastic neutron scattering measurements~\cite{Jeong2014} showed systematic changes in the spectrum which when fitted to the dispersion calculated from a linear spin wave theory allows the temperature dependence of the parameters of the Hamiltonian to be deduced. With increasing temperature, there is an apparent increase and then decrease in the energy of the minima in the scattering intensity, as shown by the right dotted line in Figure~\ref{jf6}(b). This corresponds to a change in the gap between the phason and spin flip modes due to the single ion anisotropy. In addition, the initial maxima decreases slightly with increasing temperature (shown by the left dotted line in the figure), corresponding to a slight decrease of the DM interaction. This correlates with the decrease of the cycloid period observed in neutron diffraction measurements~\cite{Sosnowska2011}. The temperature dependence of the SIA may also be explained by deformations of the FeO$_6$ octahedra. Finally, the scattered intensity falls with rising temperature in line with the measured magnetic moment. However, since the neutron scattering spectrum is quite insensitive to the effect of $\mathbf{D}_c$, its temperature dependence could not be determined.

As explained previously in Section~\ref{sec-structure}, and further in appendix~\ref{sec-ap-anharmonicity}, the anharmonicity parameter $m$ may be obtained from the parameters of the microscopic spin Hamiltonian (or equivalently the Landau-Ginzburg free energy density). Using the parameters deduced above from the neutron spectroscopies, a value of $m\approx0.65 - 0.71$ was found, from which an intensity ratio of the third harmonic to the first harmonic of $\lesssim\frac{1}{200}$ can be deduced, in good agreement with the diffraction result~\cite{Sosnowska2011} although this $m$ value is larger than that estimated there.

\subsection{Phonons} \label{sec-phonon}

Just as the magnon spectrum is sensitive to the magnetic structure, so that the presence and anharmonicity of the cycloid lifts the degeneracy of the magnon branches and alters their dispersion, the phonon spectrum is similarly sensitive to the underlying crystal structure. Phonon measurements, in combination with \emph{ab initio} calculations, can thus allow the determination of the atomic motions that contribute to the ferroelectric order, and so gives insight into the nature of the ferroelectric transition. In the case of BiFeO$_3$, the lowest energy optic phonons are due to vibrations of Bi atoms~\cite{Wang2011}. 
Initial Raman spectroscopy measurements as a function of temperature~\cite{Haumont2006,Fukumura2007} showed that some of these mode decrease in energy as temperature increases before abruptly disappearing above the ferroelectric transition. Ab initio calculations~\cite{Hermet2007} showed that the motions of the lowest energy singly-degenerate transverse optic mode $A_1$($TO_1$) has a large overlap with the ferroelectric distortion, whilst the next highest $A_1$($TO_2$), which mainly involves FeO$_6$ motions, corresponds to the AFD rotation.

The first reported Raman spectrum from single crystals~\cite{Haumont2006,Fukumura2007} and thin films~\cite{Singh2006} yielded quite different zone centre phonon frequencies, in disagreement with the others. Latter Raman spectroscopy~\cite{Kothari2008,Palai2010,Hlinka2011,Beekman2012} and IR reflectivity or THz spectroscopy~\cite{Lobo2007,Kamba2007} measurements supported that of Ref.~\cite{Fukumura2007}, but introduced some confusion in the assignments of the symmetry of the modes. The trigonal point symmetry $(C_{3v})$ of the zone centre $\Gamma$-point of the $R3c$ space group means that there are three possible irreducible representations for the 27 possible phonon modes: the 4 non-degenerate $A_1$ modes are Raman and IR active and polarised along the $[111]_{\mathrm{pc}}$ direction (defined as $z$ for most measurements); the 5 non-degenerate $A_2$ modes are silent; and the 9 doubly degenerate $E$ modes are Raman and IR active and polarised in the plane perpendicular to the electric polarisation ($xy$ plane). In polarised Raman measurements, if the incident light is along the $z$ direction, then the $A_1$ longitudinal optic (LO) modes will be seen only in parallel polarisation (denoted $z(xx)z'$ and $z(yy)z'$) and not in cross-polarisation ($z(xy)z'$ and $z(yx)z'$). The $E$ transverse optic (TO) modes will be seen in both polarisation channels. 

However, as noted by Beekman {\it et al.}~\cite{Beekman2012}, a small misalignment of the laser polarisation with respects to the $(111)_{\mathrm{pc}}$ face could yield $A_1$ intensity in the cross-polarisation channel, and it is notably difficult to polish BiFeO$_3$ with a $(111)_{\mathrm{pc}}$ face~\cite{Singh2006}. Additionally, the polishing itself could cause problems, for example by creating multiple ferroelectric domains, each with a different orientation of the polarisation vector~\cite{Fukumura2007} or by inducing strain in the material, thus altering the phonons. Finally, when the phonon propagation vector defined by the direction of the incident laser beam is not along a symmetry direction, \emph{oblique} modes appear to disperse in frequency~\cite{Hlinka2011}, which may also account for some of the variations in the reported phonon frequencies.

\begin{figure} \begin{center}
  \includegraphics[width=0.9\columnwidth]{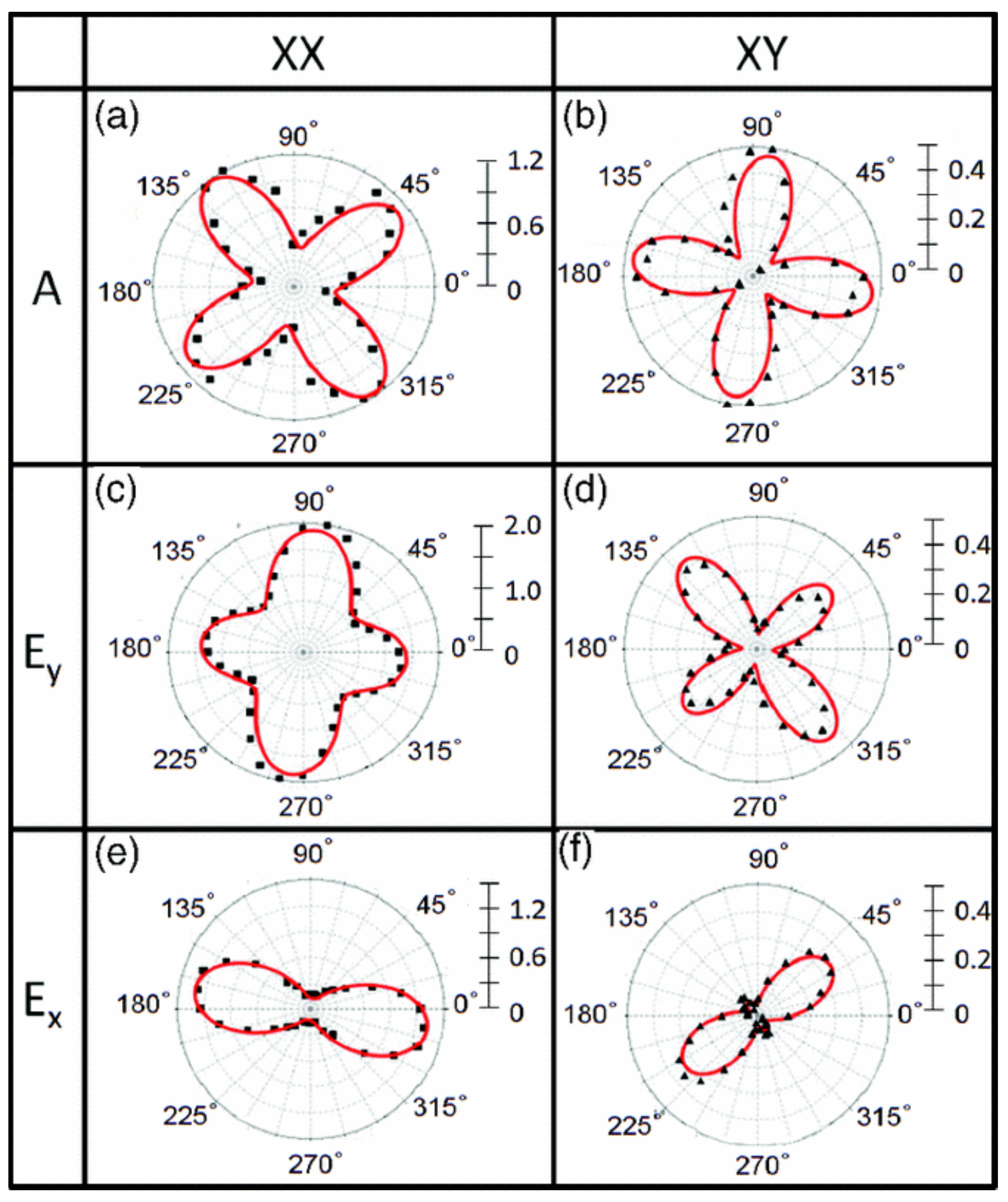} 
  \caption{Polar intensities determined from the Raman spectra as a function of polarization rotation. (a) and (b) A mode at 350 cm$^{-1}$. (c) and (d) E$_y$ mode at 140 cm$^{-1}$. (e) and (f) E$_x$ mode at 471 cm$^{-1}$. The solid red lines are fits. Reprinted with permission from~\cite{Beekman2012}. Copyright 2012 American Physical Society.
} \label{jf7}
\end{center} \end{figure}

The correct assignment of mode symmetries would allow a much better comparison between \emph{ab initio} calculations and measurements, so that conclusions about the nature of the high temperature phase transitions could be more firmly drawn. To this end, Hlinka {\it et al.}~\cite{Hlinka2011} and Beekman {\it et al.}~\cite{Beekman2012} offer two alternative methods to determine the mode symmetry. For example, Hlinka {\it et al.}~\cite{Hlinka2011} chose to take advantage of the dispersion of oblique modes by measuring the unpolarised micro Raman spectrum of over 70 randomly oriented single-crystalline grains, reasoning that the dispersion should be monotonic with the oblique angle $\phi$, and that the $E$(TO) modes do not disperse. The spectra can thus be sorted in order of the measured phonon frequencies, with the lowest set of frequencies corresponding to the beam being parallel to $[111]_{\mathrm{pc}}$ ($\phi=0^{\circ}$) and the highest frequency dispersed mode corresponding to the beam being perpendicular ($\phi=90^{\circ}$). The frequencies of the $A_1$(LO) modes can be determined from the $\phi=0^{\circ}$ spectrum whilst the $E$(LO) and $A_1$(TO) modes from that with $\phi=90^{\circ}$. Beekman {\it et al.}~\cite{Beekman2012}, on the other hand, used the azimuthal polarisation dependence of the intensities of the modes to determine their symmetry as reproduced in Figure~\ref{jf7}. This allowed them to reliably determine not just the mode symmetry but also the Raman scattering tensors associated with each zone centre phonon mode, making a direct comparison with \emph{ab initio} calculations possible. The experimental results, however, are at odds with the LSDA calculations of Hermet {\it et al.}~\cite{Hermet2007}.

\begin{figure} \begin{center}
  \includegraphics[width=0.9\columnwidth]{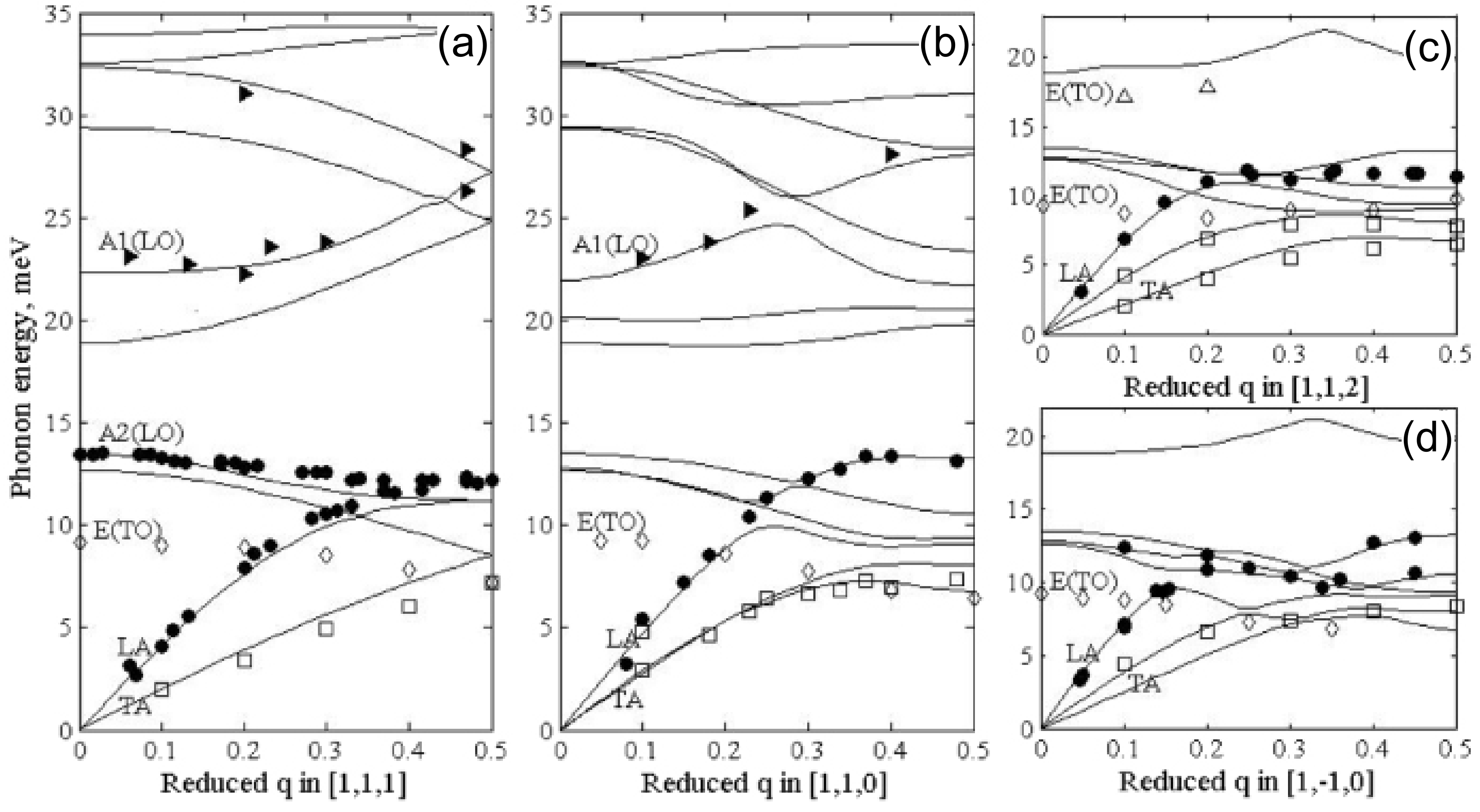} 
  \caption{Experimental phonon dispersion curves (symbols) with \it{ab initio} \rm calculation (dashed line) along (a) $[1 1 1]$, (b) $[1 1 0]$, (c) $[1 1 2]$ and (d) $[1 \bar{1} 0]$ directions. Open and full symbols correspond to transverse  and longitudinal modes, respectively. Reprinted with permission from~\cite{Borissenko2013}. Copyright 2013 IOP Publishing.} \label{jf8}
\end{center} \end{figure}

A better comparison between theory and experiments could be achieved when the full phonon dispersion is known, rather than just the zone centre frequencies that are accessible to optical spectroscopic techniques. Such a study was reported recently by Borissenko {\it et al.}~\cite{Borissenko2013}, who found good agreement between the measured inelastic X-ray scattering spectrum and calculated phonon dispersion curves (see Figure~\ref{jf8}). The LDA calculations in this case is very similar to that of Ref.~\cite{Hermet2007} using the same functionals and pseudopotentials. Both sets of calculations do not, however, reproduce exactly the experimental crystal structure, so deviations between their predictions of the zone centre phonon frequencies and intensities indicates that these are particularly sensitive to slight differences in the structure. The same group also reported that using the recently proposed Wu-Cohen (WC) exchange-correlation functional in the generalised gradient approximation (GGA) or a hybrid functional (B1-WC) which mixes an exact exchange term together with the GGA-WC functional (with mixing parameter $a_0$=0.16)~\cite{Goffinet2009} reproduces the experimental structure much more faithfully also gives better agreement with the measured Raman frequencies. 

\begin{figure} \begin{center}
  \includegraphics[width=0.9\columnwidth]{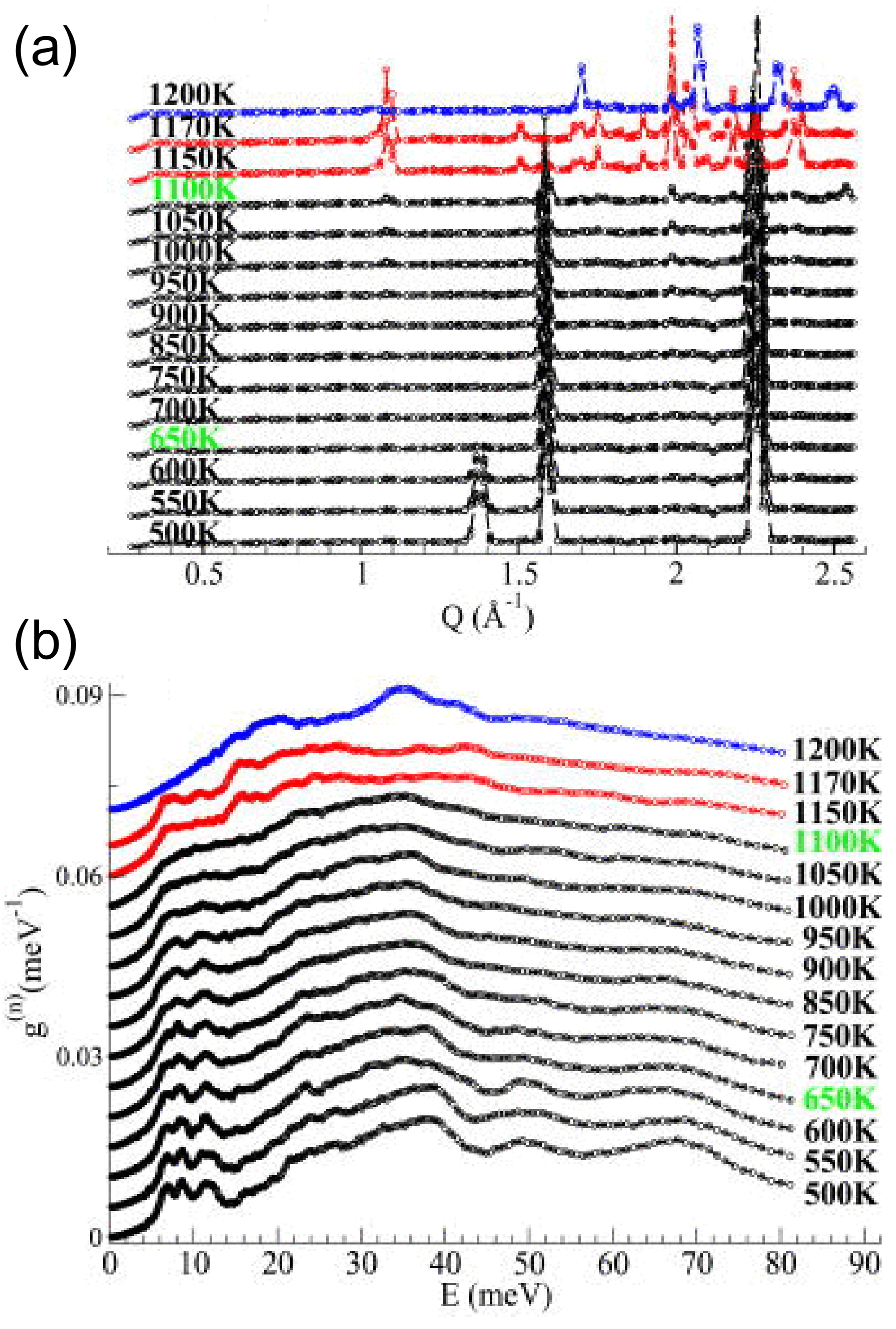} 
  \caption{(a) Temperature dependence of diffraction patterns. (b) Phonon gDOS spectra. Ferroelectirc $\alpha$-phase (black), paraelectric $\beta$- (red) and $\gamma$-phases (blue) are highlighted with different colors. Reprinted with permission from~\cite{Zbiri2012}. Copyright 2012 American Institute of Physics.} \label{jf9}
\end{center} \end{figure}

Thus, whilst the details of lattice dynamical properties of BiFeO$_3$ require further theoretical work, the good agreement between the calculated and measured dispersion obtained by Borissenko {\it et al.}~\cite{Borissenko2013} shows that broader conclusions may still be drawn from the data. 
This suggests that the low energy phonons, both those dominated by Bi vibrations and associated with the ferroelectric distortion and those at slightly higher energies associated with Fe motions and the AFD rotations~\cite{Hermet2007,Wang2012}, may lay behind a displacive first order ferroelectric transition. This view is  supported by recent inelastic neutron scattering and GGA calculations of Zbiri {\it et al.}~\cite{Zbiri2012}, who obtained very good agreement between the measured and calculated generalised phonon density of states (gDOS). Their measured gDOS as a function of temperature is reproduced in Figure~\ref{jf9}, and shows clear, abrupt, changes at the $\alpha\rightarrow\beta$ and $\beta\rightarrow\gamma$ phase transitions, indicating that these are of first order. In addition, it shows an incomplete softening of the low energy phonons around 10~meV ($\approx$80~cm$^{-1}$) as the sample is cooled in the $\beta$ phase towards $T_c$.

The data also shows the appearance of two strong peaks around 20 and 35~meV in the $\gamma$ phase which is linked to the disappearance of the oxygen modes at high energies, and was suggested to correspond to the breaking of Fe-O bonds leading to a change of iron valence from (high spin) Fe$^{3+}$ to (low spin) Fe$^{2+}$ and thus to an insulator to metal transition.

\subsection{Electromagnons and spin-lattice coupling} \label{sec-electromagnon}

In addition to the phonons discussed above, optical spectroscopic techniques can also be used to measure zone centre magnons (called \emph{magnetic resonances} in this context) which are excited by the oscillating magnetic field of the incident photon. Whilst the intensities of these modes are sometimes much weaker than phonons, they can nevertheless often be detected. In multiferroics, furthermore, as a result of the magneto-electric coupling, a mixing of magnon and phonon excitations occurs, which gives hybrid magnon-like modes which can be excited by a photon's electric field (called \emph{electromagnons}) and phonon-like modes which are excited by magnetic fields. In addition to mixing the modes, the magneto-electric coupling also shifts their frequencies with respects to the bare magnon or phonon frequencies, but in BiFeO$_3$ these energy shifts are small~\cite{DeSousa2008,Fishman2012}, so that, for example, magnetic resonances and electromagnons are coincident within experimental resolution.

Electromagnons were first reported by Singh {\it et al.}~\cite{Singh2008}, who found that the Raman spectrum of BiFeO$_3$ showed a rich structure at low frequencies ($\lesssim 70$~cm$^{-1}$) and that the temperature of the frequencies exhibited two anomalies at $\approx$140 and $\approx$200~K. These transitions were subsequently determined to affect only the surface of the crystals as explained in Section~\ref{sec-struct-3-2-3}. Subsequent measurements and careful analysis of the polarisation dependence by Cazayous {\it et al.}~\cite{Cazayous2008} showed that the modes fell into two categories, corresponding to the \emph{cyclon} (denoted $\Phi$) and \emph{extra-cyclon} (or spin-flip, denote $\Psi$) magnetic excitations associated with fluctuations of the moments within or out of the cycloidal plane, respectively. A theoretical treatment by de Sousa and Moore~\cite{DeSousa2008} using the Lifshitz invariant as the magnetoelectric coupling term was found to satisfactorily account for the Raman observations. The plethora of modes (twelve were observed by Cazayous {\it et al.}~\cite{Cazayous2008}) is due to the umklapp scattering of magnons outside the first Brillouin zone which becomes allowed due to the anharmonicity and periodicity of the cycloid. 

As noted by de Sousa and Moore~\cite{DeSousa2008} and Fishman {\it et al.}~\cite{Fishman2012}, in a harmonic cycloid, where the single ion anisotropy $\mathcal{K}$=0, only one magnon (the $\Psi_1$ mode) becomes electric dipole active. The other electromagnons arise only in the presence of anharmonicities of the cycloid (non-zero SIA), which allows the umklapp coupling between these magnons and a zero-wavevector phonon. In particular, de Sousa and Moore~\cite{DeSousa2008} determined that the $\Psi_3$ and $\Phi_2$ modes should only become electric dipole active if the cycloid has a third harmonic. This is supported by THz spectroscopy measurements~\cite{Talbayev2011} where the $\Psi_3$ mode was only observable at low temperatures where the cycloid anharmonicity is more pronounced and a third harmonic is observed~\cite{Zalesskii2002,Ramazanoglu2011}, as shown in figure~\ref{jf5}(b). In addition to making the cycloid anharmonic, the single ion anisotropy also splits the $\Psi_n$ and $\Phi_n$ modes into two $\Psi_{\pm n}$ and $\Phi_{\pm n}$ modes~\cite{Fishman2012}, with the splitting being most pronounced for $\Psi_{\pm1}$. This splitting may be the source of initial confusion in the assignment of (electro)magnon modes from Raman scattering data, as the model of de Sousa and Moore does not explicitly include the SIA. Finally, an additional effective DM interaction along $[111]_{\mathrm{pc}}$ is required to make the $\Psi_0$ and $\Phi_1$ modes electric dipole active~\cite{Fishman2013a}.

The models of both de Sousa and Moore~\cite{DeSousa2008} and Fishman {\it et al.}~\cite{Fishman2012,Fishman2013a} relied on the Dzyaloshinskii-Moriya interaction to provide the necessary magneto-electric coupling to produce electromagnons. In the case of~Ref.\cite{DeSousa2008}, this was given in the form of the Liftshitz invariant, the microscopic origin of which was shown by Zvezdin {\it et al.}~\cite{Zvezdin2012} to be from the DM interaction in conjunction with the AFD rotation and polar distortions of the FeO$_6$ octahedron and Fe polar displacements. In the theory of Fishman {\it et al.}, a coupling between the spin and electric polarisation was not explicitly included into the spin Hamiltonian, but the inverse DM mechanism was used to determine the electric dipole matrix elements (and hence oscillator strengths) of normally magnon modes. Thus, Fishman {\it et al.} do not obtain any frequency shifts of electromagnons with respects to bare magnons, whereas de Sousa and Moore do. That the measured modes from Raman, IR and THz spectroscopies agree well with the theory suggests that such shifts are small. The dynamical magneto-electric coupling considered in these two theories thus bears similarities to the original model of electromagnons proposed by Katsura, Balatsky and Nagaosa~\cite{Katsura2007}, and is distinct from the dynamical exchange-striction model~\cite{ValdesAguilar2009} which is now thought to be responsible for the dominant electromagnons seen in the perovskite manganites, such as TbMnO$_3$~\cite{Shuvaev2011}. In that case it is a modulation of the symmetric Heisenberg exchange interaction by an incident electric field which excites strong electromagnons.

Finally, as we mentioned in Section~\ref{sec-struct-3-3-2}, the magneto-electric coupling proposed by de Sousa {\it et al.}~\cite{DeSousa2013} in which an applied static electric field enhances the single-ion anisotropy, could also have strong effects on the electromagnons, by increasing the cycloid anharmonicities, potentially making more magnons electric dipole active, in addition to shifting the frequencies of the magnon modes as was observed~\cite{Rovillain2010}.

\section{Summary and Outlook} \label{sec-summary}

Multiferroic materials are a class of systems having the coexistence of more than one ferroic phases. Of particular interest have been compounds which exhibit both ferroelectric and ferro- or antiferromagnetic phase transitions in a single compound. Initially, research was mainly driven by curiosity and the desire to find new materials, which then appeared to be rather rare among naturally occurring compounds when the subject resurfaced in late 90s, and its potential applications was recognised. However, as the list of multiferroic materials grows ever larger, this initial motivation seems to have become gradually less important. Nonetheless, the lessons learnt from the physics of noncentrosymmetric materials are still relevant to several topic of current interest.

Of the known multiferroic materials, BiFeO$_3$ stands out for one reason alone: that it is a room-temperature multiferroic system. Thanks to the extensive research, summarised in part in this review, we have come to know enormous details about the bulk properties of BiFeO$_3$. As such there are only a few areas where further work appears necessary, amongst them the high temperature transitions between the paraelectric and ferroelectric phases, and the crystal structure within these phases, for which there is still some debate. The magnetic phase transition at 650 K is well studied, as is the spin dynamics such that we can now construct a full microscopic Hamiltonian, which explains both the magnon dispersion and some structural aspects of the magnetic cycloid.

Another major issue is the magnetoelectric coupling. As we demonstrated in a high-resolution neutron diffraction study~\cite{Lee2013}, there is clear evidence of a magnetoelectric coupling with a magnitude of a few hundreds nC/cm$^2$, which seems to be consistent with theoretical calculations~\cite{Lee_Fishman}. With this convincing result, it is probably a good time to think about how to maximize the effects in our favour. One area of further research may be doping experiments~\cite{Do2011,Kim2010,Wu2012}. Although there have been some attempts in this direction, there is still much room for further exploration. Finally, it will also be instructive to observe the magneto-electric coupling in the spin dynamics, such as through a change in the magnon dispersion under an electric field or the observation of an electromagnon. So far, all evidence of these excitations in BiFeO$_3$ have come from optical measurements so it will be highly complementary to observe it by inelastic neutron scattering experiment and to examine its dispersion as well as its dynamical structure factor.

\appendix
\section{Equivalence of the Landau-Ginzburg free energy and spin Hamiltonians} \label{sec-ap-equivalence}

To compare the spin Hamiltonian~Eq.~\ref{eq:lswH} with the L-G free energy~Eq.~\ref{eq:LGfe}, let us consider the classical (magnetic) ground state and the corresponding energy. When the exchange interaction is dominant and the DM term is negligible, the ground state is a collinear order. 
Considering the collinear state, the ground state energy obtained from the spin Hamiltonian and the free energy are

\begin{align}
\mathcal{H} &= \sum_{\mathbf{r}} \left[ \sum_{\Delta}\mathcal{J}_{\Delta}\mathbf{S}_\mathbf{r}\cdot\mathbf{S}_{\mathbf{r}+\Delta} - \mathcal{K} \left(\mathbf{S}_{\mathbf{r}}\cdot\hat{\mathbf{c}}\right)^2 \right] \nonumber \\
&= NS^2 \left(-3\mathcal{J}+6\mathcal{J'} - \mathcal{K}\cos^2\theta\right), \\
F &= \int dV \left[\lambda\mathbf{l}^2 - K_u l_c^2\right] = V\left(\lambda-K_u\cos^2\theta\right) ,
\end{align}

\noindent where $M_0$ is the magnitude of the magnetization vector $\mathbf{M}_i$ of sublattices, $\mathbf{l}=(\mathbf{M}_1-\mathbf{M}_2)/2M_0$ is the antiferromagnetic unit vector and $\theta$ is the angle between $\mathbf{l}$ and the $c$-axis.
Clearly, $\lambda<0$ or $\mathcal{J}>0$ for the AFM ground state and $\theta=0$ for an easy-axis anisotropy $K_u>0$ or $\mathcal{K}>0$. Here, we can find the relations

\begin{gather}
V \lambda = NS^2 \left(-3\clJ+6\clJ'\right) , \\
V K_u = NS^2 \clK .
\end{gather}

The DM interaction favours a magnetic cycloid. If the small canting out of the cycloid plane and the SIA are ignored, the ground state is a harmonic cycloid that lies on the $vc$-plane.
Considering the harmonic cycloid with

\begin{align}
\mathbf{S}_\mathbf{r} &= S\left(0, \sin\theta_{\mathbf{r}} , \cos\theta_{\mathbf{r}} \right), & \theta_{\mathbf{r}} &= \theta_0+\mathbf{Q}\cdot\mathbf{r} ,
\end{align}

\noindent where $\mathbf{Q}=\mathbf{Q}_G+\mathbf{Q}_m=(0.0045,0.0045,3)$, the ground state energies are

\begin{align}
\mathcal{H} &= \sum_{\mathbf{r},\Delta}\mathcal{J}_{\Delta}\mathbf{S}_\mathbf{r}\cdot\mathbf{S}_{\mathbf{r}+\Delta} - \sum_\mathbf{r} \mathcal{D}_u \hat{\mathbf{u}} \cdot\left(\mathbf{S}_\mathbf{r}\times\mathbf{S}_{\mathbf{r}+a\hat{\mathbf{v}}}\right) \nonumber \\
&= NS^2 \left[ -\mathcal{J}\left(1+2\cos\frac{\Delta\theta}{2}\right)\right. \nonumber \\
& \qquad\qquad \left. +\mathcal{J'}\left(1+4\cos\frac{\Delta\theta}{2}+\cos\Delta\theta\right) +\mathcal{D}_u\sin\Delta\theta \right] \nonumber \\
&\approx NS^2 \left[ \left(-3\mathcal{J}+6\mathcal{J'}\right) +\frac{\mathcal{J}-4\mathcal{J'}}{4}a^2\left(d_v\theta\right)^2 +\mathcal{D}_u a\left(d_v\theta\right)\right], \\
F &= \int dV \left[ \lambda\mathbf{l}^2 +A\sum_{i}{\left( \nabla l_i \right)^2} \right. \nonumber \\
& \qquad\qquad\qquad \left. \phantom{\sum_{i}} -\alpha\,\mathbf{P}\cdot\left[\mathbf{l}\left(\nabla\cdot\mathbf{l}\right) + \mathbf{l}\times\left(\nabla\times\mathbf{l}\right)\right]\right] \nonumber \\
&= V\left[ \lambda +A\left(d_v\theta\right)^2 -\alpha P_z \left(d_v\theta\right)\right] ,
\end{align}

\noindent where $\Delta\theta=\theta_{\mathbf{r}+a\hat{\mathbf{v}}}-\theta_{\mathbf{r}}=a(d_v\theta)$ is the angle difference between spins along the cycloid direction $\left[1~1~0\right]_\mathrm{hex}$ and $\mathbf{P}$ is the spontaneous polarization vector. Note that $\alpha>0$ or $\clD_u<0$ for the clockwise rotation and $\alpha<0$ or $\clD_u>0$ for the counterclockwise.
Now, we can find the relations

\begin{gather}
V A = NS^2 a^2 \left(\frac{\clJ-4\clJ'}{4}\right) , \\
V \alpha P_z = -NS^2 a \clD_u .
\end{gather}

The terms responsible for the canting out of the cycloid plane are

\begin{align}
\mathcal{H}_\mathrm{DM}^c &= -\sum_\mathbf{r} (-1)^{\frac{6}{c}\mathbf{r}\cdot\hat{\mathbf{c}}} \mathcal{D}_c\hat{\mathbf{c}}\cdot\left(\mathbf{S}_\mathbf{r}\times\mathbf{S}_{\mathbf{r}+\frac{c}{2}\hat{\mathbf{c}}}\right) , \\
F_\mathrm{DM} &= -\int dV 2\beta M_0 \mathbf{P}\cdot (\mathbf{m}\times\mathbf{l}) ,
\end{align}

\noindent where $\mathbf{m}=(\mathbf{M}_1+\mathbf{M}_2)/2M_0$ is the magnetization unit vector.
Note that these vanish without a local ferromagnetic moment $\mathbf{m}$. The ground state is the canted cycloid~Eq.~\ref{eq:GS3} or 

\begin{align}
\frac{\mathbf{M}_1}{M_0} &= \left( \begin{array}{c}\sin\Phi \sin\theta \\ \cos\Phi \sin\theta \\ \cos\theta\end{array}\right) , & \frac{\mathbf{M}_2}{M_0} &= \left( \begin{array}{c}\sin\Phi \sin\theta \\ -\cos\Phi \sin\theta \\ -\cos\theta\end{array} \right),
\end{align}

\noindent where $\Phi$ is the canting angle and $\theta$ is the rotation angle on the cycloid plane. Whilst the ground state energy corresponding to other terms previously discussed is also changed in this state, the differences may be ignored for small $\Phi\ll 1$. The classical energies are now

\begin{align}
\mathcal{H}_\mathrm{DM}^c &= -\sum_\mathbf{r} \clD_c \sin\left(\phi_{\mathbf{r}+\frac{c}{2}\hat{\mathbf{c}}}-\phi_{\mathbf{r}}\right) \sin^2\theta_\mathbf{r} \\
& = \sum_\mathbf{r} \clD_c \sin2\Phi \sin^2\theta_\mathbf{r}  , \\
F_\mathrm{DM} &= -\int dV 2\beta M_0 P_z \sin\Phi \cos\Phi \sin^2\theta \\
& = -\int dV \beta M_0 P_z \sin 2\Phi \sin^2\theta ,
\end{align}

\noindent so that we can find the relation

\begin{equation}
V \beta M_0 P_z = NS^2 \clD_c .
\end{equation}

\section{Anharmonicity of the spin cycloid structure} \label{sec-ap-anharmonicity}

In order to obtain the anharmonicity of the spin cycloid, we focus on the derivative ($\nabla L_i$) and the anisotropy terms, and ignore the small canting out of the cycloid plane. The corresponding Landau-Ginzburg free energy density can be then written as

\begin{align}
f &= f_{\mathrm{exch}}+f_{\mathrm{L}}+f_{\mathrm{an}} \nonumber \\
 &= A\sum_{i=x,y,z}{\left( \nabla l_i \right)^2} - \alpha\mathbf{P}\cdot\left[\mathbf{l}\left(\nabla\cdot\mathbf{l}\right) + \mathbf{l}\times\left(\nabla\times\mathbf{l}\right)\right] - K_u  l_z^2 ,
\tag{\ref{eq:structfe}}
\end{align}

\noindent where $M_0$ is the magnitude of the magnetization vector $\mathbf{M}_i$ of the sublattices, $\mathbf{P}$ is the spontaneous polarization vector, and $\mathbf{l}=(\mathbf{M}_1-\mathbf{M}_2)/2M_0$ is the antiferromagnetic unit vector.
The first term is an exchange interaction term with the inhomogeneous exchange constant $A$ (exchange stiffness), the second term is the Lifshitz invariant term with the inhomogeneous magnetoelectric (flexomagnetoelectric) interaction constant $\alpha$, and the last term is an uniaxial anisotropy term with the magnetic anisotropy constant $K_u$.

Let $\mathbf{l}=(\sin\theta\cos\phi, \sin\theta\sin\phi, \cos\theta)$, so each term can be written as

\begin{align}
f_{\mathrm{exch}} &= A\sum_{i,j} \left(\partial_i l_j \right)^2 = A\left[\left(\nabla\theta\right)^2 + \sin^2\theta\left(\nabla\phi\right)^2 \right] , \\
f_{\mathrm{L}} &= \alpha P_z \left(l_x \partial_x l_z + l_y \partial_y l_z - l_z \partial_x l_x - l_z \partial_y l_y\right) \nonumber \\
& = -\alpha P_z \left[ \cos\phi\left(\partial_x\theta\right) + \sin\phi\left(\partial_y\theta\right) \right. \nonumber \\ 
& \qquad\quad - \left.\sin\theta\cos\theta\left(\sin\phi\left(\partial_x\phi\right) - \cos\phi\left(\partial_y\phi\right) \right) \right] , \\
f_{\mathrm{an}} &= -K_u l_z^2 = - K_u\cos^2\theta .
\end{align}

The Euler-Lagrange equations for $f\left(\theta,\phi,\partial_i\theta,\partial_i\phi\right)$ can be derived as

\begin{multline} \label{eq:ap-EL1}
2A\left( {{\nabla ^2}\theta } \right) + 2\alpha {P_z}{\sin ^2}\theta \left( {\sin \phi \left( {{\partial _x}\phi } \right) - \cos \phi \left( {{\partial _y}\phi } \right)} \right) \\
- \sin 2\theta \left( {A{{\left( {\nabla \phi } \right)}^2} + {K_u}} \right) = 0 ,
\end{multline}
\begin{multline} \label{eq:ap-EL2}
2A\,\nabla  \cdot \left( {{{\sin }^2}\theta \left( {\nabla \phi } \right)} \right) \\
- 2\alpha {P_z}{\sin ^2}\theta \left( {\sin \phi \left( {{\partial _x}\theta } \right) - \cos \phi \left( {{\partial _y}\theta } \right)} \right) = 0 .
\end{multline}
 
The solutions are

\begin{gather}
\nabla \phi  = {\bf{0}},\quad \sin \phi \left( {{\partial _x}\theta } \right) - \cos \phi \left( {{\partial _y}\theta } \right) = 0 \\
\phi  = {\rm{const}} = \arctan \left( {\frac{{{\partial _y}\theta }}{{{\partial _x}\theta }}} \right) \\
2A\left( {{\nabla ^2}\theta } \right) - {K_u}\sin 2\theta  = 0 . \label{eq:ap-solution}
\end{gather}

The solution gives a cycloid spin structure on the plane defined by the $z$-axis and a direction in the $xy$-plane. The coordinate dependence of the angle $\theta$ generally exhibits a nonlinear behavior due to equation~Eq.~\ref{eq:ap-solution}. By rotating the coordinate system, so that $\theta \left( {x,y} \right) = \theta \left( r \right)$, i.e., the cycloid lies in the $rz$-plane, it can be rewritten as
\begin{equation}
2A\frac{{{d^2}\theta }}{{d{r^2}}} - {K_u}\sin 2\theta  = 0 .
\end{equation}
This is the sine-Gordon equation for a nonlinear oscillator, which has two different solutions depending on the sign of $K_u$.
\begin{equation}
\varphi  = \left\{ \begin{array}{ll} \theta , & {K_u} < 0 \\ \theta  + \frac{\pi }{2}, & {K_u} > 0 \end{array} \right. .
\end{equation}
In both cases, the equation can be reduced to
\begin{equation}
\frac{{{d^2}\varphi }}{{d{r^2}}} + \frac{\varepsilon }{2}\sin 2\varphi  = 0
\end{equation}
where $\varepsilon  = {{\left| {{K_u}} \right|} / A}$.
We then obtain the solution by integrating this, as follow

\begin{gather}
{\left( {\frac{{d\varphi }}{{dr}}} \right)^2} + \varepsilon {\sin ^2}\varphi  = C , \\
\frac{{d\varphi }}{{dr}} =  \pm \sqrt {\frac{\varepsilon }{m}} \sqrt {1 - m{{\sin }^2}\varphi } , \\ 
r\left( \varphi  \right) = \pm \sqrt {\frac{m}{\varepsilon }}{\rm F} \left( {\varphi ,m} \right) ,
\end{gather}

\noindent where $C$ is the constant of integration, $m \equiv \frac{\varepsilon }{C} = \frac{{\left| {{K_u}} \right|}}{{AC}}$ and ${\mathop{\rm F}\nolimits} \left( {\theta ,\,m} \right) = \int_0^\theta  {\frac{{d\theta '}}{{\sqrt {1 - m{{\sin }^2}\theta '} }}} $ is the incomplete elliptic integral of the first kind with the parameter $m$. The period of the cycloid is
\begin{equation} \label{eq:lambda}
\lambda  = r\left( {2\pi } \right) - r\left( 0 \right) = 4\sqrt {\frac{m}{\varepsilon }}{\rm K} \left( m \right) ,
\end{equation}
where ${\rm K} \left( m \right) = {\rm F} \left( {\frac{\pi }{2},m} \right) = \int_0^{\pi /2} {\frac{{d\theta }}{{\sqrt {1 - m{{\cos }^2}\theta } }}} $ is the complete elliptic integral of the first kind. The anharmonicity parameter $m$ can be obtained from this relation for given $\varepsilon$ and $\lambda$. Note that $m=0$ for $K_u=0$ and it approaches 1 for $\left| {{K_u}} \right| \gg A$.

The angle $\varphi$ can be expressed by the Jacobi amplitude as the inverse of the incomplete elliptic integral,
\begin{equation}
\varphi \left( r \right) = {\rm F ^{ - 1}}\left( { \pm \sqrt {\frac{\varepsilon }{m}} r,m} \right)  =  \pm {\rm am} \left( {\sqrt {\frac{\varepsilon }{m}} r,m} \right) .
\end{equation}
The two possible signs correspond to the winding direction of spins in the cycloid. Choosing a positive sign,
\begin{equation}
\theta \left( r \right) = \left\{ \begin{array}{ll}
{\rm{am}}\left( {\sqrt {\frac{{\left| {{K_u}} \right|}}{{Am}}} r,m} \right), & {K_u} < 0   \\
{\rm{am}}\left( {\sqrt {\frac{{{K_u}}}{{Am}}} r,m} \right) - \frac{\pi }{2}, & {K_u} > 0   
\end{array} \right. .
\end{equation}
The shape of the anharmonic cycloid can be obtained in terms of the Jacobi elliptic function,

\begin{gather}
\frac{L_z}{L} = \cos \varphi  = \mathrm{cn} \left( \sqrt {\frac{\varepsilon }{m}} r,m \right)   \\
\frac{L_r}{L} = \sin \varphi = \mathrm{sn} \left( \sqrt {\frac{\varepsilon }{m}} r,m \right)
\end{gather}

\noindent for the easy-plane anisotropy (${K_u} < 0$) and

\begin{gather}
\frac{L_z}{L} = \cos\left( {\varphi  - \frac{\pi }{2}} \right) = \mathrm{sn} \left( \sqrt {\frac{\varepsilon }{m}} r,m \right) \\
\frac{L_r}{L} = \sin \left( {\varphi  - \frac{\pi }{2}} \right) = -\mathrm{cn} \left( \sqrt {\frac{\varepsilon }{m}} r,m \right)
\end{gather}

\noindent for the easy-axis anisotropy (${K_u} > 0$).
It can be rewritten with $\lambda$ using the relation~Eq.~\ref{eq:lambda} as

\begin{align}
\frac{L_z}{L} &= \mathrm{cn} \left( \frac{4 \mathrm{K}(m) }{\lambda} r,m \right), & \frac{L_r}{L} &= \mathrm{sn} \left( \frac{4 \mathrm{K}(m) }{\lambda} r,m \right)
\end{align}

\noindent for the easy-plane anisotropy and

\begin{align}
\frac{L_z}{L} &= \mathrm{sn} \left( \frac{4 \mathrm{K}(m) }{\lambda} r,m \right), & \frac{L_r}{L} &= -\mathrm{cn} \left( \frac{4 \mathrm{K}(m) }{\lambda} r,m \right)
\end{align}

\noindent for the easy-axis anisotropy.

\section*{Acknowledgements}

It has been our great pleasure to work with all of our collaborators over the past few years, who have enormously enriched our understanding of this unusual material. However, we should like to acknowledge a few people in particular for their help. First, we would like to thank Sang-Wook Cheong, who not only brought BiFeO$_3$ to the attention of one of us (JGP) but also provided the high-quality single crystals which we used for the high-resolution neutron scattering experiments. Shunskuke Furukawa helped us in analyzing the spin waves and Gunsang Jeon carried out the Monte-Carlo simulation at the beginning of this project: both have played an important role in helping us build a {\it standard model} of BiFeO$_3$, which has been presented in this review. We should also thank Y. Noda for his kind reading of the manuscript and comments, and J. H. Lee and R. S. Fishman for communicating their unpublished results to us. Finally, we should acknowledge the authors of the papers from which we reproduced the figures used in this review and the publishers (APS, AIP, and IOP) for their generous permissions to reproduce such figures. Finally, whilst we have tried to be comprehensive, the large body of work on bulk BiFeO$_3$ has inevitably meant that we have missed some publications. This work was supported by the Research Center Program of IBS (Institute for Basic Science) in Korea: Grant No. EM1203.


\bibliographystyle{apsrev4-1}     
 \bibliography{bfo_review-refs}        

\begin{thebibliography}{187}
\bibitem [{not()}]{note1}%
  \BibitemOpen
  \href@noop {} {}\bibinfo {note} {Although non-standard, the
  "proper"/"improper" terminology, deriving from the description of
  ferroelectrics, is gradually coming into usage. A proper ferroelectric has
  the polarisation as the primary order parameter, whereas in an improper
  ferroelectric, the polarisation is a secondary (induced) order parameter.
  Type II multiferroics, where the polarisation is induced by magnetic order is
  clearly an improper ferroelectric, whereas type I multiferroics (the case of
  BiFeO$_3$ where neither magnetic nor electric orders induce the other) may be
  proper or improper ferroelectrics. For example, both BiFeO$_3$ and hexagonal
  RMnO$_3$ belong to the type I group although BiFeO$_3$ is often thought of as
  a proper ferroelectric whereas RMnO$_3$ is known as an improper
  ferrlelectric.}\BibitemShut {Stop}%
\end{thebibliography}


\end{document}